\documentclass[12pt,a4paper]{article}
\usepackage[T1]{fontenc}
\usepackage[sc,osf]{mathpazo}
\usepackage{a4wide}
\usepackage{amsfonts}
\usepackage{amssymb}
\usepackage{amsmath}
\usepackage{bbm}
\usepackage{booktabs} 
\usepackage{ifpdf}
\ifpdf
\usepackage[pdftex,unicode,implicit]{hyperref}
\hypersetup{%
  pdftitle    = {On the dualization of scalars into (d-2)-forms in supergravity}
  pdfkeywords = {duality, U-duality, electric-magnetic duality, p-forms,
    branes, supergravity, gaugings, embedding tensor, momentum map, coset
    spaces, symmetric spaces, non-linear sigma models},
  pdfauthor   = {Igor A. Bandos and Tomas Ortin},
  plainpages  = true,
  colorlinks  = true,
  citecolor   = blue,
  urlcolor    = red,
  linkcolor   = black
}
\newcommand{\hepth}[1]{{\tt
\href{http://www.arXiv.org/abs/hep-th/#1}{hep-th/#1}}}

\newcommand{\arxiv}[1]{{\tt
\href{http://www.arXiv.org/abs/#1}{#1}}}
\newcommand{\doi}[1]{{\tt DOI:\href{http://dx.doi.org/#1}{#1}}}
\else
  \usepackage[dvips]{graphicx}
  \usepackage[unicode,implicit]{hyperref}
  \newcommand{\hepth}[1]{{\tt hep-th/#1}}

  \newcommand{\arxiv}[1]{{\tt arXiv:#1}}
  \newcommand{\doi}[1]{{\tt DOI:#1}}
\fi
\makeatletter
\@addtoreset{equation}{section}
\makeatother

\begin{document}

\begin{flushright}
\small
IFT-UAM/CSIC-15-022\\
\texttt{arXiv:1605.05559 [hep-th]}\\
May 18\textsuperscript{th}, 2016, $\;$\\
\normalsize
\end{flushright}

\vspace{1.5cm}

\begin{center}

{\Large {\bf On the dualization of scalars into $(d-2)$-forms in supergravity}}

\vspace{.3cm}

{\large {Momentum maps, R-symmetry and gauged supergravity}}

\vspace{1cm}

\renewcommand{\thefootnote}{\alph{footnote}}
{\sl\large  Igor A.~Bandos$^{1,2}$}${}^{,}$\footnote{E-mail: {\tt Igor.Bandos [at] ehu.eus}}
{\sl\large  and Tom\'as Ort\'{\i}n$^{3}$}${}^{,}$\footnote{E-mail: {\tt Tomas.Ortin [at] csic.es}}

\setcounter{footnote}{0}
\renewcommand{\thefootnote}{\arabic{footnote}}

\vspace{1cm}

${}^{1}${\it Department of Theoretical Physics, University of the Basque Country UPV/EHU,\\
P.O. Box 644, 48080 Bilbao, Spain}
\\ [.3cm]
${}^{2}${\it IKERBASQUE, Basque Foundation for Science, 48011 Bilbao, Spain}
\\[.3cm]
${}^{3}${\it Instituto de F\'{\i}sica Te\'orica UAM/CSIC\\
C/ Nicol\'as Cabrera, 13--15,  C.U.~Cantoblanco, E-28049 Madrid, Spain}

\vspace{1cm}


{\bf Abstract}

\end{center}

\begin{quotation}
  {\small We review and investigate different aspects of scalar fields in
    supergravity theories both when they parametrize symmetric spaces and when
    they parametrize spaces of special holonomy which are not necessarily
    symmetric (K\"ahler and Qua\-ternionic-K\"ahler spaces): their r\^ole in the
    definition of derivatives of the fermions covariant under the R-symmetry
    group and (in gauged supergravities) under some gauge group, their
    dualization into $(d-2)$-forms, their role in the supersymmetry
    transformation rules (via fermion shifts, for instance) etc. We find a
    general definition of momentum map that applies to any manifold admitting
    a Killing vector and coincides with those of the holomorphic and
    tri-holomorphic momentum maps in K\"ahler and quaternionic-K\"ahler spaces
    and with an independent definition that can be given in symmetric
    spaces. We show how the momentum map occurs ubiquitously: in
    gauge-covariant derivatives of fermions, in fermion shifts, in the
    supersymmetry transformation rules of $(d-2)$-forms etc.  We also give the
    general structure of the Noether-Gaillard-Zumino conserved currents in
    theories with fields of different ranks in any dimension.
}
\end{quotation}

\newpage
\pagestyle{plain}

\tableofcontents

\newpage

\section*{Introduction}

One of the main features of supergravity theories is the presence of scalar
fields. In many cases this presence can be traced to a compactification of
some higher-dimensional supergravity theory and, then, the scalars encode a
great deal of information about the moduli space of the compactification. In
gauged supergravity theories, the scalar potential gives rise to symmetry and
supersymmetry breaking and identify possible vacua, and can be used to construct 
inflationary models. Furthermore, the gravitating solutions of supergravity
theories can (or must, depending on the case) have active scalars. For
instance, the supergravity generalizations of the Reissner-Nordstr\"om black
hole have them, giving rise to very interesting phenomena such as the
attractor mechanism
\cite{Ferrara:1995ih,Strominger:1996kf,Ferrara:1996dd,Ferrara:1996um}. Actually,
the fact that their values at infinity (which can be interpreted as their
vacuum expectation values) do not occur in the charged black-hole entropy
formula \cite{Ferrara:1997tw} is, certainly, a major indication of the
existence of a microscopic interpretation for the black-hole entropy. On top
of this, the relation of the scalars with compactification moduli plays a
fundamental r\^ole in the microscopic interpretation of the black-hole entropy
in the context of string theory \cite{Strominger:1996sh}. But there are
solutions much more directly related to scalar fields: these are the domain
walls and the $(d-3)$-brane solutions. We will discuss the latter later.

Scalar fields are, therefore, not just a nuisance one has to live with in
supergravity, but a blessing, a fantastic tool whose use one has to master, in
spite of the fact that, so far, we have only found one scalar field in Nature.

Scalar fields can be coupled non-linearly among themselves (non-linear
$\sigma$-models and scalar potentials) or to other fields (scalar-dependent
kinetic matrices) in very simple ways, without having to include terms of
higher-orders in derivatives, because their transformations do not contain
spacetime derivatives (even if they couple to a gauge field). Using this
property one can rewrite theories of higher-order in derivatives of other
fields (for instance, theories of gravity with corrections of higher-order in
curvature) as standard quadratic theories with couplings to scalar fields. A
well-known example is the equivalence between $f(R)$ theories of gravity and
Jordan-Brans-Dicke scalar-tensor theories of gravity.

This versatility of scalar fields comes at a price, though, and the
non-linearities create their own problems. In this paper we want to address
specially one of them: that of the dualization of scalars into $(d-2)$-form
potentials.

A it is well known, supergravity theories, as the low-energy, effective
field-theory limits of superstring theories, contain a great deal of
information about the $p$-dimensional extended objects (\textit{branes}) that
occur in the latter. This information is encoded in the $(p+1)$-form
potentials they electrically couple to. For $p\leq d/2-2$, these fields appear
in the supergravity action as fundamental fields. For higher values, though,
one has to consider their electric-magnetic duals.

In most cases, the supergravity theory cannot be completely reformulated in
terms of the dual supergravity fields: even their field strengths can only be
defined using the fundamental ones. We have learned to deal with all of them
at the same time using the so-called ``democratic formulations''
\cite{Bergshoeff:2001pv}
or PST-type duality-symmetric actions \cite{Bandos:1997gd,Dall'Agata:1998va}
constructed with the use of Pasti-Sorokin-Tonin approach
\cite{Pasti:1995ii,Pasti:1995tn}.\footnote{Observe that the democratic
  formulation of Ref.~\cite{Bergshoeff:2001pv} does not have manifest
  SL$(2,\mathbb{R})$ invariance in the IIB sector because only the RR 6- and
  8-forms are considered and they are part of a doublet and a triplet. In this
  sense it is incomplete. The more complicated PST-type type IIB supergravity
  action of \cite{Dall'Agata:1998va} contains the complete set of higher
  forms.}
  
The technical reason is that, typically, the supergravity action contains
potentials without derivatives and the standard procedure for dualization
requires, as a first step, the replacement of the potentials by their field
strengths as independent variables in the action. This problem is more acute
for scalar fields, because they generically appear without derivatives in the
$\sigma$-model metric and in the kinetic matrices.

When the $\sigma$-model metric admits an isometry, it is possible to make it
independent of the associated scalar coordinate by a change of variables. If
the isometry is a global symmetry of the theory, the kinetic matrices may also
be independent of that scalar too and, then, one could dualize it into a
$(d-2)$-form potential. One could repeat the procedure for additional
commuting isometries but most $\sigma$-models do not have as many commuting
isometries as scalar fields, even if they have more isometries than scalar
fields, as it happens in $\mathcal{N}>2,d=4$ supergravities, whose scalars
parametrize symmetric Riemannian spaces. How should one proceed in that case?

An additional problem is that we expect the dualization procedure to preserve
all the duality symmetries of the theory, \textit{i.e.}~all the symmetries of
the equations of motion, including those that do not leave the action
invariant. This implies that it is not enough to dualize the scalars (even if
possible), since they do not transform in linear representations of the
duality group and the dual $(d-2)$-form potentials can only transform
linearly.

The basic idea to solve this problem was proposed in
Ref.~\cite{Meessen:1998qm} for the case of the SL(2,$\mathbb{R}$)/SO(2)
$\sigma$-model that occurs in $\mathcal{N}=2B, d=10$ supergravity (as well as
in many other theories): the objects to be dualized are not the scalars but
the Noether 1-forms $j_{A}=j_{A\, \mu}dx^{\mu}$ associated to the
symmetry\footnote{Here and in what follows the indices $A,B,C,\ldots$ run over
  the adjoint representation of the whole global symmetry group.}. In a
background metric $g_{\mu\nu}$, these are related to the Noether current
densities $\mathfrak{j}_{A}^{\mu}$ satisfying on-shell the continuity equation

\begin{equation}
\partial_{\mu} \mathfrak{j}_{A}^{\mu}=0\, ,
\end{equation}

\noindent
by

\begin{equation}
j_{A\, \mu} \equiv \frac{\mathfrak{j}_{A}{}^{\nu}g_{\nu\mu}}{\sqrt{|g|}}\, .
\end{equation}

\noindent
In terms of the Noether 1-forms, the continuity equations take the form

\begin{equation}
d\star j_{A}=0\, ,
\end{equation}

\noindent
and can be locally solved by introducing $(d-2)$-form potentials $B_{A}$ so
that

\begin{equation}
\star j_{A} = dB_{A}\, .
\end{equation}

The $(d-2)$-form fields $B_{A}$ are the duals of the
scalars.\footnote{Actually, using the embedding-tensor formalism, it can be
  argued that the $(d-2)$-form potentials of any field theory transform in the
  adjoint representation of the global symmetry group
  \cite{Bergshoeff:2009ph,Hartong:2009vc}. Some of the symmetries may not act
  on the scalars at all but, to simplify matters, we focus here on the
  symmetries of the scalar $\sigma$-model.}  In the kind of theories we are
interested in, the numbers of scalars and $(d-2)$-forms do not coincide in
general because there are more global symmetries than scalars. However, there
are constraints to be taken into account that reduce the number of
independent dynamical degrees of freedom associated to the latter, as we will
see.

When the scalars couple to other fields, these must transform under the global
symmetries of the $\sigma$-model as well. Some of the transformations may be
electric-magnetic dualities and only the equations of motion will be left
invariant by them. Accordingly, there are no Noether currents for those
transformations. As shown in Ref.~\cite{Bergshoeff:2009ph}, in the
4-dimensional case it is always possible to use the Noether-Gaillard-Zumino
(NGZ) 1-forms \cite{Gaillard:1981rj} which are conserved on-shell, to define
the 2-forms $B_{A}$. We will study the higher-dimensional, higher-rank analog
of the NGZ 1-forms in Section~\ref{sec-higherdimensionshigherrank}.

In order to describe systematically the procedure, it is convenient to start
by reviewing the construction of the metrics, Killing vectors, Vielbeins and
connection 1-forms etc. in symmetric spaces, since this is the kind of target
spaces that occurs in most extended supergravities. We will do this in
Section~\ref{sec-reviewsigmamodels}. In the process we will (re-) discover
structures which appear in the gauging of the theories (specially in the
supersymmetric case) (covariant derivatives, fermion shifts etc.)  In
particular, we are going to see that in all symmetric spaces there exists a
generalization of the holomorphic and triholomorphic momentum maps associated
to the K\"ahler-Hodge and quaternionic-K\"ahler manifolds of $\mathcal{N}=1,2$
supergravities in $d=4$ dimensions (reviewed in the Appendices) which play
exactly the same r\^ole in the construction of the gauge-covariant derivatives
of fermions, in the fermion shifts of the supersymmetry transformation rules
of gauged supergravities and also in the supersymmetry transformation rules of
the $(d-2)$-forms dual to the NGZ currents. These generalizations share the
same properties and deserve to be called momentum maps as well.

Furthermore, we are also going to give an even more general definition of
momentum map (Section~\ref{sec-basicmomentummap}), valid for any manifold
admitting one isometry, showing that in symmetric spaces, K\"ahler-Hodge or
quaternionic-K\"ahler spaces our general definition is equivalent to the
standard one. This is one of the main results of this paper.

To end Section~\ref{sec-reviewsigmamodels} we will review some well-known
examples which will be useful in what follows.

In Section~\ref{sec-dualization1} we address the dualization of the scalars of
a symmetric $\sigma$-model into $(d-2)$-form potentials along the lines
explained before. Then, in Section~\ref{sec-dualizationd=4} we will consider
the case in which the scalars of the symmetric $\sigma$-model are coupled to
the vector fields of a generic 4-dimensional field theory of the kind
considered by Gaillard and Zumino in Ref.~\cite{Gaillard:1981rj}, introducing
the NGZ current 1-form and studying its dualization into 2-forms. We will also
consider there (Section~\ref{sec-susytrans}) the general form of the
supersymmetry transformations of the 2-forms and the r\^ole played in them by
the momentum map. It is because of this r\^ole that we expect the tensions of
the strings that couple to the 2-forms to be determined by the momentum map
(Section~\ref{sec-tensions}). We will also show how the momentum map occurs in
the fermion shifts of the supersymmetry transformation rules of the fermions
of 4-dimensional extended supergravities (Section~\ref{sec-fermionshifts}).

The higher-dimensional case in which the scalars are also coupled to
potentials of different and higher ranks will be considered in
Section~\ref{sec-higherdimensionshigherrank} and we will show through examples
that the equation of conservation of the generalized NGZ current 1-form has a
universal form.

Our conclusions are contained in Section~\ref{sec-conclusions}.

\section{Review of symmetric $\sigma$-models}
\label{sec-reviewsigmamodels}

Let us consider\footnote{In this review we follow
  Refs.~\cite{Castellani:1991et,Ortin:2015hya} although a big part of material
  can be also found in the classical papers
  Refs.~\cite{Coleman:1969sm,Salam:1969rq,Volkov:1973vd,Ogievetsky73}.} a
homogeneous space M on which the Lie group G acts transitively, and where
H$\subset$G, topologically closed, is the isotropy subgroup. Then M is
homeomorphic to the coset space G/H of equivalence classes under right
multiplication by elements of H $\{g{\rm H}\}$ (G acts from the left on these
equivalence classes) and can be given the structure of a manifold of dimension
dim G$-$dim H. Furthermore, G can be seen as a principal bundle with base
space M=G/H, structure group H, and projection G$\rightarrow$ G/H.

In any homogeneous space G/H, the Lie algebra of G, as a vector space, can be
decomposed as the direct sum $\mathfrak{g}=\mathfrak{h}\,\oplus\,
\mathfrak{k}$, where $\mathfrak{h}$ is the Lie subalgebra of H and
$\mathfrak{k}$ is its orthogonal complement. By definition of subalgebra

\begin{equation}
\left[\mathfrak{h},\mathfrak{h} \right] \subset \mathfrak{h}.
\end{equation}

\noindent
G/H is said to be a reductive homogenous space if

\begin{equation}
\left[\mathfrak{k},\mathfrak{h} \right] \subset \mathfrak{k},
\end{equation}

\noindent
which means that $\mathfrak{k}$ is a representation space of H.  Finally, G/H
is said to be symmetric and $(\mathfrak{k},\mathfrak{h})$ is called a
symmetric pair if it is reductive and

\begin{equation}
\left[\mathfrak{k},\mathfrak{k} \right] \subset \mathfrak{h}.
\end{equation}

The two components of a symmetric pair are mutually orthogonal with respect to
the Killing metric which is block-diagonal.

Now, if G/H is a symmetric space (G connected, and H compact) and there is a
G-invariant metric defined on it, then it is Riemannian symmetric space.

The metrics of the scalar $\sigma$-models that appear in all supergravities in
$d\geq 4$ dimensions with more than 8 supercharges are the metrics of some
Riemannian symmetric space. The metrics and the kinetic terms can be
constructed using a G/H coset representative or by using a generic element of
G and gauging an H subgroup. Let us start by reviewing the first method.

\subsection{Coset representative}

Let us introduce some notation: we denote by $\{T_{A}\}$, $\{M_{i}\}$ and
$\{P_{a}\}$ (where $A,B,\ldots =1,\ldots\!,{\rm dim}\, $G, $i,j,\ldots
=1,\ldots\!,{\rm dim}\,$H and $a,b,\ldots=1,\ldots\!,d\equiv{\rm dim}\, {\rm
  G}-{\rm dim}\, {\rm H}$), three bases of, respectively,
$\mathfrak{g},\mathfrak{h}$ and $\mathfrak{k}$ with $\{T_{A}\}=\{M_{i}\}\cup
\{P_{a}\}$ (see Ref.~\cite{Cremmer:1979up,de Wit:1981eq} and also
Ref.~\cite{Galperin:2001uw}). The structure constants are defined by

\begin{equation}
[T_{A},T_{B}] =f_{AB}{}^{C}T_{C}\, ,
\end{equation}

\noindent
and, by definition of symmetric space, the only non-vanishing components are

\begin{equation}
[M_{i},M_{j}] =f_{ij}{}^{k}M_{k}\, ,
\hspace{1cm}
[P_{a},M_{i}]=f_{ai}{}^{b}P_{b}\, ,
\hspace{1cm}
[P_{a},P_{b}]=f_{ab}{}^{i}M_{i}\, .
\end{equation}

The adjoint representation of $\mathfrak{g}$ is defined by the matrices

\begin{equation}
\Gamma_{\rm Adj}(T_{A})^{B}{}_{C} \equiv f_{AC}{}^{B} \, ,
\end{equation}

\noindent
and, obviously, their restriction to the indices $i,j$ is the adjoint
representation of $\mathfrak{h}$. Furthermore, the matrices

\begin{equation}
\Gamma (M_{i})^{a}{}_{b} = f_{ib}{}^{a}\, ,
\end{equation}

\noindent
provide another representation of $\mathfrak{h}$ with representation space
$\mathfrak{k}$.

Only the diagonal blocks of the the Killing metric

\begin{equation}
K_{AB}
\equiv
\mathrm{Tr}\left[\Gamma_{\rm Adj}(T_{A})\Gamma_{\rm Adj}(T_{A})\right]
=
f_{AC}{}^{D}f_{BD}{}^{C}\, ,
\end{equation}

\noindent
$K_{ab}$ and $K_{ij}$ are non-vanishing ($K_{ai}=0$). $K_{ij}$ is the
restriction of the Killing metric\footnote{That is: $K_{ij}=
  f_{iC}{}^{D}f_{jD}{}^{C} = f_{ik}{}^{l}f_{jl}{}^{k}
  +f_{ia}{}^{b}f_{jb}{}^{a}$.} of G to H and, for the kind of groups we are
considering, it is proportional to the Killing metric of H.\footnote{Because
  $f_{ia}{}^{b}f_{jb}{}^{a}=\mathrm{Tr}[\Gamma(M_{i})\Gamma(M_{j})] \propto
  f_{ik}{}^{l}f_{jl}{}^{k}$.} Under the adjoint action of G, defined by

\begin{equation}
\label{eq:adjointaction}
g^{-1}T_{A} g  \equiv T_{B} \Gamma_{\rm Adj}(g^{-1})^{B}{}_{A}\, ,
\end{equation}

\noindent
The Killing metric metric is invariant due to the cyclic property of the trace

\begin{equation}
K_{AB}
=
\mathrm{Tr}
\left[
\Gamma_{\rm Adj}(g^{-1}T_{A}g
g^{-1}T_{B} g)
\right]
=
K_{CD} \Gamma_{\rm Adj}(g^{-1})^{C}{}_{A} \Gamma_{\rm Adj}(g^{-1})^{D}{}_{B}\, .
\end{equation}

\noindent
Then, since the Killing metric is invertible ($K^{AB}$) for the kind of groups
we are considering, we find that

\begin{equation}
K_{BC}\Gamma_{\rm Adj}(g)^{C}{}_{D} K^{DA} = \Gamma_{\rm
  Adj}(g^{-1})^{A}{}_{B}\, .
\end{equation}

Let us denote by $u(\phi)=u(\phi^{1},\ldots\!,\phi^{d})$ a coset
representative of G/H in some local coordinate patch. In practice it will be a
matrix transforming in some representation $r$ of G. The scalar fields of the
$\sigma$-model will be mappings from spacetime to G/H expressed in these
coordinates as the functions $\phi^{m}(x)$.  Under a left transformation
$g\in$G, $u(\phi)$ transforms into another element of G, which becomes a coset
representative $u(\phi^{\prime})$ only after a right transformation with the
inverse of $h\in$H, that is (see
Refs.~\cite{Coleman:1969sm,Salam:1969rq,Volkov:1973vd,Ogievetsky73})

\begin{equation}
gu(\phi) = u(\phi^{\prime}) h\, .
\end{equation}

\noindent
For a given choice of coset representative $u$, $h$ will depend on $g$ and
$\phi$, but we will not indicate explicitly that dependence.

The left-invariant Maurer--Cartan 1-form $V \in\mathfrak{g}$ and can be
expanded as follows:\footnote{The elements of the basis of $\mathfrak{k}$ and
  $\mathfrak{h}$ will be in the representation $r$ in which $u$
  transforms. However, for the sake of simplicity, we will write $P_{a}$
  instead of $\Gamma_{r}(P_{a})$ etc. whenever this does not lead to
  confusion. }

\begin{equation}
\label{eq:M-Cdef}
V \equiv - u^{-1}du = e^{a}P_{a}+\vartheta^{i}M_{i}\, .
\end{equation}

\noindent
The $e^{a}$ components can be used as Vielbeins in G/H and the $\vartheta^{i}$
components play the role of connection\footnote{See the transformation rules
  Eqs.~(\ref{eq:transformation1}).}. The Maurer--Cartan equations satisfied by
$V$ ($dV -V\wedge V=0$) take the following form in terms of the above 1-form
components:

\begin{eqnarray}
\label{eq:MCe}
de^{a} -\vartheta^{i}\wedge e^{b}f_{ib}{}^{a}
& = &
0\, ,
\\
& & \nonumber \\
\label{eq:MCt}
d\vartheta^{i} -\tfrac{1}{2}\vartheta^{j}\wedge \vartheta^{k}f_{jk}{}^{i}
-\tfrac{1}{2}e^{b}\wedge e^{c}f_{bc}{}^{i}
& = &
0\, .
\end{eqnarray}

\noindent
Comparing the first of these equations with Cartan's structure equation with
vanishing torsion $\mathcal{D}e^{a}=de^{a} +\omega_{b}{}^{a}\wedge e^{b}=0$ we
find the connection 1-form

\begin{equation}
\label{eq:omegaconnection}
\omega_{b}{}^{a} = -\vartheta^{i}f_{ib}{}^{a}\, ,
\end{equation}

\noindent
which also justifies the identification of $\vartheta^{i}$ with a
connection. The curvature 2-form of this connection is, from the definition

\begin{equation}
\label{eq:Rtheta1}
R_{b}{}^{a}(\omega) = -R(\vartheta)^{i}f_{ib}{}^{a}\, ,
\,\,\,\,\,
\mbox{where}
\,\,\,\,\,
R(\vartheta)^{i}
\equiv
d\vartheta^{i} -\tfrac{1}{2}\vartheta^{j}\wedge \vartheta^{k}f_{jk}{}^{i}\, .
\end{equation}

\noindent
Then, Eqs.~(\ref{eq:MCt}) tell us that

\begin{equation}
\label{eq:Rtheta}
R(\vartheta)^{i}
=
\tfrac{1}{2}e^{b}\wedge e^{c}f_{bc}{}^{i}\, ,
\hspace{1cm}
R_{b}{}^{a}(\omega)
=
-\tfrac{1}{2}e^{d}\wedge e^{e}f_{de}{}^{i}f_{ib}{}^{a}\, .
\end{equation}

We have defined a Vielbein basis and an affine connection on G/H, but we have
not defined a Riemannian metric yet. We can do so if we are provided with a
metric $g_{ab}$ on $\mathfrak{k}$:

\begin{equation}
ds^{2} = g_{ab}e^{a}\otimes e^{b}=
g_{ab}e^{a}{}_{m}e^{b}{}_{n}d\phi^{m}d\phi^{n}
\equiv \mathcal{G}_{mn}(\phi)d\phi^{m}d\phi^{n}\, .
\end{equation}

\noindent
Its pullback over spacetime, conveniently normalized, can be used in the
action for the $\sigma$-model:

\begin{equation}
\label{eq:sigmamodelaction1}
S
=
\tfrac{1}{2}\int  \mathcal{G}_{mn}(\phi)
d\phi^{m}\wedge \star d\phi^{n}\, .
\end{equation}

In order to construct a Riemannian symmetric $\sigma$-model the metric
$\mathcal{G}_{mn}(\phi)$ must be invariant under the left action of G.  Under
the left multiplication by $g\in$G, $u(\phi^{\prime}) = gu(\phi)h^{-1}$, and
the components of the left-invariant Maurer--Cartan 1-form transform
in the adjoint representation of H (the $\vartheta^{i}$ as a connection of H):

\begin{equation}
\label{eq:transformation1}
\left\{
\begin{array}{rcl}
e^{a}(\phi^{\prime})
& = &
(h e(\phi) h^{-1})^{a}
= \Gamma_{\rm Adj}(h)^{a}{}_{b} e^{b}(\phi)\, ,
\\
& & \\
\vartheta^{i}(\phi^{\prime})
& = &
(h \vartheta(\phi) h^{-1})^{i} +(dhh^{-1})^{i}\, ,
\\
\end{array}
\right.
\end{equation}

\noindent
where $e(\phi)=e^{a}(\phi)P_{a}$ and
$\vartheta(\phi)=\vartheta^{i}(\phi)M_{i}$.  Infinitesimally,

\begin{equation}
h \sim 1 + \sigma^{i}(\phi)M_{i}\, ,\,\,\,\,
\Rightarrow\,\,\,\,\,
e^{a}(\phi^{\prime})
\sim
e^{a}(\phi) +\sigma^{i}(\phi)f_{ib}{}^{a}e^{b}(\phi)\, ,
\end{equation}

\noindent
and the Riemannian metric ${\cal G}_{mn}(\phi)$ will be invariant under the
left action of G if the metric $g_{ab}$ on $\mathfrak{k}$ is H-invariant:

\begin{equation}
\label{eq:HinvarianceBab}
f_{i(a}{}^{c}g_{b)c}=0\, .
\end{equation}

In all the relevant cases we can set $g_{ab}\propto K_{ab}$, the projection on
$\mathfrak{k}$ of the Killing metric and we will do so from now on. More
precisely, we will use this normalization:\footnote{We will sometimes use
  $g_{AB}= \mathrm{Tr}\left[T_{A}T_{B}\right]$.}

\begin{equation}
\label{eq:flatmetric}
g_{ab}= \mathrm{Tr}\left[P_{a}P_{b}\right]\, .
\end{equation}

Observe that the H-invariance of $g_{ab}$ Eq.~(\ref{eq:HinvarianceBab})
automatically guarantees that the torsionless connection $\omega_{b}{}^{a}$ in
Eq.~(\ref{eq:omegaconnection}) is metric-compatible and, therefore, it is the
Levi--Civita connection of the above metric.

In addition to the isometries corresponding to the left action of G, with
Killing vectors $k_{A}$, the resulting Riemannian metric is also invariant
under the right action of N(H)/H, N(H) being the normalizer of H in G. The
Killing vectors associated to the latter are just the vectors $e_{a}$ dual to
the horizontal Maurer--Cartan 1-forms in the directions of N(H)/H
\cite{Castellani:1991et}.

Our next task is to find the general expression of the Killing vectors $k_{A}$
which defines the transformation rule of the Goldstone fields. From the
infinitesimal version of $gu(\phi)=u(\phi^{\prime})h$ with

\begin{equation}
\label{eq:transformations}
 \begin{array}{rcl}
g
& = &
1 + \sigma^{A}T_{A}\, ,
\\
& & \\
h
& = &
1 -\sigma^{A}W_{A}{}^{i}M_{i}\, ,
\\
& & \\
\phi^{m\, \prime}
& = &
\phi^{m} +\sigma^{A}k_{A}{}^{m}\, ,
 \end{array}
\end{equation}

\noindent
where $W_{A}{}^{i}$ is known as the \textit{H-compensator}, we get after some
straightforward manipulations

\begin{eqnarray}
\label{eq:KV}
k_{A}{}^{a}
& = &
-\Gamma_{\rm Adj}(u^{-1}(\phi))^{a}{}_{A}\, ,
\\
& & \nonumber \\
\label{eq:Hcompensator}
W_{A}{}^{i}
& = &
-k_{A}{}^{m}\vartheta^{i}{}_{m}
-P_{A}{}^{i}\, ,
\end{eqnarray}

\noindent
where we have defined the \textit{momentum map} $P_{A}{}^{i}$

\begin{equation}
\label{eq:momentummap}
P_{A}{}^{i} \equiv \Gamma_{\rm Adj}(u^{-1}(\phi))^{i}{}_{A}\, .
\end{equation}

\noindent
It transforms as

\begin{equation}
P^{\prime}_{A}{}^{i}
=
\Gamma_{\rm Adj}(h(\phi))^{i}{}_{j} P_{B}{}^{j}
\Gamma_{\rm Adj}(g^{-1})^{B}{}_{A}\, ,
\end{equation}

\noindent
and it plays a crucial role in the gauging of the global symmetry group G, as
we are going to see below.

\subsubsection{H-covariant derivatives and the momentum map}
\label{sec-H-covariantderivative}

With the objects that we have found we can construct \textit{H-covariant
  derivatives} and \textit{H-covariant Lie derivatives}, which transform
covariantly under the compensating H transformations associated to global G
transformations.  Let us start by discussing the former.

In supergravity theories, H coincides with the R-symmetry group, under which
all the fermions transform, and, therefore, all the derivatives of fermions
must be H-covariant derivatives.  Under a global G transformation of the
scalars, these spacetime fields undergo an H scalar-dependent, compensating
transformation that can be contravariant $\xi^{\prime} = \Gamma_{s}(h)\xi$, or
covariant, $\psi^{\prime} = \psi\Gamma_{s}(h^{-1})$, in some representation
$s$ of H. For those fields, with the help of the pullback over the spacetime
of the H connection $\vartheta^{i}_{m}d\phi^{m}$,\footnote{We will denote the
  pullback with the same symbol, $\vartheta^{i}$ when this does not lead to
  confusion.} (see the second of Eqs.~(\ref{eq:transformation1})), we define
the H-covariant\footnote{In spite of the name, which is, admittedly,
  misleading, there is no true gauge symmetry in this construction. The
  H-transformations are not arbitrary functions of the spacetime
  coordinates. Neither they are arbitrary functions of the scalar fields (the
  coordinates on G/H). The only arbitrary parameters in these transformations
  are the global parameters of the G transformation that needs to be
  compensated to go back to the coset representative.} derivative
by

\begin{equation}
\label{eq:Hcovariantderivatives}
\mathcal{D}\xi\equiv d\xi
- \vartheta^{i}\Gamma_{s}(M_{i})\xi\, ,
\hspace{1cm}
\mathcal{D}\psi\equiv d\psi
+\psi\vartheta^{i}\Gamma_{s}(M_{i})\, .
\end{equation}

The H-covariant derivative satisfies the Ricci identities

\begin{equation}
\mathcal{D}^{2} \xi = -R(\vartheta)^{i}\Gamma_{s}(M_{i})\xi\, ,
\hspace{1cm}
\mathcal{D}^{2} \psi = \psi R(\vartheta)^{i}\Gamma_{s}(M_{i})\, ,
\end{equation}

\noindent
where $R(\vartheta)^{i}$ stands here for the curvature 2-form defined in
Eq.~(\ref{eq:Rtheta}).

Using

\begin{equation}
k_{A}{}^{a}k_{B}{}^{b} f_{ab}{}^{i}
+P_{A}{}^{j}P_{B}{}^{k} f_{jk}{}^{i}
=
\Gamma_{\rm Adj}(u^{-1})^{A^{\prime}}{}_{A}
\Gamma_{\rm Adj}(u^{-1})^{B^{\prime}}{}_{B}
f_{A^{\prime}B^{\prime}}{}^{i}
=
f_{AB}{}^{C}P_{C}{}^{i}\, ,
\end{equation}

\noindent
we find that  the momentum map that we have defined above satisfies the
equivariance condition

\begin{equation}
\mathcal{D}_{A}P_{B}{}^{i} -  \mathcal{D}_{B}P_{A}{}^{i}
-k_{A}{}^{a}k_{B}{}^{b} f_{ab}{}^{i}
+P_{A}{}^{j}P_{B}{}^{k} f_{jk}{}^{i}
= f_{AB}{}^{C}P_{C}{}^{i}\, ,
\end{equation}

\noindent
where $\mathcal{D}_{A} = k_{A}{}^{m}\mathcal{D}_{m}$. Using the explicit form
of the curvature Eq.~(\ref{eq:Rtheta}) it is easy to derive the following
equation, which is sometimes used as definition of the momentum map

\begin{equation}
\label{eq:DPRk}
\mathcal{D}_{m}P_{A}{}^{i}
=
-R_{mn}{}^{i}(\vartheta)k_{A}{}^{n}\, .
\end{equation}

One is often interested in gauging the global symmetry group G (or a subgroup
of G), making the supergravity theory invariant under transformations of the
form Eq.~(\ref{eq:transformations}) with parameters $\sigma^{A}$ which are
promoted to arbitrary spacetime functions $\sigma^{A}(x)$. Under these
transformations, the pullback of the second equation of
(\ref{eq:transformation1}) acquires an additional term and the (spacetime
pullback of) above H-covariant derivatives do not transform covariantly
anymore. As usual, it is necessary to introduce spacetime 1-forms $A^{A}$ and
modify the above covariant derivatives as follows:\footnote{Here we will not
  concern ourselves with the problem of matching the rank of the subgroup
  of G to be gauged with the number of 1-forms available in the supergravity
  theory. In general this requires the explicit or implicit introduction of
  the so-called \textit{embedding tensor}. We will use it in
  Section~\ref{sec-fermionshifts}.}

\begin{equation}
\label{eq:gaugecovariantderivatives}
\mathfrak{D}\xi\equiv d\xi
-\left(A^{A}P_{A}{}^{i}+\vartheta^{i}\right)\Gamma_{s}(M_{i})\xi\, ,
\hspace{1cm}
\mathfrak{D}\psi\equiv d\psi
+\psi\left(A^{A}P_{A}{}^{i}+\vartheta^{i}\right)\Gamma_{s}(M_{i})\, .
\end{equation}

The structure of these gauge covariant derivatives is identical to the
covariant derivatives that occur in gauged $\mathcal{N}=1,2$
supergravities\footnote{See, for instance,
  Refs.~\cite{Andrianopoli:1996cm,Freedman:2012zz,Ortin:2015hya}.}, even
if the scalar manifolds (K\"ahler--Hodge and quaternionic--K\"ahler
manifolds) are no coset spaces: the spinors of these theories
transform under U(1) K\"ahler transformations and the K\"ahler 1-form
connection plays the role of $\vartheta$ in
Eq.~(\ref{eq:Hcovariantderivatives}). In $\mathcal{N}=2$ theories with
hypermultiplets, the spinors also transform under SU(2) compensating
transformations and the pullback of the SU(2) connection of the
quaternionic-K\"ahler manifold plays the role of $\vartheta$.
Associated to these symmetries there are holomorphic and
tri-holomorphic momentum maps which play the same role as
$P_{A}{}^{i}$. A more detailed comparison between these structures and
the ones that arise in symmetric spaces can be found in the
appendices.

Observe that these covariant derivatives cannot be obtained by the often-used
(but generally wrong) replacement of the pullback by the ``covariant
pullback'' of the H connection $\vartheta^{i}$ in
Eqs.~(\ref{eq:Hcovariantderivatives})

\begin{equation}
\vartheta^{i}_{m}d\phi^{m}
\rightarrow
\vartheta^{i}_{m}\mathfrak{D}\phi^{m}\, ,
\end{equation}

\noindent
where

\begin{equation}
\mathfrak{D}\phi^{m}
\equiv
d\phi^{m} -A^{A}k_{A}{}^{m}\, ,
\end{equation}

\noindent
is the covariant derivative of the scalars, because, according to
Eq.~(\ref{eq:Hcompensator}),

\begin{equation}
\vartheta^{i}_{m}\mathfrak{D}\phi^{m}
=
\left(A^{A}P_{A}{}^{i}+\vartheta_{m}^{i}d\phi^{m}\right)
+A^{A}W_{A}{}^{i}\, ,
\end{equation}

\noindent
and the H-compensator does not vanish in general. Something similar happens in
the K\"ahler--Hodge and quaternionic--K\"ahler manifolds of $\mathcal{N}=1,2$,
$d=4$ supergravities \cite{Andrianopoli:1996cm,Ortin:2015hya}.

Using the identity Eq.~(\ref{eq:DPRk}) and other results derived in this
section one can compute the Ricci identities for the $\mathfrak{D}$ covariant
derivative

\begin{equation}
\begin{array}{rcl}
\mathfrak{D}^{2}\xi
& = &
-\left[
F^{A}P_{A}{}^{i}+
R(\vartheta)^{i}_{mn}\mathfrak{D}\phi^{m}\mathfrak{D}\phi^{n}
\right]
\Gamma_{s}(M_{i})\xi\, ,
\\
& & \\
\mathfrak{D}^{2} \psi
& = &
\psi
\left[
F^{A}P_{A}{}^{i}+R(\vartheta)^{i}_{mn}\mathfrak{D}\phi^{m}\mathfrak{D}\phi^{n}
\right]
\Gamma_{s}(M_{i})\, ,
\\
\end{array}
\end{equation}

\noindent
where

\begin{equation}
F^{A} = dA^{A} -\tfrac{1}{2}f_{BC}{}^{A}A^{B}\wedge A^{C}\, .
\end{equation}

\subsubsection{A more basic definition of the momentum map}
\label{sec-basicmomentummap}

Let us consider a $d$ Riemannian manifold $M$,\footnote{The signature is
  irrelevant in this discussion, which also applies to pseudo-Riemannian
  spaces.} not necessarily symmetric, but admitting a set of Killing vectors
$k_{A}{}^{a}$, that is

\begin{equation}
\nabla_{(a|}k_{A\, |b)}=0\, ,
\end{equation}

\noindent
where $\nabla_{a}$ is the Levi-Civita covariant derivative. To each
Killing vector one can associate an infinitesimal rotation in tangent
space generated by

\begin{equation}
P_{A}{}^{b}{}_{a}
\equiv
\nabla_{a}k_{A}{}^{b}\, .
\end{equation}

The antisymmetry of $P_{A\, ab}=g_{ac}P_{A}{}^{c}{}_{a}$ follows from
the Killing equation.  Let $\{M_{i}\}$ be a basis of the Lie algebra
of the holonomy group of M. Generically it will be $\mathrm{SO}(d)$
but for spaces of special holonomy it will be some subgroup
$\mathrm{H}\subset \mathrm{SO}(d)$. In particular, for Riemannian
homogeneous spaces G/H, the holonomy group is precisely H. We can, then,
decompose $P_{A}{}^{b}{}_{a}$ in that basis, defining at the same
time the coefficients as the components of the momentum map

\begin{equation}
P_{A}{}^{b}{}_{a}
\equiv
P_{A}{}^{i}\Gamma(M_{i})^{b}{}_{a}\, .
\end{equation}

It is not hard to show using the explicit expressions for the Killing vectors
and connection Eqs.~(\ref{eq:KV}) and (\ref{eq:omegaconnection}) that the
momentum map we have just defined reduces to the one defined in
Eq.~(\ref{eq:momentummap}) for symmetric spaces. Furthermore, using the
general identity for Killing vectors

\begin{equation}
\nabla_{a}\nabla_{b}k_{A}{}^{c}
=
k_{A}{}^{d}R_{dab}{}^{c}\, ,
\,\,\,\,\,
\Rightarrow
\,\,\,\,\,
\nabla_{a}P_{A}{}^{c}{}_{b}
=
k_{A}{}^{d}R_{dab}{}^{c}\, ,
\end{equation}

\noindent
and decomposing both sides of this equation in the basis of the
holonomy algebra\footnote{That is
\begin{equation}
R_{adb}{}^{c} = -R_{ad}{}^{i} \Gamma(M_{i})^{c}{}_{b}\, ,
\end{equation}
where the minus sign is chosen to match the sign of the H-curvature
$R(\vartheta)$ in symmetric spaces.
}, we get a general version of Eq.~(\ref{eq:DPRk})

\begin{equation}
\label{eq:dP=kRgeneral}
\nabla_{a}P_{A}{}^{i}
=
k_{A}{}^{d}R_{da}{}^{i}(\vartheta)\, .
\end{equation}

Finally, we can also show that the momentum map satisfies the
equivariance property

\begin{equation}
[P_{A},P_{B}]{}^{a}{}_{b}
=
f_{AB}{}^{C}P_{C}{}^{a}{}_{b}
-k_{A}{}^{c}k_{B}{}^{d}R_{cda}{}^{b}\, .
\end{equation}

Under an infinitesimal isometry

\begin{equation}
\delta_{\alpha}x^{m} = \alpha^{A}k_{A}{}^{m}\, ,
\end{equation}

\noindent
objects living in tangent space (vectors $\psi^{a}$, say\footnote{These
  variables arise naturally in $\mathcal{N}=1$ supersymmetric mechanics where
  one introduces $d$ scalar multiplets $x^{m}+\theta
  e^{m}{}_{a}\psi^{a})$. See,
  \textit{e.g.}~\cite{Gibbons:1993ap,vanHolten:1995qt} and references
  therein.}) transform as

\begin{equation}
\delta_{\alpha}\psi^{a}
=
\alpha^{A}(P_{A}{}^{a}{}_{b}+k_{A}{}^{m}\omega_{m}{}^{a}{}_{b})\psi^{b}\, ,
\end{equation}

\noindent
and, when gauging the isometry group, these compensating
transformations must be taken into account in the construction of the
covariant derivative, which is given by

\begin{equation}
\mathfrak{D}_{m}\psi^{a}
=
\nabla_{m}\psi^{a} -A^{A}{}_{m}P_{A}{}^{a}{}_{b}\psi^{b}
=
\partial_{m}\psi^{a}
-\left(A^{A}{}_{m}P_{A}{}^{a}{}_{b}+\omega_{m}{}^{a}{}_{b} \right)\psi^{b}\, .
\end{equation}

This covariant derivative reduces to the H-covariant derivative in
Eq.~(\ref{eq:gaugecovariantderivatives}) for symmetric spaces.  The
above definition can be generalized to objects transforming in other
representations of $\mathrm{SO}(n)$ in the obvious way.

\subsubsection{H-covariant Lie derivatives}

The H-compensator can be understood as the ``local'' parameter of the H
compensating transformation associated to the infinitesimal G transformation
generated by $T_{A}$ or the Killing vector $k_{A}$. To study the behavior
under G global transformations of fields transforming under these H
compensating transformations (something usually done through the Lie
derivative) it is necessary to take into account the latter. This requires
an H-covariant generalization of the standard Lie derivative with respect to
the Killing vector $k_{A}$ denoted by $\mathbb{L}_{k_{A}}$.\footnote{See
  Ref.~\cite{Ortin:2015hya}, which we follow here, and references therein. One
  could define the H-covariant Lie derivative with respect to any vector but
  the crucial Lie-algebra property Eq.~(\ref{eq:Liebracketproperty}) only
  holds for Killing vectors.} The equivariance property of the H-compensator

\begin{equation}
\label{eq:equivarianceHcompensator}
\mathcal{L}_{k_{A}}W_{B}{}^{i} -\mathcal{L}_{k_{B}}W_{A}{}^{i}
+W_{A}{}^{j}W_{B}{}^{k}f_{jk}{}^{i} = f_{AB}{}^{C}W_{C}{}^{i}\, ,
\end{equation}

\noindent
plays an essential role.

On fields transforming contravariantly $\xi^{\prime} = \Gamma_{s}(h)\xi$, or
covariantly $\psi^{\prime} = \psi\Gamma_{s}(h^{-1})$, the H-covariant
derivative is defined by

\begin{equation}
\mathbb{L}_{k_{A}}\xi
\equiv
\mathcal{L}_{k_{A}}\xi +W_{A}{}^{i}\Gamma_{s}(M_{i})\xi\, ,
\hspace{1cm}
\mathbb{L}_{k_{A}}\psi
\equiv
\mathcal{L}_{k_{A}}\psi -\psi W_{A}{}^{i}\Gamma_{s}(M_{i})\, .
\end{equation}

\noindent
The main properties satisfied by this derivative are

\begin{eqnarray}
\label{eq:Liebracketproperty}
[\mathbb{L}_{k_{A}},\mathbb{L}_{k_{B}}]
& = &
\mathbb{L}_{[k_{A},k_{B}]}\, ,
\\
& & \nonumber \\
\mathbb{L}_{k_{A}} e^{a}
& = & 0\, ,
\\
& & \nonumber \\
\label{eq:lastproperty}
\mathbb{L}_{k_{A}} u
& = &
\mathcal{L}_{k_{A}}u -u W_{A}{}^{i}M_{i} = T_{A} u\, .
\end{eqnarray}

Infinitesimally, the H connection $\vartheta^{i}$ transforms with the
H-covariant derivative of the transformation parameters. Thus, an appropriate
definition for its H-covariant Lie derivative would be

\begin{equation}
\label{eq:invarianceproperty}
\mathbb{L}_{k_{A}}\vartheta^{i}
\equiv
\mathcal{L}_{k_{A}}\vartheta^{i} +\mathcal{D}W_{A}{}^{i}\, .
\end{equation}

\noindent
Using the definitions and Eq.~(\ref{eq:DPRk}), one can show that

\begin{equation}
\mathbb{L}_{k_{A}}\vartheta^{i}=0\, .
\end{equation}

\subsubsection{Final remarks}

The H-covariant derivative of $u(\phi)$, which transforms covariantly in some
representation $r$, is, according to the definition

\begin{equation}
\label{eq:Du}
\mathcal{D}u
\equiv
du +u\vartheta^{i}M_{i}
=
u
\left[
u^{-1}du +u\vartheta^{i}M_{i}
\right]
=
-u e^{a}P_{a} \, ,
\end{equation}

\noindent
where we have used the expansion of the Maurer--Cartan 1-form
Eq.~(\ref{eq:M-Cdef}).

We can use this result to obtain a very convenient expression of the action of
the $\sigma$-model directly in terms of the coset representative $u(\phi)$,
which transforms linearly under G (with the metric defined in
Eq.~(\ref{eq:flatmetric})):\footnote{Using the cyclic property of the trace,
  it can also be written in the form
\begin{equation}
\label{eq:sigmamodelaction3}
S
=
\tfrac{1}{2}\int \mathrm{Tr}[\mathcal{D}u u^{-1}\wedge \star \mathcal{D}uu^{-1}]\, .
\end{equation}
}

\begin{equation}
\label{eq:sigmamodelaction2}
S
=
\tfrac{1}{2}\int \mathrm{Tr}[u^{-1}\mathcal{D}u \wedge \star u^{-1}\mathcal{D}u]\, .
\end{equation}

The invariance under the left action of G on the coset representative is
manifest in this form. This expression connects this approach with the
approach that we are going to review in the next section.

To end this section, when the coset space is of the kind SL($n$)/SO($n$),
there is an alternative but completely equivalent construction which is often
used in supergravity\footnote{This coset arises naturally in toroidal
  compactifications.}. One defines the symmetric and H-invariant matrix

\begin{equation}
\label{eq:M=uuT}
\mathcal{M} \equiv u u^{\rm T}\, ,
\hspace{1cm}
\mathcal{M}^{\prime}
=
\mathcal{M} (\phi^{\prime})
=
g\mathcal{M}(\phi)g^{T}\, ,
\end{equation}

\noindent
and choosing a basis in which the $P_{a}$ are symmetric matrices and the
$M_{i}$ are antisymmetric, it is not difficult to show that

\begin{equation}
\mathrm{Tr}[\mathcal{M}^{-1}d \mathcal{M} \wedge \star \mathcal{M}^{-1}d\mathcal{M}]
=
2 \mathrm{Tr}[u^{-1}\mathcal{D}u \wedge \star u^{-1}\mathcal{D}u]\, .
\end{equation}

The equations of motion from this action are obtained much more easily from
the formulation that we are going to discuss in the next section because one
does not have to deal with the scalar dependence of the connection
$\vartheta^{i}$.

\subsection{Gauging of an H subgroup}

$\sigma$-models on coset spaces G/H are often constructed by gauging a
subgroup H of a $\sigma$-model constructed on the group manifold G. The latter
has the action

\begin{equation}
\label{eq:sigmamodelactionG}
S_{\rm G}[u]
=
\tfrac{1}{2}\int \mathrm{Tr}[u^{-1}du \wedge \star u^{-1}du]\, ,
\end{equation}

\noindent
where $u(\varphi) =u(\varphi^{1},\cdots,\varphi^{\rm dim\, G})$ is now a
generic element of the group G in some representation $r$ and in some local
coordinate patch that contains the identity. This action is invariant under
the left and right (global) action of the group G and, therefore, its global
symmetry group is G$\times$G.

Now we want to gauge a subgroup H of the right symmetry group, under which
$u^{\prime}= u h^{-1}(x)$. Here $h(x)$ stands for an arbitrary function of the
spacetime coordinates that gives an element of H at each point. Such a
function can be constructed by exponentiation of a linear combination the
generators of $\mathfrak{h}$ with coefficients $\sigma^{i}(x)$ which are
arbitrary functions of the spacetime coordinates.  After gauging, the global
symmetry group will be broken to G$\times$H.

We introduce an $\mathfrak{h}$-valued spacetime gauge field
$A=A^{i}{}_{\mu}dx^{\mu}M_{i}$ transforming exactly as the $\vartheta^{i}$
components of the left-invariant Maurer--Cartan 1-form of G/H
Eq.~(\ref{eq:transformation1}) and the H-covariant derivative

\begin{equation}
\mathcal{D}u \equiv du +uA\, ,
\end{equation}

\noindent
and we simply replace the exterior derivative $d$ by $\mathcal{D}$ in the
action Eq.~(\ref{eq:sigmamodelactionG}) without adding a kinetic term for the
gauge field:

\begin{equation}
\label{eq:sigmamodelactionGauged}
S_{\rm Gauged} [u,A]
=
\tfrac{1}{2}\int
\mathrm{Tr}[u^{-1}\mathcal{D}u \wedge \star u^{-1}\mathcal{D}u]\, .
\end{equation}

The gauged action describes only $d=$dim\, G$-$dim\, H degrees of freedom: on
general grounds we expect that the gauge symmetry can be used to eliminate
dim\, H of the scalars $\varphi^{x}$, $x=1,\ldots,$dim\, G, from the action,
leaving only those that parametrize the coset space G/H, that we have denoted
by $\phi^{m}$.\footnote{The representation of the coset element by a group
  element defined modulo H-transformations is also characteristic of the
  harmonic superspace approach \cite{Galperin:1984av,Galperin:2001uw} as well
  as of the spinor moving frame formalism
  \cite{Bandos:1992np,Bandos:1993yc}. See Ref.~\cite{Bandos:2016tsm} for a
  recent application and more references.}  This is the so-called
\textit{unitary gauge}.  As we are going to see, only $d$ of the scalar
equations of motion are independent, in complete agreement with the general
expectation.

Let us first consider the equations of motion of the gauge field. These are
purely algebraic:

\begin{equation}
\frac{\delta S_{\rm Gauged}}{\delta A^{i}}
=
\star  \mathrm{Tr}[M_{i} u^{-1}\mathcal{D}u]=0\, ,
\end{equation}

\noindent
and the solution is

\begin{equation}
A^{i}= V^{i}\, ,
\end{equation}

\noindent
where $V=-u^{-1}du$ is the left-invariant Maurer--Cartan 1-form in G.  In the
unitary gauge, $V$ depends only on the physical scalars $\phi^{m}$ and
becomes, automatically, the left-invariant Maurer--Cartan 1-form in G/H, so
$V^{i}=\vartheta^{i}$.

This solution can be substituted in the above action: this substitution
$A^{i}=V^{i}(\varphi)$ and the derivation of the equations of motion for the
scalars from the action are two operations that commute and the final result
is the same. As a matter of fact, after the substitution, the scalar equations
of motion are

\begin{equation}
\frac{\delta S_{\rm Gauged}[\varphi,V(\varphi)]}{\delta \varphi^{x}}
=
\left. \frac{\delta S_{\rm Gauged}[\varphi,A]}{\delta\varphi^{x}}\right|_{A^{i}=V^{i}}
+
\left. \frac{\delta S_{\rm Gauged}[\varphi,A]}{\delta A^{i}}\right|_{A^{i}=V^{i}}
\frac{\delta  V^{i}}{\delta \varphi^{x}} \, ,
\end{equation}

\noindent
but the second term vanishes identically.

In the unitary gauge, after the substitution $A^{i}=V^{i}=\vartheta^{i}(\phi)$
we recover the action Eq.~(\ref{eq:sigmamodelaction2}). Classically, these two
formulations are completely equivalent. However, in this formulation the
scalar equations of motion are easier to derive because we have just shown
that we can ignore the variations of the connection with respect to the
scalars.

The equations of motion of the scalars $\varphi^{x}$, $x=1,\ldots,$dimG
are

\begin{equation}
\frac{\delta S_{\rm Gauged}}{\delta \varphi^{x}}
=
2 V_{x}{}^{A} \mathrm{Tr}[T_{A} \mathcal{D}\star(u^{-1}\mathcal{D}u)]
=
0\, .
\end{equation}

The invariance under local, right, H-transformations implies, according to
Noether's second theorem, the following dimH Noether identities relating the
scalar equations of motion:

\begin{equation}
\label{eq:gaugeidentities}
 \mathrm{Tr}[M_{i} \mathcal{D}\star(u^{-1}\mathcal{D}u)]=0\, .
\end{equation}

\noindent
These are off-shell identities and are also valid in the unitary gauge after
the substitution $A^{i}=\vartheta^{i}$. Therefore, they are valid in the case
discussed in the previous section.

Taking into account the gauge identities the only non-trivial equations of
motion are those of the physical scalars $\phi^{m}$, which take the form

\begin{equation}
\label{eq:equationsofmotionscalars}
\frac{\delta S_{\rm Gauged}}{\delta \phi^{m}}
=
e^{a}{}_{m} \mathrm{Tr}[P_{a} \mathcal{D}\star(u^{-1}\mathcal{D}u)]
=
e^{a}{}_{m}g_{ab}\mathcal{D}\star e^{b}
=
0\, .
\end{equation}

\subsection{Examples}

In this subsection we are going to review a few examples which we will use
repeatedly in what follows:

\begin{enumerate}
\item The $\mathrm{SL}(2,\mathbb{R})/\mathrm{SO}(2)$ coset space, which occurs in
  many supergravities: $\mathcal{N}=2B,d=10$ supergravity (the effective field
  theory of the type~IIB superstring) $\mathcal{N}=2,d=9,8$ supergravity
  (obtained from the former by toroidal dimensional reduction),
  $\mathcal{N}=4,d=4$ supergravity (the effective field theory of the
  heterotic string compactified on a six torus), and in many truncations of
  the maximal supergravities in diverse dimensions. In $\mathcal{N}=2B,d=10$
  supergravity the scalar fields that parametrize this coset are the dilaton
  $\varphi$ and RR 0-form $\chi$, combined into the axidilaton field
  $\tau=\chi+ie^{-\varphi}$ (see Eq.~(\ref{eq:axidilaton}) below). In
  $\mathcal{N}=4,d=4$ supergravity the fields are the 4-dimensional dilaton
  $\phi$ and the dual of the 4-dimensional Kalb-Ramond 2-form $a$ and
  $\tau=a+ie^{-2\phi}$.

\item The $\mathrm{SU}(1,1)/\mathrm{U}(1)$ coset, which is an often used
  completely equivalent alternative form of the
  $\mathrm{SL}(2,\mathbb{R})/\mathrm{SO}(2)$ coset space: constructed by
  Schwartz in the $\mathrm{SU}(1,1)$ formulation
  \cite{Schwarz:1983wa,Schwarz:1983qr,Howe:1983sra} and rewritten in the
  $\mathrm{SL}(2,\mathbb{R})$ formulation in which the dilaton and RR 0-form
  appear more naturally in Ref.~\cite{Bergshoeff:1995as}. The supersymmetry
  transformations of all the fields of $\mathcal{N}=2B,d=10$ supergravity,
  including the 8-forms dual to the dilaton and RR 0-form, were given in the
  $\mathrm{SU}(1,1)$ formulation in
  Ref.~\cite{Bergshoeff:2005ac,Bergshoeff:2010mv} and
  \cite{Dall'Agata:1998va}. In Section~\ref{sec-n2bd10} we are going to study
  the dualization of the scalars and the supersymmetry transformations of the
  dual 8-form fields proposed in that reference from the geometrical point of
  view taken here.

\item The $\mathrm{E}_{7(+7)}/\mathrm{SU}(8)$ coset space of
  $\mathcal{N}=8,d=4$ supergravity.

\end{enumerate}

\subsubsection{$\mathrm{SL}(2,\mathbb{R})/\mathrm{SO}(2)$}
\label{eq:SL2}

The group $\mathrm{SL}(2,\mathbb{R})$ is isomorphic to $\mathrm{SO}(2,1)$. The
Lie brackets of the three generators $\{T_{A}\}$ can be conveniently written
in the form

\begin{equation}
[T_{A},T_{B}] = -\varepsilon_{ABD}\mathsf{Q}^{DC} T_{C}\, ,
\,\,\,\,\,\,\,
\Rightarrow f_{AB}{}^{C} = -\varepsilon_{ABD}\mathsf{Q}^{DC},\,\,\,\,\,
A,B,\ldots =1,2,3\, ,
\end{equation}

\noindent
where $\mathsf{Q}= \mathrm{diag}(++-)$. A 2-dimensional representation (the
one we are going to work with) is provided by

\begin{equation}
\label{eq:SL2generators}
T_{1}= \tfrac{1}{2}\sigma^{3}\, ,\,\,\,
T_{2}= \tfrac{1}{2}\sigma^{1}\, ,\,\,\,
T_{3}= \tfrac{i}{2}\sigma^{2}\, ,
\end{equation}

\noindent
where the $\sigma^{i}$s are the standard Hermitian, traceless, Pauli matrices
satisfying $\sigma^{i}\sigma^{j}= \delta^{ij}+i\varepsilon^{ijk}\sigma^{k}$.
The Killing metric is

\begin{equation}
K_{AB}= -2 \mathsf{Q}_{AB}\, ,
\,\,\,\,\,
\mbox{and}
\,\,\,\,\,
g_{AB} = \mathrm{Tr}(T_{A}T_{B})= \tfrac{1}{2}\mathsf{Q}_{AB}\, .
\end{equation}

A convenient $\mathrm{SL}(2,\mathbb{R})/\mathrm{SO}(2)$ coset representative
is provided by

\begin{equation}
u
=
e^{-\varphi T_{1}} e^{e^{\varphi/2}\chi (T_{2}+T_{3})}
=
\left(
  \begin{array}{ll}
 e^{-\varphi/2} &   e^{\varphi/2}\chi \\
& \\
0 &  e^{\varphi/2} \\
  \end{array}
\right)\, ,
\end{equation}

\noindent
and the symmetric $\mathrm{SL}(2,\mathbb{R})$ matrix $\mathcal{M}$
constructed according to the recipe Eq.~(\ref{eq:M=uuT}) is the usual one

\begin{equation}
\label{eq:axidilaton}
\mathcal{M}
=
e^{\varphi}
\left(
  \begin{array}{ll}
| \tau |^{2} &  \chi \\
& \\
\chi &  1 \\
  \end{array}
\right)\, ,
\,\,\,\,\,
\mbox{where}
\,\,\,\,\,
\tau \equiv \chi +ie^{-\varphi}\, ,
\end{equation}

\noindent
is sometimes called the \textit{axidilaton} field.

The coset representative $u$ transforms according to the general rule
$u^{\prime}(x)= u(x^{\prime})= gu(x) h^{-1}$ where

\begin{equation}
g
=
\left(
  \begin{array}{ll}
a & b \\
& \\
c &  d \\
  \end{array}
\right) \in \mathrm{SL}(2,\mathbb{R})\, ,
\,\,\,\,\,
\mbox{so}
\,\,\,\,\,
ad-bc=1\, ,
\hspace{1cm}
h
=
\left(
  \begin{array}{ll}
\cos{\theta} & \sin{\theta} \\
& \\
-\sin{\theta} &  \cos{\theta} \\
  \end{array}
\right) \in \mathrm{SO}(2)\, .
\end{equation}

\noindent
In order to preserve the upper-triangular form of the coset representative
$u$, the compensating $h$ transformation must be such that

\begin{equation}
\tan{\theta} = \frac{-c e^{-\varphi}}{c\chi +d}\, ,
\end{equation}

\noindent
and this completely determines the transformation rules for the coordinates
$\varphi,\chi$: in terms of $\tau$ they take the usual fractional-linear form

\begin{equation}
\tau^{\prime} = \frac{a\tau +b}{c\tau +d}\, .
\end{equation}

The generators $P_{a},M$ are
\begin{equation}
P_{1,2} \equiv T_{1,2}\, ,
\hspace{1cm}
M\equiv T_{3}\, .
\end{equation}

\noindent
The components of the left-invariant MC 1-form in the above basis are given by

\begin{equation}
e^{1} = d\varphi\, ,
\hspace{1cm}
e^{2} = -e^{\varphi}d\chi\, ,
\hspace{1cm}
\vartheta =  e^{2}\, ,
\end{equation}

\noindent
and the $\mathrm{SL}(2,\mathbb{R})$-invariant metric of the coset space is

\begin{equation}
\label{eq:SL2invariantmetric}
ds^{2}
=
g_{ab}e^{a} e^{b}
=
\frac{d\tau d\tau^{*}}{2 (\Im\mathfrak{m} \tau)^{2}}\, .
\end{equation}

\begin{equation}
\label{eq:uadjointSL2}
\left(\Gamma_{\rm Adj}(u)^{A}{}_{B} \right)
 =
\left(
\begin{array}{ccc}
1                &      e^{\varphi} \chi       & -e^{\varphi} \chi \\
& & \\
-\chi & \tfrac{1}{2}e^{\varphi} (1-|\tau|^{2})+e^{-\varphi} &
-\tfrac{1}{2}e^{\varphi} (1-|\tau|^{2})\\
& & \\
-\chi & -\tfrac{1}{2}e^{\varphi} (1+|\tau|^{2})+e^{-\varphi}&
-\tfrac{1}{2}e^{\varphi} (1+|\tau|^{2})\\
\end{array}
\right)\, .
\end{equation}

The first two components of each of the three columns of the matrix

\begin{equation}
\left(\Gamma_{\rm Adj}(u^{-1})^{A}{}_{B} \right)
=
\left(
\begin{array}{ccc}
1                &      -\chi       & \chi \\
& & \\
e^{\varphi} \chi & \tfrac{1}{2}e^{\varphi} (1-|\tau|^{2})+e^{-\varphi} &
\tfrac{1}{2}e^{\varphi} (1+|\tau|^{2})-e^{-\varphi} \\
& & \\
e^{\varphi} \chi & \tfrac{1}{2}e^{\varphi} (1-|\tau|^{2})&
\tfrac{1}{2}e^{\varphi} (1+|\tau|^{2})\\
\end{array}
\right)\, .
\end{equation}

\noindent
are the two components of each of the three Killing vectors, while the third
row gives the three components of the momentum map (one for each
isometry). Using the Vierbeins, and in terms of $\partial_{\tau}=
\tfrac{1}{2}(\partial_{\chi} +ie^{\varphi}\partial_{\varphi})$ we get the
following explicit expressions for the Killing vectors, momentum map and
H-compensator

\begin{eqnarray}
k_{1} & = & \tau\partial_{\tau}+\mathrm{c.c.}\, ,
\hspace{.5cm}
k_{2} = \tfrac{1}{2}(1-\tau^{2})\partial_{\tau}+\mathrm{c.c.}\, ,
\hspace{.5cm}
k_{3} = \tfrac{1}{2}(1+\tau^{2})\partial_{\tau}+\mathrm{c.c.}\, ,
\\
& & \nonumber \\
(P_{A})
& = &
\left(e^{\varphi} \chi\, ,\,\,\, \tfrac{1}{2}e^{\varphi} (1-|\tau|^{2})\, ,\,\,\,
\tfrac{1}{2}e^{\varphi} (1+|\tau|^{2})\right)\, ,
\\
& & \nonumber \\
(W_{A})
& = &
i(\tfrac{1}{2}\, ,\,\,\, \tfrac{1}{2}\tau\, ,\,\,\, -\tfrac{1}{2}\tau)+\mathrm{c.c.}\, ,
\end{eqnarray}

\subsubsection{$\mathrm{SU}(1,1)/\mathrm{U}(1)$}
\label{sec-su11}

An $\mathrm{SU}(1,1)$ matrix can be parametrized by two complex numbers $a,b$

\begin{equation}
u
=
\left(
  \begin{array}{cc}
   a & b^{*} \\
   b & a^{*} \\
  \end{array}
\right)\, ,
\hspace{1cm}
|a|^{2}-|b|^{2} =1\, .
\end{equation}

$\mathrm{U}(1)$ acts on these two complex numbers by multiplication
$(a,b)\rightarrow e^{-i\varphi}(a,b)$. Therefore, its right action on this
matrix is

\begin{equation}
u
\longrightarrow
u \left(
  \begin{array}{cc}
   e^{-i\varphi} & 0 \\
   0 & e^{i\varphi} \\
  \end{array}
\right)\, .
\end{equation}

It is customary to introduce the vectors $V^{\alpha}{}_{\pm}$, $\alpha=1,2$
where the subindices $+,-$ refer to the $\mathrm{U}(1)$ weight

\begin{equation}
u
\equiv
\left( V^{\alpha}{}_{-} V^{\alpha}{}_{+} \right)\, ,
\,\,\,\,
\Rightarrow
\,\,\,\,
(V^{\alpha}{}_{+})^{*} = \sigma^{1\, \alpha}{}_{\beta}V^{\beta}{}_{-}\, .
\end{equation}

\noindent
In terms of these vectors the constraint $|a|^{2}-|b|^{2} =1$ takes the form

\begin{equation}
V^{\alpha}{}_{-} V^{\beta}{}_{+}
-
V^{\alpha}{}_{+} V^{\beta}{}_{-}
=
\varepsilon^{\alpha\beta}\, ,
\hspace{1cm}
\varepsilon_{\alpha\beta}V^{\alpha}{}_{-} V^{\beta}{}_{+}
=
1\, .
\end{equation}

This constraint can be solved by a complex number $z$ and a real one $\xi$

\begin{equation}
\left\{
  \begin{array}{rcl}
a & = & \cosh{\rho} +i\xi \frac{\sinh{\rho}}{\rho}\, ,\\
& & \\
b & = & z \cosh{\rho}\, ,
  \end{array}
\right.
\,\,\,\,\,
\mbox{where}
\,\,\,\,\,
\xi^{2} = |z|^{2} \frac{\rho^{2}}{\tanh^{2}{\rho}} -\rho^{2}\, .
\end{equation}

In these coordinates the general $\mathrm{SU}(1,1)$ matrix $u$ behaves near
the origin as

\begin{equation}
u \sim \mathbb{1}_{2\times 2} + \xi\, i\sigma^{3} +\Re\mathfrak{e}(z)\, \sigma^{1}
+\Im\mathfrak{m}(z)\, \sigma^{2}\, ,
\end{equation}

\noindent
from which we can read the generators and find the Lie algebra

\begin{equation}
T_{1} = \sigma^{1}\, ,\,\,\,
T_{2} = \sigma^{2}\, ,\,\,\,
T_{3} = i\sigma^{3}\, ,
\hspace{1cm}
[T_{A},T_{B}]= -2\varepsilon_{ABD}\mathsf{Q}^{DC}T_{C}\, ,
\end{equation}

\noindent
where $\mathsf{Q}=\mathrm{diag}(++-)$.  The Lie algebra is the same as that of
$\mathrm{SL}(2,\mathbb{R})$, with a different normalization and the metrics
$K_{AB},g_{AB}$ are proportional. The subgroup $\mathrm{U}(1)$ is generated by
$T_{3}$.

As we are going to see, in many instances, the adjoint index $A=1,2,3$ can be
replaced by a symmetric pair of indices $\alpha, \beta=1,2$ which can be
obtained from the corresponding Pauli matrix by left multiplication with
$\varepsilon_{\alpha\beta}$.

Using $\mathrm{U}(1)$ we can always bring $u$ to the gauge $\xi=0$ in which
all its components are written in terms of the unconstrained complex variable
$z$:

\begin{equation}
u
=
\frac{1}{\sqrt{1-|z|^{2}}}
\left(
  \begin{array}{cc}
   1 & z^{*} \\
   z & 1 \\
  \end{array}
\right)\, ,
\hspace{1cm}
z = V^{2}{}_{-}/V^{1}{}_{-}\, .
\end{equation}

Then, this coset representative $u(z)$ transforms according to the general
rule $u^{\prime}(z)=u(z^{\prime})= gu(z)h^{-1}$ with

\begin{equation}
g=\left(
  \begin{array}{cc}
   u & v^{*} \\
   v & u^{*} \\
  \end{array}
\right)  \in \mathrm{SU}(1,1)\, ,\,\,\, \mbox{so} \,\,\,
|u|^{2}-|v|^{2}=1\, ,
\hspace{1cm}
h=\left(
  \begin{array}{cc}
   e^{i\varphi} & 0 \\
   0 & e^{-i\varphi} \\
  \end{array}
\right) \in \mathrm{U}(1)\, ,
\end{equation}

\noindent
where

\begin{equation}
e^{i\varphi}
=
\frac{u+vz^{*}}{|u+vz^{*}|}\, ,
\,\,\,\,\,
\mbox{and}
\,\,\,\,\,
z^{\prime}
=
\frac{v+u^{*}z}{u+v^{*}z}\, .
\end{equation}

The generators $P_{a},M$ are

\begin{equation}
P_{1,2} \equiv T_{1,2}\, ,
\hspace{1cm}
T_{3}\equiv M\, .
\end{equation}

\noindent
Since $M$ is diagonal and the $P_{a}$ are anti-diagonal, the left-invariant
1-forms can be decomposed in terms of a complex Vielbein $e=e^{1}+ie^{2}$ and
a real connection $\vartheta$ as

\begin{equation}
-u^{-1}du
=
e^{a}P_{a}+\vartheta M
=
\left(
  \begin{array}{cc}
    i\vartheta & e^{*}        \\
     e         & -i\vartheta \\
  \end{array}
\right)\, ,
\end{equation}

\noindent
and using $u^{-1} = \left(
  \begin{smallmatrix}
    -V^{\beta}{}_{+}\varepsilon_{\beta\alpha} \\
    +V^{\beta}{}_{-}\varepsilon_{\beta\alpha} \\
  \end{smallmatrix}
\right)$ we find that they are given by

\begin{equation}
  \begin{array}{rcl}
e
& = &
-\varepsilon_{\alpha\beta}  V^{\alpha}{}_{-}dV^{\beta}{}_{-}
=
{\displaystyle\frac{dz}{1-|z|^{2}}}\, ,
\\
& & \\
\vartheta
& = &
i\varepsilon_{\alpha\beta}V^{\alpha}{}_{+}dV^{\beta}{}_{-}
=
\tfrac{i}{2}{\displaystyle\frac{zdz^{*}-z^{*}dz}{1-|z|^{2}}}\, .
\end{array}
\end{equation}

The invariant metric is\footnote{The change of variables
\begin{equation}
-i\tau = \frac{1-z}{1+z}\, ,
\end{equation}
brings the metric of the $\mathrm{SL}(2,\mathbb{R})/\mathrm{SO}(2)$ coset
Eq.~(\ref{eq:SL2invariantmetric}) to this one.}

\begin{equation}
ds^{2}= g_{ab}e^{a} e^{b}= 2\frac{dz dz^{*}}{(1-|z|^{2})^{2}}\, .
\end{equation}

The Killing vectors and momentum maps can be computed by using the general
formulae Eqs.~(\ref{eq:KV}) and (\ref{eq:momentummap}). Using the definition
of the adjoint action of a group on its Lie algebra
Eq.~(\ref{eq:adjointaction}):

\begin{equation}
T_{B}
\Gamma_{\rm Adj}(u^{-1})^{B}{}_{A}
=
u^{-1}T_{A} u\, ,
\,\,\,\,
\Rightarrow
\,\,\,\,
\Gamma_{\rm Adj}(u^{-1})^{B}{}_{A}
=
g^{BC}
\mathrm{Tr}\left[ T_{C} u^{-1}T_{A} u\right]\, ,
\end{equation}

\noindent
so

\begin{eqnarray}
k_{A}{}^{a}
& = &
-g^{ab} \mathrm{Tr}\left[P_{b} u^{-1} T_{A} u \right]\, ,
\\
& & \nonumber \\
P_{A}
& = &
\mathrm{Tr}\left[M u^{-1} T_{A} u \right]\, .
\end{eqnarray}

It is convenient to use the symmetric  pairs $\alpha\beta$ to label the
Killing vectors and momentum maps. We get, introducing a global factor of $-i$
to make them real

\begin{eqnarray}
\label{eq:KVsu11}
k^{(\alpha\beta)\, 1}+ik^{(\alpha\beta)\, 2}
& = &
iV^{\alpha}{}_{-}V^{\beta}{}_{-}\, ,
\\
& & \nonumber \\
\label{eq:mmsu11}
P^{\alpha\beta}
& = &
2V^{(\alpha}{}_{+}V^{\beta)}{}_{-}\, ,
\end{eqnarray}

\noindent
and

\begin{equation}
k_{A}{}^{a}
=
-ik^{(\alpha\beta)\, a}\varepsilon_{\alpha\gamma}T_{A}{}^{\gamma}{}_{\beta}\, ,
\hspace{1cm}
P_{A} =
-iP^{(\alpha\beta)}\varepsilon_{\alpha\gamma}T_{A}{}^{\gamma}{}_{\beta}\, .
\end{equation}

\subsubsection{$\mathrm{E}_{7(+7)}/\mathrm{SU}(8)$}

The $\mathrm{E}_{7(+7)}/\mathrm{SU}(8)$ coset representative can be written
in the convenient form of a $\mathrm{Usp}(28,28)$ matrix

\begin{equation}
\label{eq:E77SU8cosetrepresentative}
U
=
\left(
\begin{array}{cc}
u^{AB}{}_{IJ} & v^{*\, AB\, IJ} \\
& \\
v_{AB\, IJ} & u^{*}_{AB}{}^{IJ} \\
\end{array}
\right)\, ,
\hspace{1cm}
\begin{array}{rcl}
U^{\dagger}
\left(
  \begin{smallmatrix}
    \mathbb{1} & 0           \\
    0          & -\mathbb{1} \\
  \end{smallmatrix}
\right)
U
& = &
\left(
  \begin{smallmatrix}
    \mathbb{1} & 0           \\
    0          & -\mathbb{1} \\
  \end{smallmatrix}
\right)\, ,
\\
& & \\
U^{T}
\left(
  \begin{smallmatrix}
    0          &  \mathbb{1} \\
   -\mathbb{1} &  0          \\
  \end{smallmatrix}
\right)
U
& = &
\left(
  \begin{smallmatrix}
    0          &  \mathbb{1} \\
   -\mathbb{1} &  0          \\
  \end{smallmatrix}
\right)\, ,
\\
\end{array}
\end{equation}

\noindent
where all the indices are complex $\mathrm{SU}(8)$ indices raised and lowered
by complex conjugation that occur in antisymmetric pairs $AB$, $IJ$ so that

\begin{equation}
\label{eq:inverseE77SU8cosetrepresentative}
U^{-1}
=
\left(
\begin{array}{cc}
(u^{\dagger})^{IJ}{}_{AB} & -(v^{\dagger})^{IJ\, AB} \\
& \\
-(v^{T})_{IJ\, AB} & (u^{T})_{IJ}{}^{AB} \\
\end{array}
\right)\, .
\end{equation}

\noindent
The $Usp(28,28)$ condition implies for the $28\times 28$ matrices $u$ and $v$
the two conditions

\begin{equation}
  \begin{array}{rcl}
u^{\dagger}u - v^{\dagger}v & = & \mathbb{1}\, , \\
& & \\
v^{T}u +u^{T}v & = & 0\, .
\end{array}
\end{equation}

$\mathrm{E}_{7(+7)}$ acts on the $AB$ indices and the compensating
(scalar-dependent) $\mathrm{SU}(8)$ transformations act on the $IJ$ indices. A
parametrization in terms of independent scalar fields can be found, for
instance in Ref.~\cite{Kallosh:2008ic}.

Each column of the above matrix provides a set of complex vectors labeled by
the pair $IJ$ transforming in the fundamental (i.e~$\mathbf{56}$)
representation under $\mathrm{E}_{7(+7)}$. The action of this group on the
fundamental representation in this complex basis can be described as follows:
consider, for instance the complex combinations of the electric and magnetic
2-form field strengths $\mathcal{F}^{AB}$ given by

\begin{equation}
  \mathcal{F}^{\, AB}
  \equiv
  \tfrac{1}{4\sqrt{2}}\left(F^{ij}-iG_{ij} \right)\Gamma^{ij}{}_{AB}\, ,
\end{equation}

\noindent
where $ij$ are antisymmetric pairs of real $\mathrm{SL}(8)$ indices, the
$F^{ij}$ are the $28$ electric field strengths of the theory, the $G_{ij}$ are
the $28$ magnetic field strengths defined from the Lagrangian of the theory
$L$ by

\begin{equation}
G_{ij\, \mu\nu}
\equiv
\tfrac{1}{4}\star \frac{\partial L}{\partial F^{ij\, \mu\nu}}\, ,
\end{equation}

\noindent
and the $\Gamma^{ij}$s are the $\mathrm{SO}(8)$ gamma matrices. Then, the
infinitesimal action of $\mathrm{E}_{7(+7)}$ on the fundamental representation
is given by

\begin{equation}
\delta
\left(
\begin{array}{c}
\mathcal{F}^{AB} \\ \\ \mathcal{F}^{*}_{AB} \\
\end{array}
\right)
=
\left(
  \begin{array}{cc}
2\Lambda^{[A}{}_{[C}\delta^{B]}{}_{D]} & \Sigma^{ABCD} \\
& \\
\Sigma^{*}{}_{ABCD}                   & -2\Lambda^{[C}{}_{[A}\delta^{D]}{}_{B]} \\
\end{array}
\right)
\left(
\begin{array}{c}
  \mathcal{F}^{CD} \\ \\ \mathcal{F}^{*}_{CD} \\
\end{array}
\right)\, ,
\end{equation}

\noindent
where the $\Lambda^{A}{}_{B}$ are the anti-Hermitian parameters of
infinitesimal $\mathrm{SU}(8)$ transformations, (i.e.~$\Lambda^{*\, A}{}_{B}=
-\Lambda^{B}{}_{A}$ and $\Lambda^{A}{}_{A}=0$) and where the off-diagonal
infinitesimal parameters $\Sigma_{ABCD}$ are complex self-dual, that is

\begin{equation}
\label{eq:complexselfdualitySigma}
(\Sigma^{ABCD})^{*} \equiv \Sigma^{*}_{ABCD}
=
\tfrac{1}{4!}\varepsilon_{ABCDEFGH}\Sigma^{EFGH}\, .
\end{equation}

The generators of $\mathrm{E}_{7(+7)}$ in this representation are, therefore,

\begin{equation}
\left(
  \begin{array}{cc}
2\Lambda^{[A}{}_{[C}\delta^{B]}{}_{D]} & \Sigma^{ABCD} \\
& \\
\Sigma^{*}{}_{ABCD}                   & -2\Lambda^{[C}{}_{[A}\delta^{D]}{}_{B]} \\
\end{array}
\right)
=
\Lambda^{E}{}_{F}\mathcal{T}_{
  \begin{smallmatrix}
    E \\ F \\
  \end{smallmatrix}
}
+\Sigma^{EFGH}\mathcal{T}_{EFGH}
\end{equation}

\noindent
where

\begin{equation}
\label{eq:TEF}
\mathcal{T}_{
  \begin{smallmatrix}
    E \\ F \\
  \end{smallmatrix}
}
=
\left(
  \begin{array}{cc}
 T_{
  \begin{smallmatrix}
    E \\ F \\
  \end{smallmatrix}
}{}^{AB}{}_{CD} & 0 \\
& \\
0 & -T_{
  \begin{smallmatrix}
    E \\ F \\
  \end{smallmatrix}
}{}^{CD}{}_{AB}
  \end{array}
\right)\, ,
\hspace{1cm}
\mathcal{T}_{EFGH}
=
\left(
  \begin{array}{cc}
0 & T_{EFGH}{}^{ABCD} \\
& \\
T_{EFGH\, ABCD} & 0 \\
\end{array}
\right)\, ,
\end{equation}

\noindent
where, in its turn,

\begin{equation}
  \begin{array}{rcl}
T_{
  \begin{smallmatrix}
    E \\ F \\
  \end{smallmatrix}
}{}^{AB}{}_{CD}
& = &
2\left(\delta^{[A|}{}_{E}\delta^{F|B]}{}_{CD}
-\tfrac{1}{8}\delta^{F}{}_{E}\delta^{AB}{}_{CD}\right)\, ,
\\
& & \\
T_{EFGH}{}^{ABCD}
& = &
\delta_{EFGH}{}^{ABCD}\, ,
\\
& & \\
T_{EFGH\, ABCD}
& = &
\tfrac{1}{4!} \varepsilon_{EFGHABCD}\, .
\end{array}
\end{equation}

The generators $\mathcal{T}_{  \begin{smallmatrix}
    E \\ F \\
  \end{smallmatrix}
}$ of $\mathfrak{h}=\mathfrak{su}(8)$ will also be denoted
$M_{\begin{smallmatrix}
    F \\ E \\
  \end{smallmatrix}
}$ and the generators of the complement $\mathcal{T}_{EFGH}$ will be denoted
by $P_{EFGH}$.  In order to avoid confusion, in this section the indices in
the adjoint representation of $\mathrm{E}_{7(+7)}$ will be
$\mathbf{A},\mathbf{B},\ldots$ and correspond to the pairs $
\begin{smallmatrix}
  E \\ F \\
\end{smallmatrix}
$ plus the quartets $EFGH$. The metric in $\mathfrak{e}_{7(+7)}$
$g_{\mathbf{A}\mathbf{B}}$ will be

\begin{equation}
g_{\mathbf{A}\mathbf{B}}
\equiv
\mathrm{Tr}(\mathcal{T}_{\mathbf{A}}\mathcal{T}_{\mathbf{B}})\, .
\end{equation}

Using the explicit form of the generators $\mathcal{T}_{\mathbf{A}}$ written
above one can also compute explicitly the structure constants
$f_{\mathbf{A}\mathbf{B}}{}^{\mathbf{C}}$, the Killing metric
$K_{\mathbf{A}\mathbf{B}}$ and the metric $g_{\mathbf{A}\mathbf{B}}$. The
latter's and its inverse's only non-vanishing components  are

\begin{equation}
  \begin{array}{rclrcl}
g_{
  \begin{smallmatrix}
    E \\ F \\
  \end{smallmatrix}
\,
  \begin{smallmatrix}
    G \\ H \\
  \end{smallmatrix}
}
& = &
12 (\delta^{E}{}_{H}\delta^{G}{}_{F}
-\tfrac{1}{8}\delta^{E}{}_{F}\delta^{G}{}_{H})\, ,
\hspace{1cm}
&
g_{ABCD\, EFGH}
& = &
\tfrac{1}{12}\varepsilon_{ABDCEFGH}\, ,
\\
& & & & & \\
g^{
  \begin{smallmatrix}
    E \\ F \\
  \end{smallmatrix}
\,
  \begin{smallmatrix}
    G \\ H \\
  \end{smallmatrix}
}
& = &
\tfrac{1}{12} (\delta^{EG}\delta_{FH}
-\tfrac{1}{8}\delta^{E}{}_{F}\delta^{G}{}_{H})\, ,
\hspace{1cm}
&
g^{ABCD\, EFGH}
& = &
\tfrac{1}{2\cdot 4!}\varepsilon^{ABDCEFGH}\, .
\\
\end{array}
\end{equation}

The Vielbein and H-connection are defined by

\begin{equation}
-U^{-1}dU
=
e^{IJKL}P_{IJKL} +\vartheta^{
  \begin{smallmatrix}
    I \\ J \\
  \end{smallmatrix}
} M_{ \begin{smallmatrix}
    I \\ J \\
  \end{smallmatrix}
}
=
\left(
  \begin{array}{cc}
    \vartheta^{IJ}{}_{KL} & e^{IJKL} \\
& \\
e^{*}_{IJKL} & -\vartheta^{KL}{}_{IJ} \\
  \end{array}
\right)\, ,
\end{equation}

\noindent
and, using Eq.~(\ref{eq:E77SU8cosetrepresentative}) one finds that they are
given by

\begin{equation}
  \begin{array}{rcl}
\vartheta^{IJ}{}_{KL}
& = &
-(u^{\dagger})^{IJ}{}_{AB}du^{AB}{}_{KL}
+(v^{\dagger})^{IJ\, AB}dv_{AB\, KL}\, ,
\\
& & \\
e^{IJKL}
& = &
-(u^{\dagger})^{IJ}{}_{AB}dv^{*\, AB\, KL}
+(v^{\dagger})^{IJ\, AB}du^{*}_{AB}{}^{KL}\, .
  \end{array}
\end{equation}

From the Maurer-Cartan equations it follows that

\begin{eqnarray}
\mathcal{D}e^{IJKL}
& \equiv &
de^{IJKL} -\vartheta^{IJ}{}_{MN}\wedge e^{MNKL} +e^{IJMN}\wedge
\vartheta^{*}_{MN}{}^{KL}
=
0\, ,
\\
& & \nonumber \\
R^{IJ}{}_{KL}
& \equiv &
d\vartheta^{IJ}{}_{KL} -\vartheta^{IJ}{}_{mn}\wedge \vartheta^{MN}{}_{KL}
=
e^{IJMN}\wedge e^{*}_{MNKL}\, .
\end{eqnarray}

Again, in order to compute the Killing vectors and momentum maps we use the
same reasoning as in the previous example, arriving to

\begin{equation}
\label{eq:KVandmmE77}
\begin{array}{rcl}
k_{\mathbf{A}}{}^{EFGH}
& = &
-g^{EFGH\, ABCD}
\mathrm{Tr}\left[ \mathcal{T}_{ABCD} U^{-1}\mathcal{T}_{\mathbf{A}}
  U\right]\, ,
\\
& & \\
P_{\mathbf{A}}{}^{  \begin{smallmatrix}
    E \\ F \\
  \end{smallmatrix}
}
& = &
g^{  \begin{smallmatrix}
    E \\ F \\
  \end{smallmatrix}
\,
  \begin{smallmatrix}
    G \\ H \\
  \end{smallmatrix}
}
\mathrm{Tr}\left[ \mathcal{T}_{  \begin{smallmatrix}
    G \\ H \\
  \end{smallmatrix}
} U^{-1}\mathcal{T}_{\mathbf{A}}
  U\right]\, .
\end{array}
\end{equation}

\section{Noether 1-forms and dualization}
\label{sec-dualization1}

For each isometry of the metric $\mathcal{G}_{mn}(\phi)$ with Killing vector
$k_{A}{}^{m}(\phi)$, the $\sigma$-model action
Eq.~(\ref{eq:sigmamodelaction1}) has a global symmetry with $\delta_{A}
\phi^{m} = k_{A}{}^{m}(\phi)$.  According to Noether's first theorem, there is
a current density $j_{A}{}^{\mu}$ associated to each of them which is
conserved on-shell:

\begin{equation}
\mathfrak{j}_{A}{}^{\mu}
=
\sqrt{|g|} g^{\mu\nu}\mathcal{G}_{mn}k_{A}{}^{m}\partial_{\nu}\phi^{n}\, ,
\hspace{1cm}
\partial_{\mu}\mathfrak{j}_{A}{}^{\mu}
=
-k_{A}{}^{m}\frac{\delta S}{\delta \phi^{m}}\, .
\end{equation}

\noindent
The Noether 1-form  is defined by

\begin{equation}
j_{A}
=
\mathcal{G}_{mn}k_{A}{}^{m}d\phi^{n}\, ,
\end{equation}

\noindent
and, for the choice of metric Eq.~(\ref{eq:flatmetric}), and using the
expression Eq.~(\ref{eq:KV}), it can be written in the form

\begin{equation}
j_{A}
=
\mathrm{Tr} \left[T_{A} \mathcal{D} u u^{-1} \right]
=
-g_{AB}\Gamma_{\rm Adj}(u)^{B}{}_{a}\, e^{a}\, ,
\end{equation}

\noindent
which can also be obtained from the explicitly right-invariant expression
Eq.~(\ref{eq:sigmamodelaction3}) for the transformation $\delta_{A}u
=T_{A}u$.\footnote{The complete infinitesimal transformation of $u$ must
  include the compensating H-transformations that act from the right:
\begin{equation}
\delta_{A}u=T_{A}u +u W_{A}{}^{i}M_{i}\, ,
\end{equation}
but they can be safely ignored in the right-invariant expression.
}

The above expression makes it easier to show that not all of these 1-forms are
independent: they satisfy dim\, H scalar-dependent relations:

\begin{equation}
\label{eq:relationbetweencurrents}
j_{A} \Gamma_{\rm Adj}(u)^{A}{}_{i}
=
\mathrm{Tr} \left[M_{i} u^{-1} \mathcal{D} u \right]
=
0\, ,
\end{equation}

\noindent
where we have used Eq.~(\ref{eq:Du}) and the orthogonality of the basis of
$\mathfrak{h}$ and $\mathfrak{k}$ in symmetric spaces. Observe that the above
expression \textit{is not simply} $j_{i}=0$.

For the rest of the components we get the non-vanishing expression

\begin{equation}
\label{eq:jversuse}
j_{A} \Gamma_{\rm Adj}(u)^{A}{}_{a}
=
\mathrm{Tr} \left[P_{a} u^{-1} \mathcal{D} u \right]
=
-g_{ab}e^{b}\, .
\end{equation}

\noindent
This expression and the previous one appear in the equations of motion of the
scalars Eq.~(\ref{eq:equationsofmotionscalars}) and in the gauge identities
Eq.~(\ref{eq:gaugeidentities}). Thus, these can be rewritten, respectively in
the form

\begin{eqnarray}
\label{eq:equationsofmotionscalars2}
e_{m}{}^{a} \mathcal{D}\star \left[j_{A} \Gamma_{\rm Adj}(u)^{A}{}_{a}
\right]
& = &
0\, ,
\\
& & \nonumber \\
\label{eq:gaugeidentities2}
\mathcal{D}\star \left[j_{A} \Gamma_{\rm Adj}(u)^{A}{}_{i}\right]
& = &
0\, .
\end{eqnarray}

Using the explicit form of the H-covariant derivatives, we find that these
equations can be rewritten in the following form, much more directly related
to the conservation of the Noether 1-forms:

\begin{eqnarray}
\label{eq:equationsofmotionscalars3}
e_{m}{}^{a} \Gamma_{\rm Adj}(u)^{A}{}_{a} d\star j_{A}
& = &
0\, ,
\\
& & \nonumber \\
\label{eq:gaugeidentities3}
\Gamma_{\rm Adj}(u)^{A}{}_{i} d\star j_{A}
& = &
0\, .
\end{eqnarray}

Combining these equations with the explicit form of the Killing vectors
Eq.~(\ref{eq:KV}) we can easily prove the off-shell relation

\begin{equation}
\label{eq:kS}
k_{A}{}^{m}\frac{\delta S_{\rm Gauged}}{\delta \phi^{m}}
=
-d\star j_{A}\, .
\end{equation}

The moral of these results is that the equations of motion of the scalars can
be seen as combinations (projections) of some more fundamental equations: the
conservation laws of the Noether currents. The latter can completely replace
the former. But only the Noether 1-forms can be dualized.

The Noether 1-forms are closed on-shell and, on-shell, they can be dualized by
introducing as many $(d-2)$-forms $B_{A}$ related to them by

\begin{equation}\label{*jA=HA}
\star j_{A} \equiv dB_{A} \equiv H_{A}\, ,
\end{equation}

\noindent
solving locally the conservation laws. As usual, the Bianchi identities of the
original fields become the equations of motion of the dual ones. The obvious
candidate to Bianchi identity is

\begin{equation}
\mathcal{D}\left(\Gamma_{\rm Adj}(u^{-1})^{a}{}_{A}g^{AB}j_{B}\right)=0\, ,
\end{equation}

\noindent
by virtue of Eq.~(\ref{eq:jversuse}) and Cartan structure equation
$\mathcal{D}e^{a}=0$ which is equivalent to the first set of Maurer--Cartan
Eqs.~(\ref{eq:MCe}). Then, the equations of motion satisfied by the 2-forms
are\footnote{We use $\Gamma_{\rm Adj}(u)^{A}{}_{a} = g_{ab}\Gamma_{\rm
    Adj}(u^{-1})^{b}{}_{B} g^{BA}$.}

\begin{equation}
\label{eq:BAeom}
\mathcal{D}\left(\Gamma_{\rm Adj}(u)^{A}{}_{a} \star H_{A}\right)=0\, .
\end{equation}

There are only $\mathrm{dim G} -\mathrm{dim H}$ equations, which means that
we cannot solve for all the $H_{A}$. We must also use the  constraint

\begin{equation}
\label{eq:HAconstraint}
\Gamma_{\rm Adj}(u)^{A}{}_{i} H_{A} =0\, ,
\end{equation}

\noindent
which follows from Eq.~(\ref{eq:relationbetweencurrents}).

It would be desirable to have a kinetic term for the $(d-2)$-forms $B_{A}$
from which the equations of motion (\ref{eq:BAeom}) could be derived.
The simplest candidate would be

\begin{equation}
\label{eq:notophsaction}
S = \int \tfrac{1}{2}\mathfrak{M}^{AB}H_{A}\wedge \star H_{B}\, ,
\end{equation}

\noindent
where the scalar-dependent matrix $\mathfrak{M}^{AB}$ is defined by

\begin{equation}
\label{eq:MABdef}
\mathfrak{M}^{AB}
\equiv
\Gamma_{\rm Adj}(u)^{A}{}_{a}\Gamma_{\rm Adj}(u^{-1})^{a}{}_{C}g^{CB}\, ,
\end{equation}

\noindent
and is obviously singular, with $\mathrm{rank}(\mathfrak{M}^{AB})= \mathrm{dim
  G} -\mathrm{dim H}$ because $\mathfrak{M}^{AB}\Gamma_{\rm
  Adj}(u^{-1})^{i}{}_{A}=0\,\, \forall i$. This means that the combinations
$\Gamma_{\rm Adj}(u)^{A}{}_{i} H_{A}$ which are constrained to vanish, do not
enter the action. Observe that $\mathfrak{M}^{AB}$ transforms under global G
transformations according to

\begin{equation}
\mathfrak{M}^{\prime\, AB}
=
\Gamma_{\rm Adj}(g)^{A}{}_{A^{\prime}}
\Gamma_{\rm Adj}(g)^{B}{}_{B^{\prime}}
\mathfrak{M}^{A^{\prime}B^{\prime}}\, ,
\end{equation}

\noindent
so the above action is invariant.

Observe also that the $\sigma$-model action can be written in terms of the
same singular matrix as

\begin{equation}
\label{eq:jsaction}
S = \int \tfrac{1}{2}\mathfrak{M}^{AB}j_{A}\wedge \star j_{B}\, .
\end{equation}

Furthermore, observe that, if we define

\begin{equation}
\Gamma_{\rm Adj}(u)^{A}{}_{a}B_{A}\equiv \mathcal{B}_{a}\, ,
\hspace{1cm}
\Gamma_{\rm Adj}(u)^{A}{}_{i}B_{A}\equiv \mathcal{B}_{i}\, ,
\end{equation}

\noindent
and we impose the restriction $\mathcal{B}_{i}=0$, the above action can be
rewritten in the form

\begin{equation}
\label{eq:notophsaction2}
S = \int \tfrac{1}{2} g^{ab}\mathcal{H}_{a}\wedge \star \mathcal{H}_{b}\, ,
\,\,\,\,\,
\mbox{where}
\,\,\,\,\,
\mathcal{H}_{a}=\mathcal{D}\mathcal{B}_{a}\, .
\end{equation}

\noindent
Notice, however, that the restriction $\mathcal{B}_{i}=0$ is not equivalent to
Eq.~(\ref{eq:HAconstraint}).

The equations of motion that follow from the above action are simply

\begin{equation}
d\left(\mathfrak{M}^{AB}\star H_{B} \right)=0\, .
\end{equation}

Projecting them with $\Gamma_{\rm Adj}(u^{-1})^{a}{}_{A}$, we get the
equations of motion (\ref{eq:BAeom})

\begin{equation}
\label{eq:projected}
\Gamma_{\rm Adj}(u^{-1})^{a}{}_{A}
d\left(\mathfrak{M}^{AB}\star H_{B} \right)
=
g^{ab}
\mathcal{D}\left(\Gamma_{\rm Adj}(u)^{A}{}_{b} \star H_{A}\right)=0\, ,
\end{equation}

\noindent
whose solution is

\begin{equation}
g^{ab}
\Gamma_{\rm Adj}(u)^{A}{}_{b} \star H_{A}  \propto e^{a}\, .
\end{equation}

However, if we project the equations of motion with $\Gamma_{\rm
  Adj}(u^{-1})^{i}{}_{A}$ we get a non-trivial constraint:

\begin{equation}
\Gamma_{\rm Adj}(u^{-1})^{i}{}_{A}
d\left(\mathfrak{M}^{AB}\star H_{B} \right)
=
-e^{a}f_{ab}{}^{i} g^{bc} \Gamma_{\rm Adj}(u)^{A}{}_{c}\star H_{A}=0\, ,
\end{equation}

\noindent
which, upon use of the previous solutions gives a non-trivial constraint that
we do not want:

\begin{equation}
R(\vartheta)^{i}=0\, .
\end{equation}

We have not found any completely satisfactory way of solving this problem in
general.

Notice that a similar problem appeared with dualization of 3-form potential
$A_{3}$ of eleven dimensional supergravity. Its action \cite{Cremmer:1978km}
contains, besides the kinetic term of $A_{3}$ and interaction of $A_{3}$ with
fermions, the Chern-Simons type term $dA_{3}\wedge A_{3}\wedge A_{3}$, and this
makes impossible to construct a dual action including 6-form potential $A_{6}$
instead of $A_{3}$ \cite{Nicolai:1980kb,D'Auria:1982nx}.

However, there exists the duality invariant action of 11-d supergravity
including both $A_{6}$ and $A_{3}$ potentials \cite{Bandos:1997gd}.  It was
constructed using the PST (Pasti-Sorokin-Tonin) approach
\cite{Pasti:1995ii,Pasti:1995tn} and reproduce a (nonlinear) duality relation
between the (generalized) field strengths of $A_{6}$ and $A_{3}$ as a gauge
fixed version of the equations of motion. Notice that this action can be
presented formally as a sigma model action \cite{Bandos:2003et} for a
supergroup with fermionic generator associated to $A_{3}$ and bosonic
generators associated to $A_{6}$ \cite{Cremmer:1998px}.

Then it is natural to expect that the similar situation occurs in our case of
dualization of ``non-Abelian'' scalars.  Even if the non-existence of a
consistent way to write dual action in terms of only $(d-2)$ forms dual to a
scalars parametrizing a non-Abelian coset were proved, this would not prohibit
the existence of a PST-type action involving both the scalars and the $(d-2)$
forms and producing the duality equations (\ref{*jA=HA}) as a gauge fixed
version of the equations of motion. Moreover, for the particular case of
SU$(1,1)$/U$(1)$ coset such action was constructed in \cite{Dall'Agata:1998va},
where it was also incorporated in the complete action of type IIB
supergravity.

The generalization of the action from Ref.~\cite{Dall'Agata:1998va} for the
generic case of scalars parametrizing a symmetric space G/H reads

\begin{eqnarray}
\label{PST=act}
S_{\rm PST} 
& = &
\int L_{PST}\, , 
\\
& & \nonumber \\
\label{L=PST}
L_{\rm PST}
& = &
\tfrac{1}{2} g^{AB} j_{A}\wedge \star j_{B} 
+\tfrac{1}{2} g_{ij} F^{i}\wedge \star F_{j}
+\tfrac{(-1)^{d}}{2}\, g^{AB}\,  
\mathcal{H}_{A}  \wedge v \, i_{v}\star \mathcal{H}_{B}\; ,
 \qquad
  \end{eqnarray}

\noindent
where 

\begin{eqnarray} 
\label{cH=}
\mathcal{H}_{A} 
& = & 
H_{A} + \star j_{A}+ \Gamma_{\rm Adj}(u^{-1})^{i}{}_{A} \star F_{i}
\nonumber \\ 
& & \nonumber \\
& \equiv &  H_{A} + \star j_{A}+ P^{i}{}_{A} \star F_{i}\, , 
\end{eqnarray}


\noindent
the one-form $v$ is constructed from the PST scalar $a(x)$, the (would be)
auxiliary field of the PST formalism,

\begin{equation}
\label{v=da}
v
=
\frac{da(x)}{\sqrt{\partial  a \partial  a}}\; , 
\qquad 
v_{\mu}
= 
\frac{\partial_{\mu} a(x)}{\sqrt{\partial  a \partial  a}}\; ,
\end{equation}

\noindent
and $F_{i}=dx^{\mu} F_{\mu i}(x)$ is an auxiliary one-form carrying the index
of $H$-generators.  The contraction symbol is defined, as usual, by

\begin{equation}
\label{iv:=}
i_{v} j_{A} 
= 
v^\mu j_{\mu A}\; , 
\qquad 
i_{v} H_{A}
= \tfrac{1}{(d-2)!} dx^{\nu_{(D-2)}}\wedge \ldots \wedge dx^{\nu_{1}}
H_{\nu_{1} \ldots\nu_{(d-2)}\mu }\,  v^{\mu } \; .
\end{equation}

We assume the derivative of the PST scalar to be a time-like vector so that
the square root in denominator is well defined (in our mostly minus
signature), $v_{\mu} v^{\mu}=1$, and

\begin{equation}
\label{H=vivH+*}
H_{A} = v \wedge i_{v} H_{A} + \star (v \wedge i_{v} \star H_{A}) \; ,
\end{equation} 

\noindent
is  valid  for any $(d-2)$-form and, in particular, for our $H_{A}=dB_{A}$.

The study of this action and derivation of the duality conditions from its
equations of motion is out of the scope of this paper.  Here we would like to
stress the r\^ole the momentum map (\ref{eq:momentummap}) plays in it: the
auxiliary one-form $F_{i}$ always enter the action in contraction $F_{i}\,
P^{i}{}_{A}$. Indeed, this is the case for $\mathcal{H}_{A}$ (\ref{cH=}), and,
due to $j_{A} \Gamma_{\rm Adj}(u)^{A}{}_{i}=0$
(\ref{eq:relationbetweencurrents}), the first two terms of the Lagrangian
(\ref{L=PST}) can also be collected in $\tfrac{1}{2} g^{AB} (j_{A} +
P^{i}{}_{A}\,F_{i}) \wedge \star ( j_{B} + P^{j}{}_{A}\, F_{j})$,

\begin{equation}
\label{L=PST-FPP}
L_{\rm PST}
= 
\tfrac{1}{2} g^{AB} (j_{A} + P^{i}{}_{A}\,F_{i}) \wedge 
\star ( j_{B} + P^{j}{}_{A}\, F_{j})
+ \tfrac{(-1)^{d}}{2}\, g^{AB}\,  \mathcal{H}_{A}  
\wedge v \, i_{v}\star \mathcal{H}_{B} \; .
\end{equation}

We hope to return to the study the properties of this action and its
applications in supergravity context in future publications.

\subsection{Examples}

\subsubsection{$\mathrm{SL}(2,\mathbb{R})/\mathrm{SO}(2)$}

A short calculation gives

\begin{equation}
\mathcal{D} u u^{-1}
=
\tfrac{1}{2}d\mathcal{M}\mathcal{M}^{-1}
=
j_{A}B^{AB}T_{B}\, ,
\end{equation}

\noindent
where the Noether 1-forms are given by

\begin{equation}
  \begin{array}{rcl}
j_{1}
& = &
\tfrac{1}{4}e^{2\varphi}d |\tau|^{2}\, ,
\\
& & \\
j_{2}
& = &
-\tfrac{1}{4}e^{2\varphi}\chi d |\tau|^{2}
+\tfrac{1}{4}e^{2\varphi}(1+|\tau|^{2})d\chi\, ,
\\
& & \\
j_{3}
& = &
+\tfrac{1}{4}e^{2\varphi}\chi d |\tau|^{2}
+\tfrac{1}{4}e^{2\varphi}(1-|\tau|^{2})d\chi\, .
\end{array}
\end{equation}

\noindent
As expected, they are not independent: they are related by one (dim H)
relation of the form Eq.~(\ref{eq:relationbetweencurrents}) where $\Gamma_{\rm
  Adj}(u)^{A}{}_{3}$ is the third column of the $\mathrm{SO}(2,1)$ matrix in
Eq.~(\ref{eq:uadjointSL2}).

The singular matrix $\mathfrak{M}^{AB}$ in the action
Eq.~(\ref{eq:notophsaction}) is

\begin{equation}
\left(\mathfrak{M}^{AB}\right)
=
e^{2\varphi}
\left(
\begin{array}{ccc}
|\tau|^{2} &
\tfrac{1}{2}(1-|\tau|^{2}) \chi &
-\tfrac{1}{2}(1+|\tau|^{2}) \chi \\
& & \\
\tfrac{1}{2}(1-|\tau|^{2}) \chi &
e^{-2\varphi}+\tfrac{1}{4}(1-|\tau|^{2})^{2} &
-\tfrac{1}{4}(1-|\tau|^{4}) \\
& & \\
-\tfrac{1}{2}(1+|\tau|^{2}) \chi &
-\tfrac{1}{4}(1-|\tau|^{4}) &
-e^{-2\varphi}+\tfrac{1}{4}(1+|\tau|^{2})^{2}
\end{array}
\right)\, .
\end{equation}

The combinations of dual $(d-1)$-form field strengths $H_{A}$ that occur in
that action are

\begin{equation}
  \begin{array}{rcl}
\Gamma_{\rm Adj}(u)^{A}{}_{1} H_{A}
& = &
H_{1} -\chi(H_{2}+H_{3})\, ,
\\
& & \\
\Gamma_{\rm Adj}(u)^{A}{}_{2} H_{A}
& = &
e^{\varphi}\chi [H_{1} -\tfrac{1}{2}\chi(H_{2}+H_{3})]
+\tfrac{1}{2}e^{-\varphi}(H_{2}+H_{3})
+\tfrac{1}{2}e^{-\varphi}(H_{2}-H_{3})\, ,
\end{array}
\end{equation}

\noindent
and the constraint Eq.~(\ref{eq:HAconstraint}) is

\begin{equation}
\Gamma_{\rm Adj}(u)^{A}{}_{3} H_{A}
=
e^{\varphi}\chi H_{1}
+\tfrac{1}{2}e^{\varphi}(H_{2}+H_{3})
-\tfrac{1}{2}e^{\varphi}|\tau|^{2}(H_{2}-H_{3})
=
0\, .
\end{equation}

The relation between these three $(d-1)$-form field strengths and the
scalars (the pullbacks of the Vierbeins) is

\begin{equation}
\Gamma_{\rm Adj}(u)^{A}{}_{a} H_{A} = \star e^{a}\, ,
\end{equation}

\noindent
and, with the help of the above constraint we can invert it, expressing
entirely the three field strengths in terms of the scalars. The relations are
equivalent to $H_{A}= \star j_{A}$.

\subsubsection{$\mathrm{E}_{7(+7)}/\mathrm{SU}(8)$}

Using the same properties we used to find the Killing vectors and momentum
maps Eqs.~(\ref{eq:KVandmmE77}) we find the Noether 1-forms are given by
$j_{\mathbf{A}\, ABCD} e^{ABCD}$ where the $e^{ABCD}$ are the Vielbein and the
components are given by

\begin{equation}
j_{\mathbf{A}\, ABCD}
=
\mathrm{Tr}\left[ \mathcal{T}_{ABCD} U^{-1}\mathcal{T}_{\mathbf{A}}
  U\right]\, .
\end{equation}

The explicit expressions can be easily computed using the generators and
(inverse) coset representatives given above. For instance,

\begin{equation}
  \begin{array}{rcl}
j_{\begin{smallmatrix}
    E \\ F \\
  \end{smallmatrix}\, ABCD}
& = &
\tfrac{1}{12} \varepsilon_{ABCD GHMN}
\left[ (v^{\dagger})^{GH\, FJ}u^{*}_{EJ}{}^{MN} +
  (u^{\dagger})^{GH}{}_{EJ}v^{*\, FJ\, MN} \right]
\\
& & \\
& &
-2\delta_{ABCD}{}^{GHMN}
\left[ (v^{T})_{GH\, EJ}u^{FJ}{}_{MN} +
  (u^{T})_{GH}{}^{FJ}v_{EJ\, MN} \right]
\, .
\end{array}
\end{equation}

Observe that the 1-forms $j_{\begin{smallmatrix}
    E \\ F \\
  \end{smallmatrix}\, ABCD}e^{ABCD}$ are purely imaginary.


\section{NGZ 1-forms and dualization in $d=4$}
\label{sec-dualizationd=4}

The bosonic action of all 4-dimensional ungauged supergravities (and many
other interesting theories as well) is of the generic form

\begin{equation}
\label{eq:genericd4action}
\begin{array}{rcl}
S[g_{\mu\nu},A^{\Lambda}{}_{\mu},\phi^{m}]
& =  &
{\displaystyle
\int
}
 d^{4}x
\sqrt{|g|}
\left\{
R
+
\mathcal{G}_{mn}(\phi)\partial_{\mu}\phi^{m}\partial^{\mu}\phi^{n}
\right.
\\
& & \\
& &
\left.
+
2\Im\mathfrak{m}\mathcal{N}_{\Lambda\Sigma}
F^{\Lambda\, \mu\nu}F^{\Sigma}{}_{\mu\nu}
-
2\Re\mathfrak{e}\mathcal{N}_{\Lambda\Sigma}
F^{\Lambda\, \mu\nu}\, \star F^{\Sigma}{}_{\mu\nu}
\right\},
\end{array}
\end{equation}

\noindent
where the indices $\Lambda,\Sigma=1,\cdots,\overline{n}$ (the total number of
fundamental vector fields) and where $\mathcal{N}_{\Lambda\Sigma}(\phi)$ is
known as the period matrix and it is symmetric and, by convention, has a
negative definite imaginary part. The $\sigma$-model metric
$\mathcal{G}_{mn}(\phi)$ is that of a Riemannian symmetric space in all
$\mathcal{N}>2$ cases (and in many other cases as well) and this is the case
that we want to consider here in order to apply the results derived in the
previous sections.

First of all, we want to rewrite this action in differential-form language and
using the coset representative $u$:

\begin{equation}
\label{eq:genericd4action2}
S
 =
\int
\left\{
-\star R
+
\mathrm{Tr}[u^{-1}\mathcal{D}u \wedge \star u^{-1}\mathcal{D}u]
-4\Im\mathfrak{m}\mathcal{N}_{\Lambda\Sigma}
F^{\Lambda}\wedge \star F^{\Sigma}
-4\Re\mathfrak{e}\mathcal{N}_{\Lambda\Sigma}
F^{\Lambda}\wedge F^{\Sigma}
\right\}\, .
\end{equation}

Then, we define the dual vector field strengths

\begin{equation}
\label{eq:dualvectorfieldstrengths}
G_{\Lambda}
\equiv
\Im\mathfrak{m}\mathcal{N}_{\Lambda\Sigma}
\star F^{\Sigma}
+\Re\mathfrak{e}\mathcal{N}_{\Lambda\Sigma}
F^{\Sigma}\, ,\,\,\,\,\,
\mbox{or}\,\,\,\,\,
G_{\Lambda}{}^{+}
\equiv
\mathcal{N}^{*}_{\Lambda\Sigma}
F^{\Sigma\, +}\, .
\end{equation}

\noindent
The last relation is known as a \textit{(linear) twisted self-duality
  constraint}. Defining the symplectic vector of vector field strengths

\begin{equation}
\label{eq:symplectivectorfieldstrengths}
(\mathcal{F}^{M})
\equiv
\left(
  \begin{array}{c}
   F^{\Lambda} \\
G_{\Lambda} \\
  \end{array}
\right)\, ,
\end{equation}

\noindent
the equations of motion and the Bianchi identities can be written together as

\begin{equation}
\label{eq:dFM}
d \mathcal{F}^{M}=0\, ,
\end{equation}

\noindent
which can be solved locally by assuming the existence of 1-form potentials
$\mathcal{A}^{M}$

\begin{equation}
\mathcal{F}^{M} = d \mathcal{A}^{M}\, .
\end{equation}

This set of equations is invariant under linear transformations

\begin{equation}
\mathcal{F}^{\prime M} = S^{M}{}_{N}\mathcal{F}^{N}\, ,
\hspace{1cm}
S
\equiv
\left(
\begin{array}{cc}
A~ & ~B \\
C~ & ~D \\
\end{array}
\right)\, .
\end{equation}

\noindent
As it is well known, the preservation of the twisted self-duality constraint
requires the simultaneous transformation of the period matrix according to the
rule

\begin{equation}
\label{eq:transformationruleperiodmatrix}
\mathcal{N}^{\prime}
=
(C+D\mathcal{N}) (A+B\mathcal{N})^{-1}\, .
\end{equation}

\noindent
The preservation of the symmetry of $\mathcal{N}$ and of the negative
definiteness of $\Im\mathfrak{m}\mathcal{N}$ (together with the preservation
of the energy-momentum tensor) require $S$ to be an
Sp($2\overline{n},\mathbb{R}$) transformation, that is

\begin{equation}
\label{eq:OmegaMNdef}
S^{T}\Omega S = \Omega ,
\hspace{1cm}
 \Omega \equiv
  \left(
  \begin{array}{cc}
   0 & \mathbbm{1} \\
-\mathbbm{1} & 0 \\
  \end{array}
\right)\, .
\end{equation}

\noindent
Finally, if these transformations are going to be symmetries of the equations
of motion, the form of the period matrix as a function of the scalars must be
preserved, and this requires the above transformation rule for $\mathcal{N}$
to be equivalent to a transformation of the scalars:

\begin{equation}
\label{eq:Nprime}
\mathcal{N}^{\prime}(\phi) = \mathcal{N}(\phi^{\prime})\, .
\end{equation}

\noindent
This transformation of the scalars must be an isometry of the $\sigma$-model
metric. Thus, the symmetries of the equations of motion of the theory are the
group G of the isometries of the $\sigma$-model metric which act on the vector
fields embedded in the symplectic group\footnote{We are going to ignore the
  possibility of scalars which do not couple to the vector fields because this
  simply does not happen in $\mathcal{N}>2$ supergravities.}. These isometries
always leave invariant the scalars' kinetic term but only some of them may
leave invariant the whole action because many involve electric-magnetic
duality rotations.  Thus, there is a Noether 1-form for each isometry of the
scalar sector, and we are going to denote it by a $\sigma$ index,

\begin{equation}
j^{(\sigma)}_{A} = 2\mathrm{Tr} \left[T_{A} \mathcal{D} u u^{-1} \right]\, ,
\end{equation}

\noindent
but, in general, they do not have a standard completion to Noether 1-forms of
the full theory. A completion does, nevertheless, exist in all cases and it
was found by Gaillard and Zumino in Ref.~\cite{Gaillard:1981rj}. To construct
the Noether--Gaillard--Zumino (NGZ) 1-form we first need to define the
infinitesimal generators of G in the representation in which they act on the
vector fields, $\{\mathcal{T}_{A}\}$. By assumption $\mathcal{T}_{A}\in
\mathfrak{sp}(2\overline{n},\mathbb{R})$ and

\begin{equation}
\delta_{A} \mathcal{F}^{M} = \mathcal{T}_{A}{}^{M}{}_{N}\mathcal{F}^{N}\, .
\end{equation}

Then, the NGZ 1-forms are given by

\begin{equation}
j_{A}
=
j^{(\sigma)}_{A}
-2 \mathcal{T}_{A}{}^{M}{}_{N}\star (\mathcal{F}^{N}\wedge \mathcal{A}_{M})\, ,
\hspace{1cm}
\mathcal{A}_{M} = \Omega_{MN}\mathcal{A}^{N}\, .
\end{equation}

Observe that the NGZ 1-forms are not invariant under the gauge transformations
of the 1-forms $\delta_{\sigma}\mathcal{A}^{M}=d\sigma^{M}$ precisely because
the Noether 1-forms $j^{(\sigma)}_{A}$ are.  Let us check that they are
conserved on-shell. First, the conservation equation takes the form

\begin{equation}
d\star j_{A}
=
d\star j^{(\sigma)}_{A}
-2 \mathcal{T}_{A}{}^{M}{}_{N}\mathcal{F}^{N}\wedge \mathcal{F}_{M}\, ,
\end{equation}

\noindent
where we have used Eqs.~(\ref{eq:dFM}). Now we are going to see that this
equation is proportional to the projection of the scalar equations of motion
with the Killing vectors. The scalar equations of motion are

\begin{equation}
\frac{\delta S}{\delta \phi^{m}}
=
2\frac{\delta S_{\rm Gauged}}{\delta \phi^{m}}
-4F^{\Lambda}\frac{\partial G_{\Lambda}}{\partial \phi^{m}}
=
0\, ,
\end{equation}

\noindent
where $S_{\rm Gauged}$ is the $\sigma$-model action normalized as in
Eq.~(\ref{eq:sigmamodelactionGauged}). Contracting with the Killing vectors
we get

\begin{equation}
\label{eq:kS2}
k_{A}{}^{m}\frac{\delta S}{\delta \phi^{m}}
=
-2\left[
d\star j^{(\sigma)}_{A}
+2F^{\Lambda}k_{A}{}^{m}\frac{\partial G_{\Lambda}}{\partial \phi^{m}}
\right]
=
0\, ,
\end{equation}

\noindent
where we have used  Eq.~(\ref{eq:kS}). The infinitesimal transformation rule
for the period matrix follows from Eq.~(\ref{eq:Nprime}) and
Eq.~(\ref{eq:transformationruleperiodmatrix})\footnote{We use the following
infinitesimal form for  the symplectic matrix $S$:
\begin{equation}
S^{M}{}_{N} \sim \delta^{M}{}_{N} +\sigma^{A}\mathcal{T}_{A}{}^{M}{}_{N} \, ,
\,\,\,\,\,
\mbox{where}
\,\,\,\,\,
\mathcal{T}_{A}{}^{P}{}_{[M}\Omega_{N]P}=0\, .
\end{equation}
The different block components of $\mathcal{T}_{A}{}^{M}{}_{N}$ are defined by
\begin{equation}
\left(\mathcal{T}_{A}{}^{M}{}_{N} \right)
=
\left(
  \begin{array}{cc}
\mathcal{T}_{A}{}^{\Lambda}{}_{\Sigma}  &
\mathcal{T}_{A}{}^{\Lambda\Sigma} \\
& \\
\mathcal{T}_{A\, \Lambda\Sigma} &
\mathcal{T}_{A\, \Lambda}{}^{\Sigma} \\
  \end{array}
\right)\, ,
\,\,\,\,\,
\mbox{with}
\,\,\,\,\,
\mathcal{T}_{A}{}^{\Lambda}{}_{\Sigma}
=
-\mathcal{T}_{A\, \Sigma}{}^{\Lambda}\, ,
\hspace{.5cm}
\mathcal{T}_{A\, \Lambda\Sigma}
=
\mathcal{T}_{A\, \Sigma\Lambda}\, ,
\hspace{.5cm}
\mathcal{T}_{A}{}^{\Lambda\Sigma}
=
\mathcal{T}_{A}{}^{\Sigma\Lambda}\, .
\end{equation}
}

\begin{equation}
k_{A}{}^{m}\frac{\partial \mathcal{N}_{\Lambda\Sigma}}{\partial \phi^{m}}
=
\mathcal{T}_{A\, \Lambda\Sigma}
-\mathcal{N}_{\Lambda\Omega}\mathcal{T}_{A}{}^{\Omega}{}_{\Sigma}
+\mathcal{T}_{A\, \Lambda}{}^{\Omega}\mathcal{N}_{\Omega\Sigma}
-\mathcal{N}_{\Lambda}\mathcal{T}_{A}{}^{\Omega\Delta}
\mathcal{N}_{\Delta\Sigma}\, ,
\end{equation}

\noindent
and, replacing it in Eq.~(\ref{eq:kS2}), we find the conservation equations as
combinations of the equations of motion:

\begin{equation}
\label{eq:kS3}
k_{A}{}^{m}\frac{\delta S}{\delta \phi^{m}}
=
-2\left[
d\star j^{(\sigma)}_{A}
-2\mathcal{T}_{A}{}^{M}{}_{N}\mathcal{F}_{M}\wedge \mathcal{F}^{N}
\right]
=
0\, .
\end{equation}

The conservation of the NGZ 1-forms can be solved locally by the introduction
of 2-forms $B_{A}$ such that

\begin{equation}
\star j_{A}=dB_{A}\, ,
\,\,\,\,\,
\Rightarrow
\,\,\,\,\,
\star j^{(\sigma)}_{A}
=
dB_{A}
+2 \mathcal{T}_{A}{}^{M}{}_{N}\mathcal{F}^{N}\wedge \mathcal{A}_{M}
\equiv
H_{A}\, ,
\end{equation}

\noindent
where $H_{A}$ are the 3-form field strengths, gauge invariant under

\begin{equation}
\delta_{\sigma}\mathcal{A}^{M}=d\sigma^{M}\, ,
\hspace{1cm}
\delta_{\sigma}B_{A}
=
d\sigma_{A}
-2 \mathcal{T}_{A}{}^{M}{}_{N}\mathcal{F}^{N}\wedge d\sigma_{M}\, ,
\end{equation}

\noindent
and satisfying the Bianchi identities

\begin{equation}
dH_{A} -2 \mathcal{T}_{A\,   MN}\mathcal{F}^{M}\wedge \mathcal{F}^{N}
=
0\, .
\end{equation}

The equations of motion have the same form as in the general case studied in
Section~\ref{sec-dualization1}.

Observe that the NGZ currents are subject to the same constraint as the
Noether currents, Eq.~(\ref{eq:relationbetweencurrents}), because
$k_{A}{}^{m}\Gamma_{\rm Adj}(u)^{A}{}_{i}=0$ and because of
Eq.~(\ref{eq:kS3}).\footnote{Otherwise, the NGZ currents would represent too
  many degrees of freedom.} Together, they lead to the constraints

\begin{equation}
\label{eq:funnyconstraint}
\mathcal{T}_{i\, MN}\Gamma(u^{-1})^{M}{}_{P} \Gamma(u^{-1})^{N}{}_{Q}
\mathcal{F}^{P}\wedge \mathcal{F}^{Q}=0\, ,
\,\,\,\,\,
\mbox{and}
\,\,\,\,\,
H_{A}\Gamma_{\rm Adj}(u)^{A}{}_{i}\, .
\end{equation}

\subsection{Supersymmetry and the momentum map}
\label{sec-susyandmm}

\subsubsection{Supersymmetry transformations of $(d-2)$-forms}
\label{sec-susytrans}

In supersymmetric theories the $(d-2)$-form fields dual to the scalars,
$B_{A}$, must transform under supersymmetry and the algebra of the
supersymmetry transformations acting on these fields,
$\delta_{\epsilon}B_{A}$, must close on shell.

In the $\mathcal{N}=1,2$, $d=4$ cases \cite{Bergshoeff:2007ij,Hartong:2009az}
these transformations were found to have leading terms with a common
structure\footnote{The $\mathcal{N}=2$ cases, Special-K\"ahler and
  Quaternionic-K\"ahler target spaces for the scalars, are reviewed in
  Appendices~\ref{sec-KH2-forms} and \ref{sec-QK2forms}, respectively.}  that
can be generalized to all $\mathcal{N}$ and, actually, to all $d$:

\begin{equation}
\label{eq:susyBglobal}
  \begin{array}{rcl}
\delta_{\epsilon}B_{A\, \mu_{1}\cdots \mu_{(d-2)}}
& \sim &
P_{A}{}^{i}
(M_{i})^{I}{}_{J}
\bar{\epsilon}^{J}\gamma_{[\mu_{1}\cdots\mu_{(d-3)}}\psi_{\mu_{(d-2)}]\, I}
\\
& & \\
& &
+
\mathcal{D}_{m}P_{A}{}^{i}
(M_{i})^{I}{}_{J}
\bar{\epsilon}^{J}\gamma_{\mu_{1}\cdots\mu_{(d-2)}}\lambda^{m}{}_{I}
+\cdots
\end{array}
\end{equation}

\noindent
In this expression $I,J$ are R-symmetry indices (that is, a representation of
H), $\psi_{\mu I}$ are the gravitini, $\lambda^{m}{}_{I}$ are dilatini or,
more generally, the supersymmetric partners of the scalars, labeled here by
$m,n,p$, $(M_{i})^{I}{}_{J}$ are the generators of the Lie algebra of H in the
same representation, and the $P_{A}{}^{i}$ are the momentum maps of the
isometries of the coset space G/H or the holomorphic and tri-holomorphic
momentum maps of K\"ahler-Hodge and quaternionic-K\"ahler spaces in in
$\mathcal{N}=1,2$, $d=4$ theories.\footnote{These cases are reviewed in
  Appendices~\ref{sec-KH} and \ref{sec-QK}.} Henceforth, the index $A$ is a
``global'' adjoint G index. $\mathcal{D}$ is the $H$-covariant derivative
acting on the momentum map (or the K\"ahler- or $\mathcal{SU}(2)$-covariant
derivatives in the K\"ahler-Hodge and quaternionic-K\"ahler cases and,
according to the previous observation, only the $i$ index has to be
covariantized for. The additional terms in this supersymmetry transformation
rule are proportional to other $p$-forms of the theory and are associated to
the Chern-Simons terms in the $(d-1)$-form field strengths $H_{A}$.

The second term in this proposal adopts slightly different forms depending on
the theory under consideration. First of all, one can always apply the main
property of the momentum map Eq.~(\ref{eq:dP=kRgeneral}) which also appears in
different guises: Eqs.~(\ref{eq:DPRk}) in coset spaces G/H,
(\ref{eq:holomorphicmomentummap}) and (\ref{eq:dP=k}) in K\"ahler-Hodge spaces
and, finally, (\ref{eq:k=DP}) in quaternionic-K\"ahler spaces. Then, one can
use different properties of the H-curvature so that it does not appear
explicitly: Eq.~(\ref{eq:Rtheta}) in coset spaces G/H, the K\"ahler-Hodge
condition that identifies the K\"ahler 2-form $\mathcal{J}$ with the curvature
of the complex bundle in K\"ahler-Hodge spaces, and the condition that relates
the curvature of the $\mathrm{SU}(2)$ connection to the hyperK\"ahler
structure Eq.~(\ref{eq:FproptoJ}) in quaternionic-K\"ahler manifolds.

Furthermore, observe that in the second term of
Eq.~(\ref{susytrafohyper2form}), the hyperini $\zeta^{\alpha}$ carry a single
$\mathrm{Sp}(2n_{h})$ index $\alpha$ but their product with the Quadbein
$\mathsf{U}_{\alpha I}{}^{u}$ carries an R-symmetry index $I$ plus a
hyperscalar index $u$, according to the general expectation.

The example in Section~\ref{sec-n2bd10} provides additional confirmation of
the universality of the above supersymmetry transformation rule.

The above proposal looks different from the exact result obtained in
superspace for the $\mathcal{N}=8,d=4$ theory in Ref.~\cite{Bandos:2015ila},
but one has to take into account that the 2-forms and their 3-form field
strengths in that reference carry ``local'' (H=SU$(8)$) indices instead of
``global'' adjoint E$_{7(+7)}$ indices. The relation between these two sets of
variables is

\begin{equation}
\mathcal{B}_{A}\equiv B_{B}\Gamma_{\rm Adj}(u)^{B}{}_{A}\, ,
\end{equation}

\noindent
and it is not difficult to see that the supersymmetry transformation rules
Eq.~(\ref{eq:susyBglobal}) split into

\begin{eqnarray}
\delta_{\epsilon}\mathcal{B}_{i\, \mu_{1}\cdots \mu_{(d-2)}}
& \sim &
(M_{i})^{I}{}_{J}
\bar{\epsilon}^{J}\gamma_{[\mu_{1}\cdots\mu_{(d-3)}}\psi_{\mu_{(d-2)}]\, I}
+\cdots
\\
& & \nonumber \\
\delta_{\epsilon}\mathcal{B}_{a\, \mu_{1}\cdots \mu_{(d-2)}}
& \sim &
f_{ab}{}^{i}
(M_{i})^{I}{}_{J}
\bar{\epsilon}^{J}\gamma_{\mu_{1}\cdots\mu_{(d-2)}}e^{b}{}_{m}\lambda^{m}{}_{I}
+\cdots
\end{eqnarray}

In the particular case of $\mathcal{N}=8,d=4$ supergravity we can make use of
the results of Ref.~\cite{Bandos:2015ila}. Using Weyl spinor notation and
taking into account that

\begin{enumerate}
\item The superpartners of scalar fields $\lambda_{\underline{\alpha}\,
    I}^{ABCD}$ are split on two Weyl spinors of the form

\begin{equation}
\lambda_{\alpha}^{ABCD\; I}
=
\tfrac{2}{4!} \epsilon^{ABCDIJKL} \chi_{\alpha\, JKL}\, ,
\,\,\,\,
\mbox{and}
\,\,\,\,
\bar{\lambda}_{\dot\alpha}^{ABCD}{}_{I}
=
-2 \delta_{I}^{[A}\bar{\chi}_{\dot{\alpha} }{}^{BCD]}\, .
\end{equation}

\item The structure constants corresponding to the commutators
  $[\mathfrak{h},\mathfrak{h}]\in \mathfrak{k}$, generically denoted by
  $f_{ab}{}^{i}$ in our review, take in this case the form

\begin{equation}
f_{ABCD\, EFGH}{}^{I}{}_{J}
=
\tfrac{1}{72}
(\delta_{[A}^{I}\epsilon_{BCD]EFGHJ}-\delta_{[E}^{I}\epsilon_{FGH]ABCDJ})\, ,
\end{equation}

\noindent
and the $\mathfrak{h}$ generators and $M_{i}{}^{I}{}_{J}
\mapsto T_{
  \begin{smallmatrix}
    K \\ L \\
  \end{smallmatrix}
}{}^{I}{}_{J}= \delta^{K}{}_{J} \delta^{I}{}_{L}-
\tfrac{1}{8}\delta^{K}{}_{L}\delta^{I}{}_{J}$,

\end{enumerate}

\noindent
we find that the above generic equations, that can be extracted from the
superfield results of Ref.~\cite{Bandos:2015ila}, take the explicit form

\begin{eqnarray}
\delta_{\epsilon} \mathcal{B}_{
  \begin{smallmatrix}
    I \\ J \\
  \end{smallmatrix}\, \, \mu\nu
}
& \propto &
T_{
  \begin{smallmatrix}
    I \\ J \\
  \end{smallmatrix}
}{}^{K}{}_{L}
i(\epsilon_{K}\sigma_{[\mu}\bar{\psi}_{\nu]}^{L}
+\psi_{K[\mu}\sigma_{\nu]}\bar{\epsilon}^{L})\, ,
\\
& & \nonumber \\
\delta_{\epsilon} \mathcal{B}_{IJKL\, \mu\nu}
& \propto &
\epsilon_{[I}\sigma_{\mu\nu}\chi_{JKL]}
-\tfrac{1}{4!}  \epsilon_{IJKLABCD}\bar{\epsilon}^{A}\tilde{\sigma}_{\mu\nu}
\bar{\chi}^{BCD} \, .
\end{eqnarray}

On the other hand, the corresponding relation for the 3-form field strengths

\begin{equation}
\mathcal{H}_{A}\equiv H_{B}\Gamma_{\rm Adj}(u)^{B}{}_{A}\, ,
\end{equation}

\noindent
together with Eq.~(\ref{eq:funnyconstraint}) explain, from a technical point
of view, why the $\mathcal{H}_{i}$ were found in Ref.~\cite{Bandos:2015ila} to
be dual to fermion bilinears.

\subsubsection{Tensions of  supersymmetric $(d-1)$-branes}
\label{sec-tensions}

The supersymmetry transformations of the $(d-2)$-forms into the gravitini
determine the tension of the $1/2$-supersymmetric $(d-3)$-branes that couple
to them in a $\kappa$-symmetric action (see, for instance,
Ref.~\cite{Bergshoeff:2007ij}), if any. The explicit construction of the
U-duality-\textit{invariant} and $\kappa$-symmetric actions of the
$1/2$-supersymmetric $(d-3)$-branes, in the same spirit as the construction of
the SL$(2,\mathbb{R})$-\textit{invariant} actions for all branes in type~IIB
$d=10$ supergravity in Ref.~\cite{Bergshoeff:2006gs} or for $0$-branes in
$\mathcal{N}=2,8,d=4$ supergravity in Ref.~\cite{Billo:1999ip} has only been
carried out for the $\mathcal{N}=2,d=4$ case
\cite{Bergshoeff:2007ij}. Nevertheless, some general lessons can be learned
from those results and from the general form of the supersymmetry
transformations of $(d-2)$-form potentials Eq.~(\ref{eq:susyBglobal}).

On general grounds, $(d-3)$-branes will be characterized by charges in the
adjoint representation of G, $q^{A}$ and the Wess-Zumino term in their
effective world-volume action will contain the leading
term\footnote{T-duality-invariant Wess-Zumino terms for the $\kappa$-symmetric
  world-volume effective actions of all branes in maximal supergravity in any
  dimension have been proposed in Ref.~\cite{Bergshoeff:2011zk}.}

\begin{equation}
q^{A}\int B_{A}\, .
\end{equation}

Then, the supersymmetry transformation rule Eq.~(\ref{eq:susyBglobal})
requires the presence of a scalar-dependent factor in the kinetic term that
can be identified with the tension $\mathcal{T}_{(d-3)}$:

\begin{equation}
\int d^{(d-3)}\xi \mathcal{T}_{(d-3)} \sqrt{|g_{(d-3)}|}\, ,
\end{equation}

\noindent
which we conjecture to be of the form

\begin{equation}
\mathcal{T}_{(d-3)} = \sqrt{|q^{A}q^{B}P_{A}{}^{i}P_{B}{}^{j}g_{ij}|}\, .
\end{equation}

Observe that the rank dim$H$ matrix $P_{A}{}^{i}P_{B}{}^{j}g_{ij}$ is related
to the matrix $\mathfrak{M}^{AB}$ defined in Eq.~(\ref{eq:MABdef}) by

\begin{equation}
P_{A}{}^{i}P_{B}{}^{j}g_{ij} = g_{AB} -g_{AC}g_{BD}\mathfrak{M}^{CD}\, ,
\end{equation}

\noindent
and for $(d-3)$-brane charges in the conjugacy class $q^{A}q^{B}g_{AB}=0$
(which is the conjugacy class of the D7-and S7-branes of $\mathcal{N}=2B,d=10$
supergravity \cite{Bergshoeff:2010xc})

\begin{equation}
\mathcal{T}_{(d-3)} = \sqrt{|q_{A}q_{B}\mathfrak{M}^{CD}|} \, ,
\end{equation}

\noindent
which is the expression one would have guessed from Eq.~(\ref{eq:jsaction}).

Clearly, more work is needed in order to find the complete $\kappa$-invariant
worldvolume actions, find the U-duality-invariant $(d-3)$-brane tensions and,
eventually, prove the above conjecture, but we think that our arguments
concerning the general structure of the supersymmetry transformation
Eq.~(\ref{eq:susyBglobal}) give some support to it.

\subsubsection{Fermion shifts}
\label{sec-fermionshifts}

The holomorphic and triholomorphic momentum maps (resp.  $\mathcal{P}_{A}$ and
$\mathsf{P}_{A}{}^{x}$) also appear naturally in the so-called \textit{fermion
  shifts} of the supersymmetry transformations of the fermions of gauged
$\mathcal{N}=1,2,d=4$ supergravities.  For the standard gauging (using the
fundamental vectors $A^{\Lambda}$ as gauge fields for perturbative symmetries
of the action), the supersymmetry transformations of the gravitini, gaugini
and hyperini of $\mathcal{N}=2,d=4$ supergravity can be written in the
form:\footnote{See, for instance,
  Refs.~\cite{Andrianopoli:1996cm,Freedman:2012zz,Ortin:2015hya}. The momentum
  maps carry an index $\Lambda$ here which associates them to the vector field
  that gauges the corresponding global symmetry. It is understood in this
  notation that only the momentum maps associated to the gauge symmetries
  occur in these expressions. This notation is considerably improved by the
  introduction of the embedding tensor, as awe are going to see.}

\begin{equation}
\left\{
\begin{array}{rcl}
\delta_{\epsilon}\psi_{I\, \mu}
& = &
\mathfrak{D}_{\mu}\epsilon_{I}
+
\left[
T^{+}{}_{\mu\nu}\varepsilon_{IJ}
-\tfrac{1}{2}S^{x}\eta_{\mu\nu}\varepsilon_{IK}(\sigma^{x})^{K}{}_{J}
\right]
\gamma^{\nu}\epsilon^{J}\, ,
\\
& & \\
\delta_{\epsilon}\lambda^{Ii}
& = &
i\not\!\!\mathfrak{D} Z^{i}\epsilon^{I}
+
\left[
\left(\not\!G^{i\, +} +W^{i}\right)\varepsilon^{IJ}
+\tfrac{i}{2}W^{i\, x}\, (\sigma^{x})^{I}{}_{K}\varepsilon^{KJ}
\right]
\epsilon_{J}\, ,\\
& & \\
\delta_{\epsilon}\zeta_{\alpha} & = &
i\mathsf{U}_{\alpha I\, u}\not\!\!\mathfrak{D}q^{u}\epsilon^{I}
+N_{\alpha}{}^{I}\epsilon_{I}\, ,
\end{array}
\right.
\end{equation}

\noindent
where the fermion shifts are given by

\begin{equation}
\left\{
\begin{array}{rcl}
S^{x}
& = &
{\textstyle\frac{1}{2}} g
\mathcal{L}^{\Lambda} \mathsf{P}_{\Lambda}{}^{x}\, ,
\\
& & \\
W^{i}
& = &
{\textstyle\frac{1}{2}}g\mathcal{L}^{*\, \Lambda}k_{\Lambda}{}^{i}
=
-{\textstyle\frac{i}{2}}
g\mathcal{G}^{ij^{*}}f^{*\Lambda}{}_{j^{*}} \mathcal{P}_{\Lambda}\, ,
\\
& & \\
W^{i\,x}
& = &
g\mathcal{G}^{ij^{*}} f^{*\,\Lambda}{}_{j^{*}}
\mathsf{P}_{\Lambda}{}^{x}\, ,
\\
& & \\
N_{\alpha}{}^{I}
& = &
g\mathsf{U}_{\alpha}{}^{I}{}_{u}\mathcal{L}^{*\,\Lambda}
\mathsf{k}_{\Lambda}{}^{u}\, .
\end{array}
\right.
\end{equation}

As usual in $\mathcal{N}>1$, the scalar potential is given by an expression
quadratic in the fermion shifts:

\begin{equation}
\label{eq:potgen}
\begin{array}{rcl}
V(Z,Z^{*},q)
& = &
-6S^{*\,x}S^{x}
+2\mathcal{G}_{ij^{*}}W^{i} W^{*j^{*}}
+\tfrac{1}{2}\mathcal{G}_{ij^{*}}W^{i\ x} W^{*j^{*}\ x}
+2N_{\alpha}{}^{I} N^{\alpha}{}_{I}
\\
& & \\
& = &
g^{2}\left[
-{\textstyle\frac{1}{4}}
\Im\mathfrak{m}\mathcal{N}^{\Lambda\Sigma}\mathcal{P}_{\Lambda}\mathcal{P}_{\Sigma}
+{\textstyle\frac{1}{2}} \mathcal{L}^{*\,\Lambda} \mathcal{L}^{\Sigma}
(4\mathsf{H}_{uv}\mathsf{k}_{\Lambda}{}^{u} \mathsf{k}_{\Sigma}{}^{v}
-3\mathsf{P}_{\Lambda}{}^{x} \mathsf{P}_{\Sigma}{}^{x})
\right.
\\
& & \\
& &
\hspace{.5cm}\left.
+{\textstyle\frac{1}{2}}\mathcal{G}^{ij^{*}}f^{\Lambda}{}_{i}f^{*\, \Sigma}{}_{j^*}
\mathsf{P}_{\Lambda}{}^{x} \mathsf{P}_{\Sigma}{}^{x}\ \right]\, .
\end{array}
\end{equation}

The presence of the momentum map on all those terms is due to their
transformations properties under global duality transformations. To make this
fact manifest and gain more insight in the structure of these terms, it is
convenient to use the \textit{embedding tensor}\footnote{The embedding tensor
  and its associated formalism were introduced in
  Refs.~\cite{Cordaro:1998tx,Nicolai:2000sc,Nicolai:2001sv}. They were 
  developed in the context of the maximal 4-dimensional supergravity in
  Refs.~\cite{deWit:2002vt,deWit:2005ub}, but its use is by no means
  restricted to that context (see Chapter~2 in Ref.~\cite{Ortin:2015hya} and
  references therein.} $\vartheta_{M}{}^{A}$. This object relates each
symmetry generator ($A$ index) to the vector field of the theory that gauges
it ($M$ index). Introducing the embedding tensor in the fermion shifts
restores the (formal) symplectic invariance of the theory\footnote{The details
  of such a general gauged theory have not yet been worked out in the
  literature.}:\footnote{The gauge coupling constant $g$ is also replaced by
  the embedding tensor, since it can describe several gauge groups with
  different coupling constants.}

\begin{equation}
\left\{
\begin{array}{rcl}
S^{x}
& = &
{\textstyle\frac{1}{2}}
\mathcal{V}^{M}\vartheta_{M}{}^{A} \mathsf{P}_{A}{}^{x}\, ,
\\
& & \\
W^{i}
& = &
-{\textstyle\frac{i}{2}}\mathcal{D}^{i}\mathcal{V}^{*M}\vartheta_{M}{}^{A}
 \mathcal{P}_{A}\, ,
\\
& & \\
W^{i\,x}
& = &
\mathcal{D}^{i}\mathcal{V}^{*M}\vartheta_{M}{}^{A}
\mathsf{P}_{A}{}^{x}\, ,
\\
& & \\
N_{\alpha}{}^{I}
& = &
g\mathsf{U}_{\alpha}{}^{I}{}_{u}\mathcal{V}^{*\, M}\vartheta_{M}{}^{A}
\mathsf{k}_{A}{}^{u}\, ,
\end{array}
\right.
\end{equation}

\noindent
while the scalar potential must take the form

\begin{equation}
\begin{array}{rcl}
V(Z,Z^{*},q)
& = &
-{\textstyle\frac{1}{4}}
\mathcal{M}^{MN}\vartheta_{M}{}^{A}\vartheta_{N}{}^{B}
\mathcal{P}_{A}\mathcal{P}_{B}
+{\textstyle\frac{1}{2}} \mathcal{V}^{*\, M} \mathcal{V}^{N}
\vartheta_{M}{}^{A}\vartheta_{N}{}^{B}
(4\mathsf{H}_{uv}\mathsf{k}_{A}{}^{u} \mathsf{k}_{B}{}^{v}
-3\mathsf{P}_{A}{}^{x} \mathsf{P}_{B}{}^{x})
\\
& & \\
& &
\hspace{.5cm}
+{\textstyle\frac{1}{2}}\mathcal{G}^{ij^{*}}
\mathcal{D}_{i}\mathcal{V}^{M}\mathcal{D}_{j^{*}}\mathcal{V}^{*\, N}
\vartheta_{M}{}^{A}\vartheta_{N}{}^{B}
\mathsf{P}_{A}{}^{x} \mathsf{P}_{B}{}^{x}\, .
\end{array}
\end{equation}

Our general definition of momentum map shares the same transformation
properties and, therefore, the momentum maps should occur in all the fermion
shifts of all theories. The expressions given in the literature, though, are
written in a different language which obscures this point. Here we are going
to show in several examples how the momentum map allows one to rewrite the
fermion shifts in a universal way if one makes use of the embedding tensor.

Let us consider first the $\mathcal{N}>2,d=4$ theories with vector multiplets
(whenever possible). It is convenient to use the formulation of
Ref.~\cite{Andrianopoli:1996ve} that can describe all these theories
simultaneously and in a language very close to that of the $\mathcal{N}=2,d=4$
theories coupled to vector multiplets.\footnote{So far, this formalism has
  been used only in ungauged supergravities. Our proposals for the fermion
  shifts should help to extend this formulation to the most general gauged
  theories.}

We just need to know some details of this formulation: the $\mathcal{N}=2$
symplectic section $\mathcal{V}^{M}$ that describes the scalars in the vector
multiplets and its K\"ahler-covariant derivative
$\mathcal{D}_{i}\mathcal{V}^{M}$ are now generalized to $\mathcal{V}^{M}_{IJ}=
-\mathcal{V}^{M}_{JI}$ and $\mathcal{V}^{M}_{i}$ where the indices
$I,J=1,\cdots,\mathcal{N}$ and $i,j=1,\cdots, n_{V}$ (the number of vector
multiplets). The fermions in the supergravity multiplet are $\psi_{\mu\,
  I},\chi_{IJK},\chi^{IJKLM}$ (antisymmetric in all the SU$(\mathcal{N})$
indices, and $\chi^{IJKLM}=\tfrac{1}{3!}\varepsilon^{IJKLMNOPQ}\chi_{OPQ}$ for
$\mathcal{N}=8$).\footnote{$\chi^{IJKLM}$ is only relevant as an independent
  field for $\mathcal{N}=5$, because it is also related to another field for
  $\mathcal{N}=6$, as awe are going to see.} The fermions in the generic
vector supermultiplet are $\lambda_{iI}$ and $\lambda_{i}{}^{IJK}$ (again,
antisymmetric in all the SU$(\mathcal{N})$ indices, and $\lambda_{i}{}^{IJK} =
\varepsilon^{IJKL}\lambda_{iL}$ for $\mathcal{N}=4$). There are no vector
multiplets for $\mathcal{N}>4$. However, in this formalism, several fields of
the $\mathcal{N}=6$ theory are treated as if belonging to a vector
supermultiplet and one has two additional relations that relate them to fields
in the supergravity multiplet
$\lambda_{I}=\tfrac{1}{5!}\varepsilon_{IJ_{1}\cdots J_{5}}\chi^{J_{1}\cdots
  J_{5}}$ and $\lambda^{IJK}=\tfrac{1}{3!}\varepsilon^{IJKLMN}\chi_{LMN}$. In
practice, in $\mathcal{N}=6$, it is easier to work with $\lambda_{I}$ and
$\chi_{IJK}$, which fit in the general pattern.

Combining this knowledge with the fermion shifts of the $\mathcal{N}=2$
theories written above,\footnote{Evidently, the fermion shifts in the hyperini
  will not be generalized, as there are no hypermultiplets in $\mathcal{N}\neq
  2,d=4$ theories.} it is not difficult to guess the form of the generic
fermion shifts:

\begin{eqnarray}
\delta_{\epsilon}\psi_{\mu\, I}
& \sim &
\cdots
+\mathcal{V}^{M}{}_{IK}\vartheta_{M}{}^{A}
P_{A}{}^{\textbf{i}}(M_{\textbf{i}})^{K}{}_{J}
\gamma_{\mu} \epsilon^{J}\, ,
\\
& & \nonumber \\
\delta_{\epsilon}\chi_{IJK}
& \sim &
\cdots
+\mathcal{V}^{M}{}_{[IJ|}\vartheta_{M}{}^{A}
P_{A}{}^{\textbf{i}}(M_{\textbf{i}})^{L}{}_{|K]}\epsilon_{L}
\\
& & \nonumber \\
\delta_{\epsilon}\lambda_{iI}
& \sim &
\cdots
+\mathcal{V}^{M}{}_{i}\vartheta_{M}{}^{A}
P_{A}{}^{\textbf{i}}(M_{\textbf{i}})^{J}{}_{I}\epsilon_{J}\, ,
\end{eqnarray}

\noindent
where we have boldfaced the $H$ indices to distinguish them from those
labeling the vector supermultiplets. For the $\mathcal{N}=3,5$ cases there are
additional fermion fields which are independent of $\psi_{\mu\,
  I},\chi_{IJK},\lambda_{i\, I}$ and whose fermion shifts are more difficult
to guess. We have found the following possibilities:\footnote{We thank Mario
  Trigiante for enlightening conversations on this point.}

\begin{enumerate}
\item For the SU$(3)$ singlets $\lambda_{i} =
  \tfrac{1}{3!}\varepsilon_{IJK}\lambda_{i}{}^{IJK}$ of $\mathcal{N}=3$

\begin{equation}
  \delta_{\epsilon}\lambda_{i}
  \sim
  \cdots
  + \varepsilon_{IJK}\mathcal{V}^{M}_{i}\vartheta_{M}{}^{A}
  P_{A}{}^{\textbf{i}}(M_{\textbf{i}})^{I}{}_{L}\delta^{LJ}\epsilon^{K}\, .
\end{equation}

\item For the SU$(5)$ singlet $\chi = \tfrac{1}{5!}\varepsilon_{I_{1}\cdots
    I_{5}}\chi^{I_{1}\cdots I_{5}}$ of $\mathcal{N}=5$

\begin{equation}
  \delta_{\epsilon}\chi
  \sim
  \cdots
  +
  \varepsilon^{I_{1}I_{2}I_{3}I_{4}I_{5}}
  \mathcal{V}^{M}{}_{I_{1}I_{2}}\vartheta_{M}{}^{A}
  P_{A}{}^{\textbf{i}}\delta_{I_{3}J}(M_{\textbf{i}})^{J}{}_{I_{4}}
\epsilon_{I_{5}}\, .
\end{equation}
\end{enumerate}

In many gauged supergravities (see, for instance the $\mathcal{N}=3,d=4$
theories \cite{Castellani:1985ka,Castellani:1985wk} or the $\mathcal{N}=2,d=8$
theories \cite{Salam:1984ft,AlonsoAlberca:2003jq}), the fermion shifts are
given in terms of the ``dressed structure constants'' of the gauge group. In
the SO$(3)$-gauged $\mathcal{N}=2,d=8$ theory of Ref.~\cite{Salam:1984ft}, and
in the conventions used there, these are defined by

\begin{equation}
f_{ij}{}^{k} \equiv L_{i}{}^{m}  L_{j}{}^{n} L_{p}{}^{k}f_{mn}{}^{p}\, ,
\,\,\,\,\,
\mbox{where}
\,\,\,\,\,
f_{mn}{}^{p}=\epsilon_{mnp}\, ,
\end{equation}

\noindent
and where $L_{i}{}^{m}$ is the SL$(3,\mathbb{R})/$SO$(3)$ coset representative
($m,n,p=1,2,3$ are indices in the fundamental (vector) representation of
SL$(3,\mathbb{R})$ and $i,j,k=1,2,3$ are indices in the fundamental
representation of SO$(3)$) and $L_{m}{}^{i}=(L^{-1})_{m}{}^{i}$ is its
inverse.\footnote{In our notation $L_{i}{}^{m}=u^{m}{}_{i}$, the transposed.}

The dressed structure constants can be rewritten in terms of the three
momentum maps $P_{m}{}^{n}$ and the three Killing vectors $k_{m}{}^{a}$ (where
$a$ labels the generators of the coset) using the definition of the adjoint
action of the group on the algebra, adapted to these conventions:

\begin{equation}
f_{ij}{}^{k}
=
L_{i}{}^{m}  L_{j}{}^{n} (T_{m})_{n}{}^{p} (L^{-1})_{p}{}^{k}
=
L_{i}{}^{m}  \Gamma_{\rm Adj}(L^{-1})_{m}{}^{\alpha} (T_{\alpha})_{j}{}^{k}
=
L_{i}{}^{m} \left[ P_{m}{}^{l} (T_{l})_{j}{}^{k}
- k_{m}{}^{a} (T_{a})_{j}{}^{k} \right]\, ,
\end{equation}

\noindent
where we have used the (transposed of the) definition of the adjoint action of
the group on the algebra Eq.~(\ref{eq:adjointaction}).

Similar identities can be used in other supergravities and we hope the use of
the momentum map can be of help in writing all gauged supergravity theories in
a homogenous language.

\subsection{Examples}

\subsubsection{$\mathcal{N}=4,d=4$ supergravity}

The bosonic fields of pure $\mathcal{N}=4,d=4$ supergravity are the metric,
the axidilaton $\tau=\chi+ie^{-2\varphi}$,\footnote{Observe that here
  $\Im\mathfrak{m}\, \tau= e^{-2\varphi}$ instead of $e^{-\varphi}$ as in
  Section~\ref{eq:SL2}.} and six vector fields $A^{\Lambda}{}_{\mu}$
$\Lambda=1,\cdots,6$. The bosonic action is

\begin{equation}
S= \int d^{4}x\sqrt{|g|}
\left\{
R+ \frac{\partial_{\mu}\tau\partial^{\mu}\tau^{*}}{2(\Im\mathfrak{m}\,
  \tau)^{2}}
-\tfrac{1}{4}e^{-2\varphi}F^{\Lambda}{}_{\mu\nu}F^{\Lambda\, \mu\nu}
+\tfrac{1}{4}\chi F^{\Lambda}{}_{\mu\nu}\star F^{\Lambda\, \mu\nu}
\right\}\, ,
\end{equation}

\noindent
and, comparing with the generic action Eq.~(\ref{eq:genericd4action}) we find
that the period matrix is given by

\begin{equation}
\mathcal{N}_{\Lambda\Sigma} =-\tfrac{1}{8}\tau\delta_{\Lambda\Sigma}\, .
\end{equation}

The action of this theory is invariant under SO$(6)$ rotations of the vector
fields, whose symplectic infinitesimal generators, labeled by $a$ are

\begin{equation}
(\mathcal{T}_{a}{}^{M}{}_{N})
=
\left(
  \begin{array}{cc}
   T_{a}{}^{\Lambda}{}_{\Sigma} &     0    \\
                0            &  T_{a\, \Lambda}{}^{\Sigma} \\
  \end{array}
\right)\, ,
\,\,\,\,\,
\mbox{where}
\,\,\,\,\,
T_{a}{}^{\Lambda}{}_{\Sigma} = -T_{a\, \Sigma}{}^{\Lambda}\, .
\end{equation}

\noindent
The equations of motion are also invariant under the SL$(2\mathbb{R})$ group
of simultaneous electric-magnetic rotation of all the electric field strengths
$F^{\Lambda}$ into the dual magnetic ones $G_{\Lambda}$ defined in
Eq.~(\ref{eq:dualvectorfieldstrengths}). The symplectic generators associated
to these transformations are the tensor products of those in
Eq.~(\ref{eq:SL2generators}) by the identity in 6 dimensions. More explicitly

\begin{equation}
  \begin{array}{rcl}
(\mathcal{T}_{1}{}^{M}{}_{N})
& = &
\tfrac{1}{2}
\left(
  \begin{array}{cc}
   \delta^{\Lambda}{}_{\Sigma} &     0    \\
                0            &  -\delta_{\Lambda}{}^{\Sigma} \\
  \end{array}
\right)\, ,
\hspace{1.5cm}
(\mathcal{T}_{2}{}^{M}{}_{N})
=
\tfrac{1}{2}
\left(
  \begin{array}{cc}
   0   &     \delta^{\Lambda\Sigma}    \\
   \delta_{\Lambda\Sigma} &    0       \\
  \end{array}
\right)\, ,
\\
& & \\
(\mathcal{T}_{3}{}^{M}{}_{N})
& = &
\tfrac{1}{2}
\left(
  \begin{array}{cc}
   0   &     \delta^{\Lambda\Sigma}    \\
   -\delta_{\Lambda\Sigma} &    0       \\
  \end{array}
\right)\, .
\end{array}
\end{equation}

We will denote the generators of SL$(2,\mathbb{R})$ with the label $\alpha$ to
distinguish them from those of the SO$(6)$ group. Observe that there are no
scalars associated to SO$(6)$. The only scalar, the axidilaton, is invariant
under SO$(6)$. Since this group is a symmetry of the action, the NGZ current
coincides with the Noether current, has no scalar contribution and is given by

\begin{equation}
j_{a} = -2\mathcal{T}_{a}{}^{M}{}_{N}\star (\mathcal{F}^{N}\wedge
\mathcal{A}_{M})\, .
\end{equation}

Using Maxwell equations and Bianchi identities of the vector fields we find
that

\begin{equation}
d\star j_{a} =
-2\mathcal{T}_{a}{}^{M}{}_{N}\mathcal{F}^{N}\wedge\mathcal{F}_{M}\, ,
\end{equation}

\noindent
which vanishes identically due to the antisymmetry of the SO$(6)$
generators. In this case, evidently, there is nothing to be dualized and there
are no 2-forms $B_{a}$.

The NGZ currents of the SL$(2,\mathbb{R})$ electric-magnetic duality group are
non-trivial, though:

\begin{equation}
j_{\alpha} = j^{(\sigma)}_{\alpha}
-2\mathcal{T}_{\alpha}{}^{M}{}_{N}\star (\mathcal{F}^{N}\wedge
\mathcal{A}_{M})\, ,
\,\,\,\,\,
\alpha=1,2,3\, .
\end{equation}

\noindent
It is necessary to use the equations of motion of the scalars (and not just
the Maxwell equations and Bianchi identities) to show that they are conserved
on-shell. They are dualized into 3 2-forms $B_{\alpha}$ according to the
general prescription. We will not give the details here.

\subsubsection{$\mathcal{N}=8,d=4$ supergravity}

The above general scheme can be applied to $\mathcal{N}=8,d=4$ supergravity
even if we use a complex basis for the vector fields. Thus, we have

\begin{eqnarray}
H_{{\begin{smallmatrix}
    E \\ F \\
  \end{smallmatrix}}
}
& = &
dB_{\begin{smallmatrix}
    E \\ F \\
  \end{smallmatrix}}
-4
\left[
\mathcal{F}^{EA}\mathcal{A}_{FA}
+\mathcal{A}^{EA}\mathcal{F}_{FA}
-\tfrac{1}{8}\delta^{E}{}_{F}
\left(\mathcal{F}^{AB}\mathcal{A}_{AB}
+\mathcal{A}^{AB}\mathcal{F}_{AB}\right)
\right]\, ,
\\
& & \nonumber \\
H_{EFGH}
& = &
dB_{EFGH}
-2\left(\mathcal{F}_{[EF}\mathcal{A}_{GH]}
+\tfrac{1}{4!}\varepsilon_{EFGHABCD}\mathcal{F}^{AB}\mathcal{A}^{CD}\right)\, .
\end{eqnarray}

\section{The higher-dimensional, higher-rank NGZ 1-forms and dualization}
\label{sec-higherdimensionshigherrank}


In $d>4$ dimensions supergravity theories may contain dynamical fields which
are differential forms of rank $p>1$. The global symmetries of the theory can
act on these field as rotations or, when the rank and dimension allow it
($d=2(p+1)$), as electric-magnetic transformations. The latter are not
symmetries of the action but, nevertheless, as in the 4-dimensional case, a
generalized Noether-Gaillard-Zumino (NGZ) current 1-form $j_{A}$ which is
conserved on shell can be defined for each and all the generators of the full
duality group.

The equation that expresses this conservation can be written in a universal
form: let $F^{I},H_{m},G^{a},\ldots$ be, respectively, the 2-, 3-, 4-, \ldots
form field strengths of the $n_{1},n_{2},n_{3},\dots$ fundamental 1-, 2-, 3-,
\ldots fields of the theory and let
$\tilde{F}_{I},\tilde{H}^{m},\tilde{G}_{a},\ldots$ their dual $(d-2)$-,
$(d-3)$-, $(d-4)$-, \dots form field strengths. As the indices chosen show, if
the fundamental field strengths $F^{I},H_{m},G^{a},\ldots$ transform linearly
under the duality group as $\delta_{A}F^{I}=T_{A}{}^{I}{}_{J}F^{J}$,
$\delta_{A}H_{m}=- T_{A}{}^{n}{}_{m}H_{n}$,
$\delta_{A}G^{a}=T_{A}{}^{a}{}_{b}G^{b}$, \ldots, the dual field strengths
must transform in the conjugate representations, that is
$\delta_{A}\tilde{F}_{I}= -T_{A}{}^{J}_{I}\tilde{F}_{J}$,
$\delta_{A}\tilde{H}^{m}=T_{A}{}^{m}{}_{n}\tilde{H}^{n}$,
$\delta_{A}\tilde{G}_{a}=-T_{A}{}^{b}{}_{a}\tilde{G}_{b}$, \ldots The only
exception to these transformation rules are the electric-magnetic
transformations. In $d=4$, for instance, they relate $F^{I}$ to
$\tilde{F}_{I}$ and the pair $(\mathcal{F}^{M}) \equiv \left(
  \begin{smallmatrix}
  F^{I} \\ \tilde{F}_{I} \\
  \end{smallmatrix}
\right)$ transforms as a Sp$(2n_{1},\mathbb{R})$ vector according to
$\delta_{A}\mathcal{F}^{M} = T_{A}{}^{M}{}_{N}\mathcal{F}^{N}$ with $T_{A}{}^{M}{}_{N}\in \mathfrak{sp}(2n_{1},\mathbb{R})$. In $d=6$,
electric-magnetic duality transformations relate $H_{m}$ to $\tilde{H}^{m}$
and the pair $(\mathcal{H}^{M}) \equiv \left(
  \begin{smallmatrix}
  \tilde{H}^{m} \\ H_{m} \\
  \end{smallmatrix}
\right)$ transforms as a SO$(n_{2},n_{2})$ vector according to
$\delta_{A}\mathcal{H}^{M} = T_{A}{}^{M}{}_{N}\mathcal{H}^{N}$ with
$T_{A}{}^{M}{}_{N}\in \mathfrak{so}(n_{1},n_{1})$ etc.

It is not difficult to see through the 5- and 8-dimensional examples we are
going to present next that the equation satisfied by the Noether current
1-forms is always, up to conventional coefficients, of the form

\begin{equation}
-k_{A}{}^{x}
\frac{\delta S}{\delta \phi^{x}}
=
d\star j_{A} +T_{A}{}^{I}{}_{J}F^{J}\wedge \tilde{F}_{I}
+T_{A}{}^{m}{}_{n}\tilde{H}^{n}\wedge H_{m}\cdots =0\, ,
\end{equation}

\noindent
and in the exceptional cases mentioned above, one should replace
$T_{A}{}^{I}{}_{J}F^{J}\wedge \tilde{F}_{I}$ by
$\tfrac{1}{2} T_{A}{}^{M}{}_{N}\mathcal{F}_{M}\wedge \mathcal{F}^{N}$,
$T_{A}{}^{m}{}_{n}\tilde{H}^{n}\wedge H_{m}$ by
$\tfrac{1}{2}T_{A}{}^{M}{}_{N}\mathcal{H}_{M}\wedge \mathcal{H}^{N}$ etc.

On-shell, the above equation would take the form

\begin{equation}
d\star j^{NGZ}_{A}=0\, ,
\end{equation}

\noindent
but it is not possible to give a general form of this current because, in each
theory, the field strengths contain different Chern-Simons terms, all of them
duality-invariant. In the 5-dimensional example that follows, we have found
the explicit form, but in the 8-dimensional one, we have not.

The dualization of the NGZ current 1-forms  into $(d-2)$-form potentials
proceeds as in the 4-dimensional case.

\subsection{Examples}

\subsubsection{$\mathcal{N}=1,d=5$  supergravities}
\label{sec-higherdimensionshigherrankexample1}

The bosonic action of any 5-dimensional ungauged supergravity-like theory with
scalars $\phi^{x}$ and Abelian vector fields $A^{I}$ (in particular,
$\mathcal{N}=1,d=5$ supergravities with vector supermultiplets) can be written
in the form \cite{Hartong:2009az}

\begin{equation}
S
=
\int
\left\{\star R +\tfrac{1}{2}\mathcal{G}_{xy}d\phi^{x}\wedge \star d\phi^{y}
-\tfrac{1}{2}a_{IJ}F^{I}\wedge \star F^{J}
+\tfrac{1}{3}C_{IJK}F^{I}\wedge F^{J} \wedge A^{K}
\right\}\, ,
\end{equation}

\noindent
where $\mathcal{G}_{xy}(\phi)$ is the $\sigma$-model metric, $a_{IJ}(\phi)$ is
the kinetic matrix of the vector fields and $C_{IJK}$ is a constant, symmetric
tensor. In supergravity theories these three couplings are related in a very
precise way, but we will not need to use this structure for our purposes.

The equations of motion of the vector fields are

\begin{equation}
d(a_{IJ}\star F^{J}- C_{IJK}F^{J}\wedge A^{K})
=
0\, ,
\end{equation}

\noindent
and can be solved locally by

\begin{equation}
a_{IJ}\star F^{J}- C_{IJK}F^{J}\wedge A^{K} \equiv d\tilde{A}_{I}\, ,
\end{equation}

\noindent
where the $\tilde{A}_{I}$ are the magnetic 2-forms dual to the vector fields. Their
gauge-invariant field strengths are

\begin{equation}
\tilde{F}_{I} = d\tilde{A}_{I}+C_{IJK}F^{J}\wedge A^{K}\, ,
\,\,\,\,
\Rightarrow
\,\,\,\,
d\tilde{F}_{I}=C_{IJK}F^{J}\wedge F^{K}\, ,
\end{equation}

\noindent
and are related to the vector field strengths by

\begin{equation}
\tilde{F}_{I}= a_{IJ}\star F^{J}\, .
\end{equation}

The equations of motion of the scalars are

\begin{equation}
-\frac{\delta S}{\delta \phi^{z}}
=
\mathcal{G}_{zw}
\left[
d\star d\phi^{w}+\Gamma_{xy}{}^{w}d\phi^{x}\wedge \star d\phi^{y}
\right]
+\tfrac{1}{2}\partial_{z}a_{IJ} F^{I}\wedge \star F^{J}\, .
\end{equation}

If the action is invariant under the global transformations generated by

\begin{equation}
  \begin{array}{rcl}
\delta_{A}\phi^{x}
& = &
k_{A}{}^{x}(\phi)\, ,
\\
& & \\
\delta_{A}A^{I}
& = &
T_{A}{}^{I}{}_{J}A^{J}\, ,
\end{array}
\end{equation}

\noindent
which implies that the functions $k_{A}{}^{x}(\phi)$ are Killing vectors of
the $\sigma$-model metric $\mathcal{G}_{xy}$, the kinetic matrix satisfies

\begin{equation}
k_{A}{}^{x} \partial_{x}a_{IJ} = -2T_{A}{}^{K}{}_{(I}a_{J)K}\, ,
\end{equation}

\noindent
and the symmetric tensor satisfies

\begin{equation}
\label{eq:TC}
T_{A}{}^{L}{}_{(I}C_{JK)L}=0\, ,
\end{equation}

\noindent
we find that

\begin{equation}
-k_{A}{}^{z}\frac{\delta S}{\delta \phi^{z}}
=
d\star j_{A}^{(\sigma)}
-T_{A}{}^{K}{}_{I}a_{JK} F^{I}\wedge \star F^{J}\, .
\end{equation}

In order to dualize the Noether currents, we first have to replace the Hodge dual of
the vector field strengths by the $\tilde{F}_{I}$:

\begin{equation}
-k_{A}{}^{z}\frac{\delta S}{\delta \phi^{z}}
=
d\star j_{A}^{(\sigma)}
-T_{A}{}^{K}{}_{I}F^{I}\wedge \tilde{F}_{K}\, ,
\end{equation}

\noindent
and, then, using the invariance of the  $C_{IJK}$ tensor Eq.~(\ref{eq:TC}) we
get

\begin{equation}
-k_{A}{}^{z}\frac{\delta S}{\delta \phi^{z}}
=
d\left[\star j_{A}^{(\sigma)}
-\tfrac{1}{3}T_{A}{}^{K}{}_{I}(A^{I}\wedge \tilde{F}_{K}+2F^{I}\wedge \tilde{A}_{K})
\right]
=
0\, .
\end{equation}

As usual, we solve locally this equation by introducing 3-form potentials $D_{A}$

\begin{equation}
\star j_{A}^{(\sigma)}
-\tfrac{1}{3}T_{A}{}^{K}{}_{I}(A^{I}\wedge \tilde{F}_{K}+2F^{I}\wedge \tilde{A}_{K})
\equiv
d D_{A}\, ,
\end{equation}

\noindent
with gauge-invariant field strengths and duality relation

\begin{equation}
K_{A}= d D_{A}-\tfrac{1}{3}T_{A}{}^{K}{}_{I}(A^{I}\wedge \tilde{F}_{K}+2F^{I}\wedge
\tilde{A}_{K})\, ,
\hspace{1cm}
\star j_{A}^{(\sigma)} = K_{A}\, .
\end{equation}

\subsubsection{$\mathcal{N}=2,d=8$  supergravity}
\label{sec-higherdimensionshigherrankexample2}

This example is based on the results found in Ref.~\cite{Andino:2016bwk}.  The
possible electric fields in an 8-dimensional theory are scalars $\phi^{x}$,
1-forms $A^{I}$, 2-forms $B_{m}$, and 3-forms $C^{a}$.  The most general
Abelian, massless, ungauged supergravity-like theory in 8 dimensions with this
field content can be written in the form

\begin{equation}
\label{eq:d8Abelianmasslessgeneralaction}
\begin{array}{rcl}
S
& = &
{\displaystyle\int}
\left\{
-\star R +\tfrac{1}{2} \mathcal{G}_{xy}d\phi^{x}\wedge \star d\phi^{y}
+\frac{1}{2}\mathcal{M}_{IJ}F^{I}\wedge \star F^{J}
+\frac{1}{2}\mathcal{M}^{mn} H_{m}\wedge \star H_{n}
\right.
\\
& & \\
& &
-\frac{1}{2}\Im\mathfrak{m}\mathcal{N}_{ab} G^{a} \wedge \star G^{b}
-\frac{1}{2}\Re\mathfrak{e}\mathcal{N}_{ab} G^{a} \wedge G^{b}
\\
& & \\
& &
-dC^{a}\wedge \Delta G_{a} -\tfrac{1}{2}\Delta G^{a}\wedge \Delta G_{a}
-\tfrac{1}{6}d^{mnp}B_{m}\wedge dB_{n}\wedge dB_{p}
+\tfrac{1}{2}d^{mnp}B_{m}\wedge H_{n} \wedge H_{p}
\\
& & \\
& &
\left.
+\tfrac{1}{24}d^{i}{}_{I}{}^{m}d_{iJ}{}^{n} A^{I} \wedge A^{J}\wedge \Delta
H_{m} \wedge dB_{n}
\right\}\, .
\end{array}
\end{equation}

\noindent
where $\mathcal{G}_{xy},\mathcal{M}_{IJ},\mathcal{M}^{mn}, \mathcal{N}_{ab}$
are scalar-dependent kinetic matrices ($\mathcal{N}_{ab}$ complex and the rest
real), the field strengths are defined by

\begin{eqnarray}
F^{I}
& = &
d A^{I}\, .
\\
& & \nonumber \\
H_{m}
& = &
dB_{m} - d_{mIJ}F^{I}\wedge A^{J}\, ,
\\
& & \nonumber \\
G^{a}
& = &
dC^{a} +d^{a}{}_{I}{}^{m}F^{I}\wedge B_{m}
-\tfrac{1}{3} d^{a}{}_{I}{}^{m}d_{mJK} A^{I}\wedge F^{J}\wedge A^{K}\, ,
\end{eqnarray}

\noindent
$d_{mIJ},d^{a}{}_{I}{}^{m}$ being constant deformation parameters and $\Delta
G^{a}$ etc. denote all the terms in the corresponding field strength but
$dC^{a}$ etc.

The 3-forms can be dualized in 3-forms $C_{a}$ with field strengths and
duality relations

\begin{equation}
  \begin{array}{rcl}
G_{a}
& \equiv &
dC_{a} +d_{aI}{}^{m}F^{I}\wedge B_{m}
-\tfrac{1}{3} d_{aI}{}^{m}d_{mJK} A^{I}\wedge F^{J}\wedge A^{K}\, ,
\\
& & \\
G_{a}
& = &
-\Im\mathfrak{m}\mathcal{N}_{ab} G^{a} \wedge \star G^{b}
-\Re\mathfrak{e}\mathcal{N}_{ab} G^{a} \wedge G^{b} \equiv R_{a}\, ,
\end{array}
\end{equation}

\noindent
where the $d_{aI}{}^{m}$ are constant independent parameters. The electric and
magnetic 3-forms and the deformation parameters can be collected in symplectic
vectors:

\begin{equation}
(C^{i})
\equiv
\left(
\begin{array}{c}
C^{a} \\ C_{a} \\
\end{array}
\right)\, ,
\hspace{1cm}
(d^{i}{}_{I}{}^{m})
\equiv
\left(
\begin{array}{c}
d^{a}{}_{I}{}^{m} \\ d_{aI}{}^{m} \\
\end{array}
\right)\, ,
\end{equation}

\noindent
with the symplectic indices $i,j$ to be raised and lowered with the symplectic
metric $(\Omega_{ij})=(\Omega^{ij}) = \left(
  \begin{smallmatrix}
    0 & 1 \\ -1 & 0 \\
  \end{smallmatrix}
\right)$.

The 2-forms $B_{n}$ can be dualized into 4-forms $\tilde{B}^{m}$ with field
strength and duality relation

\begin{equation}
  \begin{array}{rcl}
\tilde{H}^{m}
& \equiv &
d\tilde{B}^{m}
+d^{i}{}_{I}{}^{m} C_{i}\wedge F^{I}
+d^{mnp}B_{n}\wedge (H_{p}+\Delta H_{p})
+\tfrac{1}{12}d^{i}{}_{I}{}^{m}d_{iJ}{}^{n} A^{I}\wedge A^{J}\wedge \Delta H_{n}\, ,
\\
& & \\
\tilde{H}^{m}
& =  &
\mathcal{M}^{mn}\star H_{n}\, ,
\end{array}
\end{equation}

\noindent
where the new deformation $d^{mnp}=d^{[mnp]}$ must be related to
the other deformations by

\begin{equation}
\label{eq:relationdddd}
d^{i}{}_{(I|}{}^{m}d_{i|J)}{}^{n} = -2d^{mnp}d_{pIJ}\, .
\end{equation}

Finally, the 1-forms $A^{I}$ can be dualized into 6-forms $\tilde{A}_{I}$ with
field strength and duality relation

\begin{equation}
  \begin{array}{rcl}
\tilde{F}_{I}
& \equiv &
d\tilde{A}_{I} + \cdots\, ,
\\
& & \\
\tilde{F}_{I}
& = &
\mathcal{M}_{IJ}\star F^{J}\, ,
\end{array}
\end{equation}

\noindent
where the dots stand for a very long expression that can be found in
Ref.~\cite{Andino:2016bwk}.

As in the previous example, let us assume that the equations of motion are
invariant under the global transformations generated by\footnote{Observe that
  the transformations involving the 3-forms include electric-magnetic
  rotations. 3-forms in 8 dimensions transform as the 1-forms in 4-dimensions
  with groups which must be embedded in the symplectic group.}

\begin{equation}
  \begin{array}{rclrcl}
\delta_{A}\phi^{x}
& = &
k_{A}{}^{x}(\phi)\, ,\hspace{2cm}
&
\delta_{A}A^{I}
& = &
T_{A}{}^{I}{}_{J}A^{J}\, ,
\\
& & & & & \\
\delta_{A}B_{m}
& = &
-T_{A}{}^{n}{}_{m}B_{n}\, ,
&
\delta_{A}C^{i}
& = &
T_{A}{}^{i}{}_{j}C^{j}\, ,
\\
\end{array}
\end{equation}

\noindent
where the matrices $T_{A}{}^{I}{}_{J},T_{A}{}^{m}{}_{n}$ and the matrices

\begin{equation}
\left( T_{A}{}^{i}{}_{j}   \right)
\equiv
\left(
  \begin{array}{cc}
T_{A}{}^{a}{}_{b} & T_{A}{}^{ab} \\
T_{A\, ab} & T_{A\, a}{}^{b} \\
  \end{array}
\right)\, ,
\end{equation}

\noindent
which must be generators of the symplectic group,

\begin{equation}
T_{A}{}^{i}{}_{[j}\Omega_{k]i} = 0\, ,
\end{equation}

\noindent
are different representations of the same Lie
algebra as the one generated by the vectors $k_{A}{}^{x}(\phi)$:

\begin{equation}
[T_{A},T_{B}] = f_{AB}{}^{C}T_{C}\, ,
\hspace{1cm}
[k_{A},k_{B}] = -f_{AB}{}^{C}k_{C}\, .
\end{equation}

As in the previous case, this implies that the functions $k_{A}{}^{x}(\phi)$
are Killing vectors of the $\sigma$-model metric $\mathcal{G}_{xy}$, the
kinetic matrices satisfy\footnote{The transformation rule of the period matrix
is unusual because our definition of the dual 4-form field strength differs by
a sign from the usual one.}

\begin{equation}
\label{eq:kdM}
  \begin{array}{rcl}
k_{A}{}^{x}\partial_{x}\mathcal{M}_{IJ}
& = &
-2T_{A}{}^{K}{}_{(I}\mathcal{M}_{J)K}\, ,
\\
& & \\
k_{A}{}^{x}\partial_{x}\mathcal{M}^{mn}
& = &
2T_{A}{}^{(m}{}_{p}\mathcal{M}^{n)p}\, ,
\\
& & \\
k_{A}{}^{x}\partial_{x} \mathcal{N}_{ab}
& = &
-T_{A\, ab} -\mathcal{N}_{ac}T_{A}{}^{c}{}_{b}
+T_{A\, a}{}^{c}\mathcal{N}_{cb}
+\mathcal{N}_{ac}T_{A}{}^{cd}\mathcal{N}_{db}\, ,
\end{array}
\end{equation}

\noindent
and the deformation tensors $d_{mIJ},d^{i}{}_{I}{}^{m}, d^{mnp}$ are invariant
under the $\delta_{A}$ transformations:

\begin{equation}
\label{eq:deltaAd}
\begin{array}{rcl}
\delta_{A}d_{mIJ}
& = &
-T_{A}{}^{n}{}_{m} d_{nIJ} -2T_{A}{}^{K}{}_{(I} d_{n|J)K}=0\, ,
\\
& & \\
\delta_{A}d^{i}{}_{I}{}^{m}
& = &
T_{A}{}^{i}{}_{j} d^{j}{}_{I}{}^{m}-T_{A}{}^{J}{}_{I} d^{i}{}_{J}{}^{m}
+T_{A}{}^{m}{}_{n} d^{i}{}_{I}{}^{n}
=0\, ,
\\
& & \\
\delta_{A} d^{mnp}
& = &
3T_{A}{}^{[m}{}_{q}d^{np]q}=0\, .
\end{array}
\end{equation}

Since, in general,  these symmetries are not symmetries of the action, we
proceed as in the 4-dimensional case, contracting the equations of motion of
the scalars, given by

\begin{equation}
  \begin{array}{rcl}
-{\displaystyle\frac{\delta S}{\delta \phi^{x}}}
& = &
\mathcal{G}_{xy}
\left[
d\star d\phi^{y}+\Gamma_{zw}{}^{y}d\phi^{z}\wedge \star d\phi^{w}
\right]
\\
& & \\
& &
-\tfrac{1}{2}\partial_{x}\mathcal{M}_{IJ}F^{I}\wedge \star F^{J}
-\tfrac{1}{2}\partial_{x}\mathcal{M}^{mn}H_{m}\wedge \star H_{n}
-G^{a}\partial_{x} R_{a}\, ,
\end{array}
\end{equation}

\noindent
with the Killing vectors of the $\sigma$-model metric
$\mathcal{G}_{xy}(\phi)$, $k_{A}{}^{x}(\phi)$. Using the Killing equation, we
get

\begin{equation}
-k_{A}{}^{x}\frac{\delta S}{\delta \phi^{x}}
=
d\star j_{A}^{(\sigma)}
-\tfrac{1}{2}k_{A}{}^{x}\partial_{x}\mathcal{M}_{IJ}F^{I}\wedge \star F^{J}
-\tfrac{1}{2}k_{A}{}^{x}\partial_{x}\mathcal{M}^{mn}H_{m}\wedge \star H_{n}
-G^{a}k_{A}{}^{x}\partial_{x} R_{a}\, .
\end{equation}

Using now Eqs.~(\ref{eq:kdM}) and the duality relations for the field
strengths, we arrive to

\begin{equation}
-k_{A}{}^{x}\frac{\delta S}{\delta \phi^{x}}
 =
d\star j_{A}^{(\sigma)} +T_{A}{}^{J}{}_{I}F^{I}\wedge \tilde{F}_{J}
+T_{A}{}^{m}{}_{n}\tilde{H}^{n}\wedge H_{m}
+\tfrac{1}{2}T_{A}{}^{i}{}_{j}G^{j}\wedge G_{j}=0\, ,
\end{equation}

\noindent
on shell. It is not difficult to see that the exterior derivative of the
expression in the l.h.s.~of the equation vanishes due to the Bianchi
identities satisfied by the field strengths and due to the invariance of the
deformation tensors expressed in the relations Eqs.~(\ref{eq:deltaAd}). This
means that it should be possible to rewrite this equation as the conservation
of a higher-dimensional NGZ current, that is

\begin{equation}
d\star j_{A}^{NGZ} =0\, ,
\hspace{1cm}
j_{A}^{NGZ} \equiv j^{(\sigma)}_{A} + \Delta j_{A}\, ,
\end{equation}

\noindent
where $\Delta j_{A}$ has a very long a complicated form.

A local solution of this conservation equation is provided by
$\star[j^{(\sigma)}_{A}+\Delta j_{A}] = d D_{A}$ where $D_{A}$ is a 6-form potential
$D_{A}$. Then, reasoning as in the 4-dimensional case, the gauge-invariant
7-form field strength $K_{A}$ and its duality relation will be given by

\begin{equation}
K_{A} \equiv dD_{A} +\star\Delta j_{A}\, ,
\hspace{1cm}
K_{A} = \star j^{(\sigma)}_{A}\, ,
\end{equation}

\noindent
and its Bianchi identity will adopt the universal form

\begin{equation}
dK_{A}
=
T_{A}{}^{I}{}_{J}F^{J} \wedge \tilde{F}_{I}
+T_{A}{}^{m}{}_{n}\tilde{H}^{n} \wedge H_{m}
+\tfrac{1}{2}T_{A}{}^{i}{}_{j}G^{j}\wedge G_{i}\, .
\end{equation}

\subsubsection{$\mathcal{N}=2B,d=10$ supergravity}
\label{sec-n2bd10}

Our last example concerns the dualization of the scalars of
$\mathcal{N}=2B,d=10$ supergravity \cite{Schwarz:1983wa,Schwarz:1983qr,Howe:1983sra}, the effective
field theory of the type~IIB superstring. They are the dilaton $\varphi$ and
the RR $0$-form $\chi$ and, combined in the axidilaton $\tau=\chi+ie^{-\varphi}$
they parametrize the $\mathrm{SL}(2,\mathbb{R})/\mathrm{SO}(2)$ described in
Section~\ref{eq:SL2}. They are dualized into 3 8-form potentials satisfying a
constraint \cite{Meessen:1998qm,Dall'Agata:1998va} according to the general rules and
the field strengths, whose form depends very strongly on conventions, satisfy
a Bianchi identity of the universal form proposed above.

Here we want to focus on the supersymmetry transformation rules of the
8-forms, 
constructed in Refs.~\cite{Dall'Agata:1998va,Bergshoeff:2005ac}
in the 
$\mathrm{SU}(1,1)/\mathrm{U}(1)$ formulation used in \cite{Schwarz:1983qr} and
studied in Section~\ref{sec-su11}. We want to compare them with the general
form proposed in Section~\ref{sec-susyandmm}. They are given by

\begin{equation}
  \begin{array}{rcl}
\delta_{\epsilon} A^{\alpha\beta}{}_{\mu_{1}\cdots \mu_{8}}
& = &
8V^{(\alpha}{}_{+}V^{\beta)}{}_{-}\bar{\epsilon}\gamma_{[\mu_{1}\cdots\mu_{7}}\psi_{\mu_{8}]}
+\mathrm{c.c.}
\\
& & \\
& &
-i V^{\alpha}{}_{+}V^{\beta}{}_{+}\bar{\epsilon}\gamma_{\mu_{1}\cdots \mu_{8}}\lambda_{C}
+\mathrm{c.c.}
\\
& & \\
& &
+\cdots
\\
\end{array}
\end{equation}

\noindent
where we have omitted terms proportional to other $p$-form fields, which are
related to the Chern-Simons terms in the 9-form field strengths. Comparing now
with Eqs.~(\ref{eq:KVsu11}) and (\ref{eq:mmsu11}) we see that the terms
constraining the gravitini are multiplied by the momentum map while the terms
containing the dilatini are proportional to the Killing vectors, as expected
according to our general arguments.

\section{Conclusions}
\label{sec-conclusions}

In this paper we have reviewed the general problem of dualizing the scalars of
a $d$-dimensional theory into $(d-2)$-form potentials preserving the dualities
of the theory in a manifest form and taking into account their possible
couplings to the potentials of the theory. We have not considered the
dualization in presence of a scalar potential, since doing this properly,
requires the full tensor hierarchy machinery, which lies outside of the scope
of this paper.\footnote{The general $d=8$ case studied in Ref.~\cite{Andino:2016bwk}
  provides a quite complete example of how to proceed in that case.}

In general, the dualization procedure has to be necessarily incomplete: the
non-linearly interacting scalars cannot be replaced completely by the
$(d-2)$-form potentials, as often happens in supergravity theories with most
potentials. Nevertheless, one may hope to find a PST-like formulation for
them. For the particular case of scalars parametrizing the coset SU(1,1)/U(1)
the PST-type action was constructed in Ref.~\cite{Dall'Agata:1998va} as a part
of type IIB supergravity action. Here we have presented the generalization of
the action of Ref.~\cite{Dall'Agata:1998va} for the generic symmetric space
G/H; the properties of this action and its applications will be considered
elsewhere.

Since we need to dualize conserved charges and some of the symmetries one has
to consider in supergravity theories leave invariant the equations of motion
but not the action, it is necessary to consider the Noether-Gaillard-Zumino
current, whose generalization to theories in higher dimensions and with
higher-rank potentials we have studied.

During this study we have found it necessary to extend the concept of momentum
map to all symmetric spaces. The holomorphic and triholomorphic momentum maps
defined in K\"ahler and quaternionic-K\"ahler spaces play a very important
r\^ole in $\mathcal{N}=1,2,d=4$ supergravities: they occur in fermion shifts
(and, therefore, in the scalar potentials, where they often appear
disguised as ``dressed structure constants'' or ``T-tensors''), in the
supersymmetry transformations of the 2-forms dual to the scalars (and,
therefore, in the tensions of the strings that couple to them) and in the
covariant derivatives of the fermions in gauged supergravities. We have shown
through examples that the generalized momentum map satisfies similar equations
and plays exactly the same r\^ole in $\mathcal{N}>2$ and $d>4$ supergravities
and we have explored the general form of the supersymmetry transformation
rules of the $(d-2)$-forms dual to the scalars and the fermion shifts.

In $\mathcal{N}=1$ supersymmetric mechanics one can consider general manifolds
with no special holonomy properties. When they admit isometries and we gauge
them, the covariant derivatives of the fermions contain an object that plays
the same r\^ole as the momentum map. We have shown that it satisfies analogous
equations and that, when the manifold has special holonomy (K\"ahler,
quaternionic-K\"ahler or symmetric space), this object is the (generalized)
momentum map. We have, therefore, proposed a more fundamental definition for
the momentum map that encompasses all the previous ones.

Supergravity theories have very different forms for different values of
$\mathcal{N},d$, mostly because of historical reasons: some of them have been
constructed by dimensional reduction, some others in superspace or using other
approaches. This complicates unnecessarily working with them and establishing
relations between them via compactifications, truncations, gaugings etc. As a
theories of dynamical supergeometry, it should be possible to describe them in
a more $\mathcal{N}$- and $d$-independent form. A big step in this direction
was taken in Ref.~\cite{Andrianopoli:1996ve}, specially for 4-dimensional
theories, which were described in an almost $\mathcal{N}$-independent fashion,
but neither the gaugings nor the higher-rank form fields were considered
there. We hope the extension of the concept of momentum map proposed here and
its systematic use (specially in the construction of fermion shifts and
scalars potentials) will be useful to rewrite all gauged supergravities in a
more homogeneous form.

\section*{Acknowledgments}

TO would like to thank M.~Trigiante for interesting and stimulating
conversations during the Workshop {\sl Theoretical Frontiers in Black Holes
  and Cosmology} held at the International Institute of Physics in Natal
(Brazil) in June 2015. BI is thankful to MIAPP and to the organizers of the
MIAPP program ''Higher spin theory and duality'' for the hospitality in Munich
at the final stage of this project.  This work has been supported in part by
the Spanish Ministry of Science and Education grant FPA2012-35043-C02-01 (IB
and TO), the Centro de Excelencia Severo Ochoa Program grant SEV-2012-0249
(TO), the Spanish Consolider-Ingenio 2010 program CPAN CSD2007-00042 (TO) and
by the Basque Country University program UFI 11/55 (IB).  TO wishes to thank
M.M.~Fern\'andez for her permanent support.

\appendix

\section{K\"ahler--Hodge manifolds in $\mathcal{N}=1,2$, $d=4$ supergravity}
\label{sec-KH}

In this Appendix we want to review briefly the definition of the holomorphic
momentum map and other structures which have their parallel in the main text
in the context of K\"ahler--Hodge (KH) manifolds, which are not necessarily
symmetric or even homogenous spaces. We adopt the notation and conventions of
Refs.~\cite{Andrianopoli:1996cm,Ortin:2015hya}.

A K\"ahler manifold is a complex, Hermitian manifold whose fundamental 2-form
$\mathcal{J}$

\begin{equation}
\label{eq:fundamental2form}
\mathcal{J}
\equiv
\mathcal{J}_{ij^{*}}dZ^{i} \wedge dZ^{*\, j^{*}}
=
2i \mathcal{G}_{ij^{*}}dZ^{i} \wedge dZ^{*\, j^{*}}\, ,
\end{equation}

\noindent
is closed

\begin{equation}
\label{eq:dJ=0}
d\mathcal{J}=0\, .
\end{equation}

\noindent
This equation implies the vanishing of the torsion, the identification of the
Hermitian connection with the Levi-Civita connection and the local existence
of a real function, the K\"ahler potential $\mathcal{K}(Z,Z^{*})$, such that

\begin{equation}
\tfrac{1}{2i}\mathcal{J}_{ij^{*}}
=
\mathcal{G}_{ij^{*}}
=
\partial_{i}\partial_{j^{*}}\mathcal{K}\, .
\end{equation}

\noindent
$\mathcal{K}$ is defined up to K\"ahler transformations, which have the form

\begin{equation}
\label{eq:Kpotentialtransformation}
\mathcal{K}^{\prime}
=
\mathcal{K}+\lambda(Z)+\lambda^{*}(Z^{*}),
\end{equation}

\noindent
were $\lambda(Z)$ is an arbitrary holomorphic function of the complex coordinates $Z^{i}$.

In $\mathcal{N}=1,2$, $d=4$ supergravity there are complex scalar field
parametrizing K\"ahler manifolds and the K\"ahler metric $\mathcal{G}_{ij^{*}}$
plays the role of the $\sigma$-model metric.

The K\"ahler (connection) 1-form is defined by

\begin{equation}
\label{eq:K1form}
\mathcal{Q}
\equiv
\tfrac{1}{2i}(\partial_{i}\mathcal{K} dZ^{i} -\mathrm{c.c.})\, ,
\end{equation}

\noindent
transforms under K\"ahler transformations as a $\mathrm{U}(1)$ connection

\begin{equation}
\mathcal{Q}^{\prime} =\mathcal{Q}
+\tfrac{1}{2i}(\partial \lambda -\partial^{*}\lambda^{*})\, ,
\end{equation}

\noindent
and the K\"ahler 2-form can be seen as its K\"ahler-invariant curvature

\begin{equation}
\mathcal{J} \equiv 2 d\mathcal{Q}\, .
\end{equation}

A K\"ahler--Hodge manifold is a K\"ahler manifold $\mathcal{M}$ on which a
complex line bundle $L^{1}\rightarrow\mathcal{M}$ has been defined such that
its first Chern class (given by the Ricci 2-form $\mathfrak{R}$ of the fiber's
Hermitian metric) is equal to the K\"ahler 2-form $\mathcal{J}$.

As we are going to show, in the KH manifolds of $\mathcal{N}=1,2$, $d=4$
supergravity, the K\"ahler 1-form connection $\mathcal{Q}$ and its curvature
$\mathcal{J}$ play the same as the H$=\mathrm{U}(1)$ connection $\vartheta$
and its curvature $R(\vartheta)$ defined in Eqs.~(\ref{eq:M-Cdef}) and
(\ref{eq:Rtheta}):

\begin{equation}
\label{eq:correspondence1}
\mathcal{Q} \rightarrow -\tfrac{1}{2}\vartheta\, ,
\hspace{1cm}
\mathcal{J} \rightarrow - R(\vartheta)\, ,
\end{equation}

\noindent
even though there is no coset structure. The requirement that the K\"ahler
manifold is actually K\"ahler--Hodge is crucial.

The fermionic fields of $\mathcal{N}=1,2$ supergravity are sections of the
associated $\mathrm{U}(1)$ bundle, which means that, under K\"ahler
transformations Eq.~(\ref{eq:Kpotentialtransformation}), they transform as

\begin{equation}
\label{eq:qqbarfield}
\psi^{\prime} = e^{-\frac{q}{2}(\lambda -\lambda^{*})}  \psi\, ,
\end{equation}

\noindent
if their weight is the real number $q$. The K\"ahler-covariant derivative on
fields of K\"ahler weight $q$ is given by

\begin{equation}
\label{eq:Kcovariantderivative2}
\mathcal{D}\psi= d\psi+iq\mathcal{Q}\psi\, ,
\end{equation}%

\noindent
where here $\mathcal{Q}$ is the spacetime pullback of the K\"ahler
1-form. This definition should be compared with that of the H-covariant
derivative Eq.~(\ref{eq:Hcovariantderivatives}).

Let us now assume that the theory we are considering has some global symmetry
transformation group acting on the scalars. These transformations must
necessarily be holomorphic isometries of the K\"ahler metric generated by
Killing vectors $K_{A}\equiv k_{A}{}^{i}(Z)\partial_{i}
+k^{*}_{A}{}^{i^{*}}(Z^{*})\partial_{i^{*}}$ but they must also preserve the
entire KH structure.

First of all, this implies that the transformations generated by the Killing
vectors will leave the K\"ahler potential invariant up to K\"ahler
transformations:

\begin{equation}
\label{eq:Kconservation}
\mathcal{L}_{k_{A}}\mathcal{K}
\equiv
k_{A}{}^{i}\partial_{i}\mathcal{K}
+
k^{*}_{A}{}^{i^{*}}\partial_{i^{*}}\mathcal{K}
=
\lambda_{A}(Z)+ \lambda^{*}_{A}(Z^{*})\, ,
\end{equation}

\noindent
for certain holomorphic functions $\lambda_{A}(Z)$.  This, in its turn,
implies that all the fields which transform as in Eq.~(\ref{eq:qqbarfield})
will transform as

\begin{equation}
\psi^{\prime} = e^{-\frac{q}{2}(\lambda_{A} -\lambda_{A}^{*})}  \psi\, ,
\end{equation}

\noindent
under the transformation generated by $K_{A}$. This is similar to the H
compensating transformations described in
Section~\ref{sec-H-covariantderivative} and it is clear that the imaginary
part of the holomorphic functions $\lambda_{A}$ plays the same role as the
H-compensator defined in Eq.~(\ref{eq:Hcompensator})

\begin{equation}
\tfrac{1}{2i}(\lambda_{A} -\lambda_{A}^{*}) \rightarrow -2 W_{A}\, .
\end{equation}

Taking another Lie derivative in Eq.~(\ref{eq:Kconservation}) we find the
following equivariance property

\begin{equation}
\label{eq:lambdaalgebra}
\mathcal{L}_{k_{A}}\lambda_{B}
-
\mathcal{L}_{k_{B}}\lambda_{A}
=
-f_{AB}{}^{C}\lambda_{C}\, ,
\end{equation}

\noindent
which is identical to that of the H-compensator
Eq.~(\ref{eq:equivarianceHcompensator}).

Secondly, the  K\"ahler 2-form $\mathcal{J}$ must also be preserved

\begin{equation}
\label{eq:Jconservation}
\mathcal{L}_{k_{A}}\mathcal{J}= i_{k_{A}}d\mathcal{J}+d(i_{k_{A}}\mathcal{J})=0\, .
\end{equation}

\noindent
Eq.~(\ref{eq:dJ=0}) and the above equation imply the local existence of real
functions $\mathcal{P}_{A}$ (the holomorphic momentum maps) such that

\begin{equation}
\label{eq:holomorphicmomentummap}
i_{k_{A}}\mathcal{J}= d\mathcal{P}_{A}\, .
\end{equation}

\noindent
Comparing this equation with Eq.~(\ref{eq:DPRk}) and taking into account the
correspondences Eq.~(\ref{eq:correspondence1}) we find that the holomorphic
momentum map plays the same role as the momentum map defined in
Eq.~(\ref{eq:momentummap}):

\begin{equation}
\mathcal{P}_{A} \rightarrow \tfrac{1}{2}P_{A}\, .
\end{equation}

A local solution of Eq.~(\ref{eq:holomorphicmomentummap}) is

\begin{equation}
i\mathcal{P}_{A}
=
k_{A}{}^{i}\partial_{i}\mathcal{K} -\lambda_{A}
=
-(k^{*}_{A}{}^{i^{*}}\partial_{i^{*}}\mathcal{K}
-\lambda^{*}_{A})
=
i_{k_{A}}\mathcal{Q}
-{\textstyle\frac{1}{2i}}(\lambda_{A}-\lambda^{*}_{A})\, ,
\end{equation}

\noindent
where we have taken into account Eq.~(\ref{eq:Kconservation}). This equation
should be compared with Eq.~(\ref{eq:Hcompensator}) that relates the
H-connection, the H-compensator and the momentum map.

Furthermore, the holomorphic Killing vectors can be obtained form the momentum
map (\textit{Killing prepotential})

\begin{equation}
\label{eq:dP=k}
\partial_{i}\mathcal{P}_{A} = i k^{*}_{A\, i}\, .
\end{equation}

In $\mathcal{N}=2,d=4$ supergravity theories, the Special K\"ahler structure
allows us to find a general expression for the holomorphic momentum map in
terms of the covariantly holomorphic symplectic section $\mathcal{V}^{M}$ and
the symplectic generators $\mathcal{T}_{A}{}^{M}{}_{N}$:

\begin{equation}
\label{eq:PAN2}
\mathcal{P}_{A}
=
\langle\, \mathcal{V}^{*} \mid \mathcal{T}_{A} \mathcal{V}\, \rangle
=
\mathcal{T}_{A}^{M}{}_{N} \mathcal{V}^{*}_{M}\mathcal{V}^{N}\, .
\end{equation}

If we now gauge the group of holomorphic isometries generated by the Killing
vectors $k_{A}{}^{i}$ we can follow the same rules as in symmetric spaces to
construct the gauge-covariant derivatives, adding to the pullback of the
H-connection (K\"ahler connection) the product $A^{A}\mathcal{P}_{A}$ where
$A^{A}$ is the spacetime gauge field:

\begin{equation}
\mathcal{Q} \rightarrow \mathcal{Q}-gA^{A}\mathcal{P}_{A}
=
\mathcal{Q}_{i}\mathfrak{D}Z^{i}
+\mathcal{Q}_{i^{*}}\mathfrak{D}Z^{*\, i^{*}}
-gA^{A}\Im\mathfrak{m} \lambda_{A}
\, .
\end{equation}

The momentum map also occurs in the fermion shifts of the fermions'
supersymmetry transformation rules. The details depend on the theory and its
R-symmetry group and can be found, for instance, in Ref.~\cite{Ortin:2015hya}.

\subsection{2-form potentials from the K\"ahler--Hodge manifolds of $\mathcal{N}=1,2$, $d=4$ supergravity}
\label{sec-KH2-forms}

The dualization of the complex scalars of $\mathcal{N}=1$ and
$\mathcal{N}=2,d=4$ supergravities belonging to chiral and vector
supermultiplets can be performed following the general procedure outlined in
Section~\ref{sec-dualizationd=4}. To finish the job, though, a supersymmetry
transformation rule must be provided for the dual 2-form fields, at least to
lowest (zeroth) order in fermions. The supersymmetry algebra must close on
shell and up to duality relations between the magnetic and electric vector
fields and between the 2-forms and the NGZ currents.

This was first done in the $\mathcal{N}=2,d=4$ theories in
Ref.~\cite{Bergshoeff:2007ij}. After the use of the expression for the
momentum maps Eq.~(\ref{eq:PAN2}), the supersymmetry transformation rules
found there can be written in the form

\begin{equation}
\label{2formsusyvectorcase}
\delta_{\epsilon}B_{A\, \mu\nu}
=
-i\mathcal{P}_{A}\bar{\epsilon}^{I}\gamma_{[\mu}\psi_{I\nu]}
-\tfrac{i}{2}k^{*}_{A\, i}
\bar{\epsilon}_{I}\gamma_{\mu\nu}\lambda^{iI}+\text{c.c.}
+8\mathcal{T}_{A}{}^{M}{}_{N}\mathcal{A}_{M\,[\mu|}
\delta_{\epsilon}\mathcal{A}_{N\, |\nu]}\, .
\end{equation}

\noindent
The commutator of two of these supersymmetry transformations gives

\begin{equation}
\label{commutatoralgebra}
[\delta_{\eta},\delta_{\epsilon}]=
\delta_{\text{g.c.t.}}(\xi)+\delta_{\text{gauge}}(\Lambda)
+\delta_{\text{gauge}}(\Lambda_{1})\, .
\end{equation}

\noindent
where $\xi^{\mu}$ are the parameters of general coordinate transformations,
$\Lambda^{M}$ are the 0-form parameters of the gauge transformations of the
gauge fields $\mathcal{A}^{M}{}_{\mu}$ and $\Lambda_{1\, A}$ are the 1-form
parameters of the gauge transformations of the 2-form fields $B_{A\,
  \mu\nu}$. Their explicit expressions can be found in
Ref.~\cite{Bergshoeff:2007ij}.

In the actual computation of the commutator, the derivative of the momentum
map, which gives the corresponding Killing vector and the scalar part of the
NGZ current appears naturally. Upon dualization, that term gives the contraction
of the 3-form field strength $H_{A}$ with $\xi^{\mu}$, which is a general
coordinate transformation of $B_{A}$ up to a gauge transformation.

The supersymmetry transformation rule for the 2-forms of $\mathcal{N}=1,d=4$
supergravity was given in Ref.~\cite{Hartong:2009az} and fits into the same
pattern (the differences are basically due to the different conventions)

\begin{equation}
\label{eq:susyBA}
\delta_{\epsilon}B_{A\, \mu\nu}
=
\tfrac{i}{2} \mathcal{P}_{A}\bar{\epsilon}^{*}\gamma_{[\mu}\psi_{\nu]}
+\tfrac{1}{4}\partial_{i}\mathcal{P}_{A}\bar{\epsilon}\gamma_{\mu\nu}\chi^{i}
+\mathrm{c.c.}
-2\mathcal{T}_{A}{}^{M}{}_{N}\mathcal{A}_{M\,[\mu}\delta_{\epsilon}
\mathcal{A}^{N}{}_{\nu]}
\, .
\end{equation}

\section{Quaternionic--K\"ahler  manifolds in $\mathcal{N}=2$, $d=4$ supergravity}
\label{sec-QK}

The structures constructed for symmetric spaces can also be generalized to
quaternionic-K\"ahler (QK) spaces, $4m$-dimensional Riemannian spaces whose
holonomy group is $\mathrm{SU}(2)\times\mathrm{Sp}(2m)$.

A QK manifold is a $4m$-dimensional Riemannian manifold that satisfies the
following properties:

\begin{enumerate}
\item It admits a triplet of complex structures $\mathsf{J}^{x}{}_{m}{}^{n}$,
  $x=1,2,3$ satisfying the algebra of the unit imaginary quaternions

\begin{equation}
\mathsf{J}^{x}{}_{m}{}^{p}\mathsf{J}^{y}{}_{p}{}^{n}
=
-\delta^{xy}\delta_{m}{}^{n} +\varepsilon^{xyz}\mathsf{J}^{z}{}_{m}{}^{n}\, .
\end{equation}

(Observe that this property implies the property that characterizes complex
structures $(\mathsf{J}^{x})^{2}= -1\, ,\,\,\, \forall x$.)

\item The Riemannian metric $g_{mn}$ is Hermitian with respect to the three
  complex structures:

  \begin{equation}
  g_{mn}=\mathsf{J}^{(x)}{}_{m}{}^{p} \mathsf{J}^{(x)}{}_{n}{}^{q}g_{pq}\, .
  \end{equation}

  We can define a triplet of K\"ahler 2-forms (hyperK\"ahler 2-form)

\begin{equation}
\mathsf{J}^{x}{}_{mn} \equiv \mathsf{J}^{x}{}_{m}{}^{p}g_{np}\, .
\end{equation}

\item There is a $\mathrm{SU}(2)$ bundle over the QK space with connection
  1-form $\mathsf{A}^{x}{}_{m}d\phi^{m}$ and it is required that the
  hyperK\"ahler 2-form is covariantly constant with respect to it:

\begin{equation}
\mathsf{D}_{m}\mathsf{J}^{x}{}_{np}
\equiv
\nabla_{m}(\omega)\mathsf{J}^{x}{}_{np}
+\varepsilon^{xyz}\ \mathsf{A}^{y}{}_{m} \mathsf{J}^{z}{}_{np}
=
0\, ,
\end{equation}

\noindent
where $\nabla_{m}(\omega)$ stands for the covariant derivative with the
Levi-Civita connection $\omega$.

\item The $\mathrm{SU}(2)$ curvature, defined by

\begin{equation}
\label{eq:FversusJ}
\mathsf{F}^{x}
\equiv
d\mathsf{A}^{x}
+ {\textstyle\frac{1}{2}}
\varepsilon^{xyz}  \mathsf{A}^{y} \wedge \mathsf{A}^{z}\, ,
\end{equation}

\noindent
is proportional to the hyperK\"ahler structure

\begin{equation}
\label{eq:FproptoJ}
\mathsf{F}^{x} = \varkappa\ \mathsf{J}^{x}\, .
\end{equation}

In $\mathcal{N}=2,d=4$ $\varkappa=-1$. (If $\varkappa=0$ the manifold is a
just a hyperK\"ahler manifold).

\end{enumerate}

This last property of QK manifolds combined with the relation between the
$\mathrm{SU}(2)$ component of the curvature 2-form of the Levi-Civita
connection (obtained through the projection with the hyperK\"ahler structure)
and the hyperK\"ahler 2-form

\begin{equation}
R_{mn}{}^{p}{}_{q}(\omega)\mathsf{J}^{x}{}_{p}{}^{q}
=
-2m \varkappa \mathsf{J}^{x}{}_{mn}\, ,
\end{equation}

\noindent
plays the same role as the relation between
the K\"ahler 2-form and the Ricci 2-form in K\"ahler-Hodge manifolds. It
establishes a bridge between symmetric spaces and QK spaces: here,
$\mathsf{A}$ will play the role of the $H=\mathrm{SU}(2)$ connection
$\vartheta$ and the hyperK\"ahler structure $\mathsf{J}^{x}$ will play the
role of the curvature $R(\vartheta)$ thanks to the above property.

\begin{equation}
\mathsf{A}^{x}\rightarrow \vartheta^{x}\, ,
\hspace{1cm}
\mathsf{J}^{x} \rightarrow \frac{1}{\varkappa}R^{x}(\vartheta)\, .
\end{equation}

The fermionic fields of $\mathcal{N}=2,d=4$ supergravity are sections of the
$\mathrm{SU}(2)$ bundle and, under an transformation with infinitesimal
parameter $\lambda^{x}$ they transform as\footnote{Here we ignore the
  $\mathrm{U}(1)$ component of the R-symmetry group, which we have discussed
  in the previous Appendix.}

\begin{equation}
\delta_{\lambda}\psi^{I} = \tfrac{i}{2}\lambda^{x}\sigma^{I}{}_{J}\psi^{J}\, ,
\end{equation}

\noindent
and the $\mathrm{SU}(2)$-covariant derivative acting on them is given by

\begin{equation}
\mathcal{D}\psi^{I}
\equiv
d\psi^{I}-\tfrac{i}{2}\mathsf{A}^{x}\sigma^{I}{}_{J}\psi^{J}\, .
\end{equation}

\noindent
Compare this definition with that of the H-covariant derivative
Eq.~(\ref{eq:Hcovariantderivatives}).

Now let us assume the existence of an isometry group of the QK manifold
preserving the hyperK\"ahler structure $\mathsf{J}^{x}$. This means that the
transformations generated by the corresponding Killing vectors $k_{A}$ (known
as triholomorphic Killing vectors) leave invariant $\mathsf{J}^{x}$ up to an
$\mathsf{SU}(2)$ transformation

\begin{equation}
\mathcal{L}_{k_{A}} \mathsf{J}^{x}
=
-\delta_{\lambda_{A}} \mathsf{J}^{x}\, ,
\,\,\,\,\,
\mathrm{or}
\,\,\,\,\,
\mathbb{L}_{k_{A}} \mathsf{J}^{x}
\equiv
\mathcal{L}_{k_{A}} \mathsf{J}^{x}
-\varepsilon^{xyz}\lambda_{A}^{y}\mathsf{J}^{z}
=0\, ,
\end{equation}

\noindent
for some $\mathrm{SU}(2)$ infinitesimal parameters $\lambda^{x}_{A}$ which
play the role of the  H-compensators defined in Eq.~(\ref{eq:Hcompensator})

\begin{equation}
\lambda_{A}^{x} \rightarrow  W_{A}{}^{x}\, .
\end{equation}

In order to determine $\lambda_{A}^{x}$ we observe that the H-compensator has
to be \textit{universal}: all the objects that define the QK geometry must be
invariant under the action of the isometry and the same compensating
$\mathrm{SU}(2)$ transformation. In particular, for the $\mathrm{SU}(2)$
connection

\begin{equation}
\mathbb{L}_{k_{A}} \mathsf{A}^{x}{}_{m}
=
\mathcal{L}_{k_{A}} \mathsf{A}^{x}{}_{m}
+\mathcal{D}_{m}\lambda_{A}^{x}
=
k_{A}{}^{n}\mathsf{F}^{x}{}_{nm}
+\mathcal{D}_{m}\left(k_{A}{}^{n}\mathsf{A}^{x}{}_{n}+\lambda_{A}^{x}\right)
=
0\, .
\end{equation}

This equation implies that $k_{A}{}^{n}\mathsf{F}^{x}{}_{nm}$ is the
$\mathrm{SU}(2)$-covariant derivative of an object that we can identify with
the (triholomorphic) momentum map:

\begin{eqnarray}
\label{eq:k=DP}
k_{A}{}^{n}\mathsf{F}^{x}{}_{nm}
& = &
\varkappa \mathcal{D}_{m}\mathsf{P}_{A}{}^{x}\, ,
\\
& & \nonumber \\
\varkappa \mathsf{P}_{A}{}^{x}
& = &
k_{A}{}^{n}\mathsf{A}^{x}{}_{n}+\lambda_{A}^{x}\, .
\end{eqnarray}

\noindent
The last equation should be compared with Eq.~(\ref{eq:Hcompensator}) while
the first should be compared with Eq.~(\ref{eq:DPRk}). Using the relation
between the curvature $\mathsf{F}^{x}$ and the hyperK\"ahler structure
$\mathsf{J}^{x}$ Eq.~(\ref{eq:FversusJ}) one can multiply both sides of the
first equation by $\mathsf{J}^{x}$ and obtain

\begin{equation}
\mathsf{k}_{A}{}^{m}
=
-\frac{1}{3\varkappa}\
\mathsf{J}^{x\, mn}\mathsf{D}_{n}\mathsf{P}_{A}{}^{x}\, .
\end{equation}

\noindent
In this equation the triholomorphic momentum map plays the role of
triholomorphic Killing prepotential.

The construction of gauge-covariant derivatives using the momentum map follows
the same pattern as in symmetric spaces (see, for instance,
Ref.~\cite{Ortin:2015hya}). The momentum map also appears in all the fermion
shift terms of the supersymmetry transformation rules of the fermions of the
$\mathcal{N}=2,d=4$ theories except in those of the hyperinos. Again, the
details can be found in Ref.~\cite{Ortin:2015hya}.

\subsection{2-form potentials from the Quaternionic--K\"ahler  manifolds in $\mathcal{N}=2$, $d=4$ supergravity}
\label{sec-QK2forms}

Since the hyperscalars do not couple to the vector fields, their dualization
is specially simple: the NGZ currents are equal to the Noether current of the
$\sigma$-model. The supersymmetry transformation rules for the dual 2-forms
are given by \cite{Bergshoeff:2007ij}

\begin{equation}
\label{susytrafohyper2form}
\delta_{\epsilon} B_{A\, \mu\nu}
=
-4\mathsf{P}_{A}{}^{J}{}_{I}\bar{\epsilon}^{I}\gamma_{[\mu}\psi_{J\vert\nu]}
+{\textstyle\frac{8i}{3}} \mathsf{U}_{\alpha J}{}^{u}
\mathfrak{D}_{u}\mathsf{P}_{A}{}^{J}{}_{I}
\bar{\epsilon}^{I}\gamma_{\mu\nu}\zeta^{\alpha}
+\mathrm{c.c.}\, ,
\end{equation}

\noindent
where $u,v$ label the real coordinates of the QK manifold (the hyperscalars
$q^{u}$, $\mathsf{U}_{\alpha J}{}^{u}$ are the inverse Vielbein of the QK
manifold (the tangent space index being splint into a $\mathrm{SU}(2)$ index
$J$ and an $\mathrm{Sp}(2m)$ index $\alpha$ and where

\begin{equation}
\mathsf{P}_{A}{}^{J}{}_{I}
\equiv
\tfrac{i}{2}\mathsf{P}_{A}{}^{x}(\sigma^{x}){}^{I}{}_{J}\, .
\end{equation}

\noindent
Again, these supersymmetry transformation rules fit into the general pattern
proposed in Section~\ref{sec-dualizationd=4}, once one takes into account the
relation between the derivative of the triholomorphic momentum map and the
triholomorphic Killing vectors Eq.~(\ref{eq:k=DP}).



\begin{thebibliography}{99}


\bibitem{Ferrara:1995ih}
S.~Ferrara, R.~Kallosh and A.~Strominger,
``N=2 extremal black holes,''
Phys.\ Rev.\ D {\bf 52} (1995) 5412.
\doi{10.1103/PhysRevD.52.R5412}.
[\hepth{9508072}].

\bibitem{Strominger:1996kf}
A.~Strominger,
``Macroscopic entropy of N=2 extremal black holes,''
Phys.\ Lett.\ B {\bf 383} (1996) 39.
\doi{10.1016/0370-2693(96)00711-3}.
[\hepth{9602111}].

\bibitem{Ferrara:1996dd}
S.~Ferrara and R.~Kallosh,
``Supersymmetry and attractors,''
Phys.\ Rev.\ D {\bf 54} (1996) 1514.
\doi{10.1103/PhysRevD.54.1514}.
[\hepth{9602136}].

\bibitem{Ferrara:1996um}
S.~Ferrara and R.~Kallosh,
``Universality of supersymmetric attractors,''
Phys.\ Rev.\ D {\bf 54} (1996) 1525.
\doi{10.1103/PhysRevD.54.1525}.
[\hepth{9603090}].

\bibitem{Ferrara:1997tw}
S.~Ferrara, G.~W.~Gibbons and R.~Kallosh,
``Black holes and critical points in moduli space,''
Nucl.\ Phys.\ B {\bf 500} (1997) 75.
\doi{10.1016/S0550-3213(97)00324-6}.
[\hepth{9702103}].

\bibitem{Strominger:1996sh}
A.~Strominger and C.~Vafa,
``Microscopic origin of the Bekenstein-Hawking entropy,''
Phys.\ Lett.\ B {\bf 379} (1996) 99.
\doi{10.1016/0370-2693(96)00345-0}.
[\hepth{9601029}].

\bibitem{Bergshoeff:2001pv}
E.~Bergshoeff, R.~Kallosh, T.~Ort\'{\i}n, D.~Roest and A.~Van Proeyen,
``New formulations of D = 10 supersymmetry and D8 - O8 domain walls,''
Class.\ Quant.\ Grav.\  {\bf 18} (2001) 3359.
\doi{10.1088/0264-9381/18/17/303}.
[\hepth{0103233}].

\bibitem{Bandos:1997gd}
I.~A.~Bandos, N.~Berkovits and D.~P.~Sorokin,
``Duality symmetric eleven-dimensional supergravity and its coupling to M-branes,''
Nucl.\ Phys.\ B {\bf 522} (1998) 214.
\doi{10.1016/S0550-3213(98)00102-3}.
[\hepth{9711055}].

\bibitem{Dall'Agata:1998va}
G.~Dall'Agata, K.~Lechner and M.~Tonin,
``D = 10, N = IIB supergravity: Lorentz invariant actions and duality,''
JHEP {\bf 9807} (1998) 017.
\doi{10.1088/1126-6708/1998/07/017}.
[\hepth{9806140}].

\bibitem{Bandos:2003et}
I.~A.~Bandos, A.~J.~Nurmagambetov and D.~P.~Sorokin,
``Various faces of type IIA supergravity,''
Nucl.\ Phys.\ B {\bf 676} (2004) 189
\doi{10.1016/j.nuclphysb.2003.10.036}
[\hepth{0307153}].

\bibitem{Pasti:1995ii}
P.~Pasti, D.~P.~Sorokin and M.~Tonin,
``Note on manifest Lorentz and general coordinate invariance in duality symmetric models,''
Phys.\ Lett.\ B {\bf 352} (1995) 59.
\doi{10.1016/0370-2693(95)00463-U}.
[\hepth{9503182}].

\bibitem{Pasti:1995tn}
P.~Pasti, D.~P.~Sorokin and M.~Tonin,
``Duality symmetric actions with manifest space-time symmetries,''
Phys.\ Rev.\ D {\bf 52} (1995) 4277
\doi{10.1103/PhysRevD.52.R4277}.
[\hepth{9506109}].

\bibitem{Meessen:1998qm}
P.~Meessen and T.~Ort\'{\i}n,
``An Sl(2,Z) multiplet of nine-dimensional type II supergravity theories,''
Nucl.\ Phys.\ B {\bf 541} (1999) 195.
\doi{10.1016/S0550-3213(98)00780-9}.
[\hepth{9806120}].

\bibitem{Bergshoeff:2009ph}
E.~A.~Bergshoeff, J.~Hartong, O.~Hohm, M.~Huebscher and T.~Ort\'{\i}n,
``Gauge Theories, Duality Relations and the Tensor Hierarchy,''
JHEP {\bf 0904} (2009) 123.
\doi{10.1088/1126-6708/2009/04/123}.
[\arxiv{0901.2054} [hep-th]].

\bibitem{Hartong:2009vc}
J.~Hartong and T.~Ort\'{\i}n,
``Tensor Hierarchies of 5- and 6-Dimensional Field Theories,''
JHEP {\bf 0909} (2009) 039.
\doi{10.1088/1126-6708/2009/09/039}.
[\arxiv{0906.4043} [hep-th]].

\bibitem{Gaillard:1981rj}
M.~K.~Gaillard and B.~Zumino,
``Duality Rotations for Interacting Fields,''
Nucl.\ Phys.\ B {\bf 193} (1981) 221.
\doi{10.1016/0550-3213(81)90527-7}.

\bibitem{Castellani:1991et}
L.~Castellani, R.~D'Auria and P.~Fr\'e,
``Supergravity and superstrings: A Geometric perspective. Vol. 1: Mathematical foundations,''
Singapore, Singapore: World Scientific (1991) 1-603

\bibitem{Ortin:2015hya}
T.~Ort\'{\i}n,
``Gravity and strings,'' 2nd Edition
Cambridge Unversity, Cambridge University Press, 2015

\bibitem{Coleman:1969sm}
S.~R.~Coleman, J.~Wess and B.~Zumino,
``Structure of phenomenological Lagrangians. 1.,''
Phys.\ Rev.\  {\bf 177} (1969) 2239.
\doi{10.1103/PhysRev.177.2239}

\bibitem{Salam:1969rq}
A.~Salam and J.~A.~Strathdee,
``Nonlinear realizations. 1: The Role of Goldstone bosons,''
Phys.\ Rev.\  {\bf 184} (1969) 1750.
\doi{10.1103/PhysRev.184.1750}

\bibitem{Volkov:1973vd}
D.~V.~Volkov,
``Phenomenological Lagrangians,''
Preprint ITF 65-75, Kiev, 1969,
Fiz.\ Elem.\ Chast.\ Atom.\ Yadra {\bf 4} (1973) 3.

\bibitem{Ogievetsky73}
V.I. Ogievetsky, Procs. of X-th Winter School of Theoretical Physics in
Karpacz, v.1, p.107, Wroclaw 1974.

\bibitem{Cremmer:1979up}
E.~Cremmer and B.~Julia,
``The SO(8) Supergravity,''
Nucl.\ Phys.\ B {\bf 159} (1979) 141.
\doi{10.1016/0550-3213(79)90331-6}.

\bibitem{de Wit:1981eq}
B.~de Wit and H.~Nicolai,
``N=8 Supergravity with Local SO(8) x SU(8) Invariance,''
Phys.\ Lett.\ B {\bf 108} (1982) 285.
\doi{10.1016/0370-2693(82)91194-7}

\bibitem{Galperin:2001uw}
A.~S.~Galperin, E.~A.~Ivanov, V.~I.~Ogievetsky and E.~S.~Sokatchev,
``Harmonic superspace,''
Cambridge, UK: Univ. Pr. (2001) 306 p

\bibitem{Andrianopoli:1996cm}
L.~Andrianopoli, M.~Bertolini, A.~Ceresole, R.~D'Auria, S.~Ferrara,
P.~Fr\'e and T.~Magri,
``{\sl N=2 supergravity and N=2 superYang-Mills theory on general scalar manifolds: Symplectic covariance, gaugings and the momentum map,}''
J.\ Geom.\ Phys.\  {\bf 23} (1997) 111.
\doi{10.1016/S0393-0440(97)00002-8}.
[\hepth{9605032}].

\bibitem{Freedman:2012zz}
D.~Z.~Freedman and A.~Van Proeyen,
``{\sl Supergravity,}''
Cambridge, UK: Cambridge Univ. Pr. (2012) 607 p

\bibitem{Gibbons:1993ap}
G.~W.~Gibbons, R.~H.~Rietdijk and J.~W.~van Holten,
``SUSY in the sky,''
Nucl.\ Phys.\ B {\bf 404} (1993) 42.
\doi{10.1016/0550-3213(93)90472-2}.
[\hepth{9303112}].

\bibitem{vanHolten:1995qt}
J.~W.~van Holten,
``D = 1 supergravity and spinning particles,''
In *Jancewicz, B. (ed.): Sobczyk, J. (ed.): From field theory to
quantum groups* 173-189, and Amsterdam NIKHEF - NI
[\hepth{9510021}].

\bibitem{Galperin:1984av}
A.~Galperin, E.~Ivanov, S.~Kalitsyn, V.~Ogievetsky and E.~Sokatchev,
``Unconstrained N=2 Matter, Yang-Mills and Supergravity Theories in Harmonic Superspace,''
Class.\ Quant.\ Grav.\  {\bf 1} (1984) 469
Erratum: [Class.\ Quant.\ Grav.\  {\bf 2} (1985) 127].
\doi{10.1088/0264-9381/1/5/004}.

\bibitem{Bandos:1992np}
I.~A.~Bandos and A.~A.~Zheltukhin,
``Green-Schwarz superstrings in spinor moving frame formalism,''
Phys.\ Lett.\ B {\bf 288} (1992) 77.
\doi{10.1016/0370-2693(92)91957-B}.

\bibitem{Bandos:1993yc}
I.~A.~Bandos and A.~A.~Zheltukhin,
``Eleven-dimensional supermembrane in a spinor moving repere formalism,''
Int.\ J.\ Mod.\ Phys.\ A {\bf 8} (1993) 1081.
\doi{10.1142/S0217751X93000424}.

\bibitem{Bandos:2016tsm}
I.~Bandos,
``BCFW-type recurrent relations for tree amplitudes of D=11 supergravity,''
\arxiv{1605.00036} [hep-th].

\bibitem{Schwarz:1983wa}
J.~H.~Schwarz and P.~C.~West,
``Symmetries and Transformations of Chiral N=2 D=10 Supergravity,''
Phys.\ Lett.\ B {\bf 126} (1983) 301.
\doi{10.1016/0370-2693(83)90168-5}.

\bibitem{Schwarz:1983qr}
J.~H.~Schwarz,
``Covariant Field Equations of Chiral N=2 D=10 Supergravity,''
Nucl.\ Phys.\ B {\bf 226} (1983) 269.
\doi{10.1016/0550-3213(83)90192-X}.

\bibitem{Howe:1983sra}
P.~S.~Howe and P.~C.~West,
``The Complete N=2, D=10 Supergravity,''
Nucl.\ Phys.\ B {\bf 238} (1984) 181.
\doi{10.1016/0550-3213(84)90472-3}.

\bibitem{Bergshoeff:1995as}
E.~Bergshoeff, C.~M.~Hull and T.~Ort\'{\i}n,
``Duality in the type II superstring effective action,''
Nucl.\ Phys.\ B {\bf 451} (1995) 547.
\doi{10.1016/0550-3213(95)00367-2}.
[\hepth{9504081}].

\bibitem{Bergshoeff:2005ac}
E.~A.~Bergshoeff, M.~de Roo, S.~F.~Kerstan and F.~Riccioni,
``IIB supergravity revisited,''
JHEP {\bf 0508} (2005) 098.
\doi{10.1088/1126-6708/2005/08/098}
[\hepth{0506013}].

\bibitem{Bergshoeff:2010mv}
E.~A.~Bergshoeff, J.~Hartong, P.~S.~Howe, T.~Ort\'{\i}n and F.~Riccioni,
``IIA/IIB Supergravity and Ten-forms,''
JHEP {\bf 1005} (2010) 061.
\doi{10.1007/JHEP05(2010)061}
[\arxiv{1004.1348} [hep-th]].



\bibitem{Kallosh:2008ic}
R.~Kallosh and M.~Soroush,
``Explicit Action of E(7)(7) on N=8 Supergravity Fields,''
Nucl.\ Phys.\ B {\bf 801} (2008) 25.
\doi{10.1016/j.nuclphysb.2008.04.006}.
[\arxiv{0802.4106}]


\bibitem{Cremmer:1978km}
  E.~Cremmer, B.~Julia and J.~Scherk,
  ``Supergravity Theory in Eleven-Dimensions,''
  Phys.\ Lett.\ B {\bf 76} (1978) 409.
  \doi{10.1016/0370-2693(78)90894-8}


\bibitem{Nicolai:1980kb}
  H.~Nicolai, P.~K.~Townsend and P.~van Nieuwenhuizen,
  ``Comments On Eleven-dimensional Supergravity,''
  Lett.\ Nuovo Cim.\  {\bf 30} (1981) 315.
  \doi{10.1007/BF02817085}

\bibitem{D'Auria:1982nx}
  R.~D'Auria and P.~Fre,
  ``Geometric Supergravity in d = 11 and Its Hidden Supergroup,''
  Nucl.\ Phys.\ B {\bf 201} (1982) 101
   Erratum: [Nucl.\ Phys.\ B {\bf 206} (1982) 496].
  \doi{10.1016/0550-3213(82)90376-5},  \doi{10.1016/0550-3213(82)90281-4}

\bibitem{Cremmer:1998px}
  E.~Cremmer, B.~Julia, H.~Lu and C.~N.~Pope,
  ``Dualization of dualities. 2. Twisted self-duality of doubled fields, and superdualities,''
  Nucl.\ Phys.\ B {\bf 535} (1998) 242
  \doi{10.1016/S0550-3213(98)00552-5}
  [hep-th/9806106].

\bibitem{Bergshoeff:2007ij}
E.~A.~Bergshoeff, J.~Hartong, M.~H\"ubscher and T.~Ort\'{\i}n,
``Stringy cosmic strings in matter coupled N=2, d=4 supergravity,''
JHEP {\bf 0805} (2008) 033.
\doi{10.1088/1126-6708/2008/05/033}.
[\arxiv{0711.0857} [hep-th]].

\bibitem{Hartong:2009az}
J.~Hartong, M.~H\"ubscher and T.~Ort\'{\i}n,
``The Supersymmetric tensor hierarchy of N=1,d=4 supergravity,''
\doi{10.1088/1126-6708/2009/06/090}.
JHEP {\bf 0906} (2009) 090.
[\arxiv{0903.0509} [hep-th]].

\bibitem{Bandos:2015ila}
I.~Bandos and T.~Ort\'{\i}n,
``Tensor gauge fields of N=8 supergravity,''
Phys.\ Rev.\ D {\bf 91} (2015) 085031.
\doi{10.1103/PhysRevD.91.085031}.
[\arxiv{1502.00649} [hep-th]].

\bibitem{Bergshoeff:2006gs}
E.~A.~Bergshoeff, M.~de Roo, S.~F.~Kerstan, T.~Ort\'{\i}n and F.~Riccioni,
``SL(2,R)-invariant IIB Brane Actions,''
JHEP {\bf 0702} (2007) 007.
\doi{10.1088/1126-6708/2007/02/007}.
[\hepth{0611036}].

\bibitem{Billo:1999ip}
M.~Bill\'o, S.~Cacciatori, F.~Denef, P.~Fr\'e, A.~Van Proeyen and D.~Zanon,
``The $0$-brane action in a general D = 4 supergravity background,''
Class.\ Quant.\ Grav.\  {\bf 16} (1999) 2335.
\doi{10.1088/0264-9381/16/7/313}.
[\hepth{9902100}].

\bibitem{Bergshoeff:2011zk}
E.~A.~Bergshoeff and F.~Riccioni,
``String Solitons and T-duality,''
JHEP {\bf 1105} (2011) 131.
\doi{10.1007/JHEP05(2011)131}.
[\arxiv{1102.0934} [hep-th]].

\bibitem{Bergshoeff:2010xc}
E.~A.~Bergshoeff and F.~Riccioni,
``D-Brane Wess-Zumino Terms and U-Duality,''
JHEP {\bf 1011} (2010) 139.
\doi{10.1007/JHEP11(2010)139}.
[\arxiv{1009.4657} [hep-th]].

\bibitem{Cordaro:1998tx}
F.~Cordaro, P.~Fr\'e, L.~Gualtieri, P.~Termonia and M.~Trigiante,
``N=8 gaugings revisited: An Exhaustive classification,''
Nucl.\ Phys.\ B {\bf 532} (1998) 245.
\doi{10.1016/S0550-3213(98)00449-0}.
[\hepth{9804056}].

\bibitem{Nicolai:2000sc}
H.~Nicolai and H.~Samtleben,
``Maximal gauged supergravity in three-dimensions,''
Phys.\ Rev.\ Lett.\  {\bf 86} (2001) 1686.
\doi{10.1103/PhysRevLett.86.1686}.
[\hepth{0010076}].

\bibitem{Nicolai:2001sv}
H.~Nicolai and H.~Samtleben,
``Compact and noncompact gauged maximal supergravities in three-dimensions
,''
JHEP {\bf 0104} (2001) 022.
\doi{10.1088/1126-6708/2001/04/022}
[\hepth{0103032}].

\bibitem{deWit:2002vt}
B.~de Wit, H.~Samtleben and M.~Trigiante,
``On Lagrangians and gaugings of maximal supergravities,''
Nucl.\ Phys.\ B {\bf 655} (2003) 93.
\doi{10.1016/S0550-3213(03)00059-2}
[\hepth{0212239}].

\bibitem{deWit:2005ub}
B.~de Wit, H.~Samtleben and M.~Trigiante,
``Magnetic charges in local field theory,''
JHEP {\bf 0509} (2005) 016.
\doi{10.1088/1126-6708/2005/09/016}.
[\hepth{0507289}].

\bibitem{Andrianopoli:1996ve}
L.~Andrianopoli, R.~D'Auria and S.~Ferrara,
``U duality and central charges in various dimensions revisited
,''
Int.\ J.\ Mod.\ Phys.\ A {\bf 13} (1998) 431.
\doi{10.1142/S0217751X98000196}.
[\hepth{9612105}].

\bibitem{Castellani:1985ka}
L.~Castellani, A.~Ceresole, S.~Ferrara, R.~D'Auria, P.~Fr\'e and E.~Maina,
``The Complete $N=3$ Matter Coupled Supergravity,''
Nucl.\ Phys.\ B {\bf 268} (1986) 317.
\doi{10.1016/0550-3213(86)90157-4}

\bibitem{Castellani:1985wk}
L.~Castellani, A.~Ceresole, R.~D'Auria, S.~Ferrara, P.~Fr\'e and E.~Maina,
``$\sigma$ Models, Duality Transformations and Scalar Potentials in
Extended Supergravities,''
Phys.\ Lett.\ B {\bf 161} (1985) 91.
\doi{10.1016/0370-2693(85)90615-X}.

\bibitem{Salam:1984ft}
A.~Salam and E.~Sezgin,
``d = 8 Supergravity,''
Nucl.\ Phys.\ B {\bf 258} (1985) 284.
\doi{10.1016/0550-3213(85)90613-3}.

\bibitem{AlonsoAlberca:2003jq}
N.~Alonso Alberca, E.~Bergshoeff, U.~Gran, R.~Linares, T.~Ort\'{\i}n and D.~Roest,
``Domain walls of D = 8 gauged supergravities and their D = 11 origin,''
JHEP {\bf 0306} (2003) 038.
\doi{10.1088/1126-6708/2003/06/038}.
[\hepth{0303113}].

\bibitem{Andino:2016bwk}
O.~L.~Andino and T.~Ort\'{\i}n,
``The tensor hierarchy of 8-dimensional field theories,''
\arxiv{1605.05882} [hep-th].


\end{thebibliography}
\end{document}